\documentclass[final,3p,times]{elsarticle}
\usepackage{graphicx}
\usepackage{amssymb,amsmath}
\biboptions{sort,compress}
\journal{Annals of Physics}

\begin{document}

\begin{frontmatter}

\title{Similarity Renormalization Group Evolution of Chiral Effective Nucleon-Nucleon Potentials in the Subtracted Kernel Method Approach}

\author[1]{S. Szpigel}
\author[2]{V. S. Tim\'oteo}
\author[1]{F. de O. Dur\~aes}

\address[1]{Centro de Ci\^encias e Humanidades, Universidade Presbiteriana Mackenzie,\\
01302-907, S\~ao Paulo - SP - Brazil}
\address[2]{Faculdade de Tecnologia, Universidade Estadual de Campinas, \\
13484-332, Limeira - SP - Brazil}

\begin{abstract}

Methods based on Wilson's renormalization group have been successfully applied in the context of nuclear physics to analyze the scale dependence of effective nucleon-nucleon ($NN$) potentials, as well as to consistently integrate out the high-momentum components of phenomenological high-precision $NN$ potentials in order to derive phase-shift equivalent softer forms, the so called $V_{low-k}$ potentials. An alternative renormalization group approach that has been applied in this context is the Similarity Renormalization Group (SRG), which is based on a series of continuous unitary transformations that evolve hamiltonians with a cutoff on energy differences. In this work we study the SRG evolution of a leading order (LO) chiral effective $NN$ potential in the $^1 S_0$ channel derived within the framework of the Subtracted Kernel Method (SKM), a renormalization scheme based on a subtracted scattering equation.

\end{abstract}


\end{frontmatter}

\section{Introduction}
\label{INTRO}

Chiral Effective Field Theory (ChEFT) offers a promising framework for a systematic and model-independent description of hadronic and nuclear systems and their interactions in the low-energy regime that is consistent with Quantum Chromodynamics (QCD), the fundamental theory of strong interactions (for detailed reviews, see e.g. Refs. \cite{bedaque1,epelbaum1,machleidt1,epelbaum2}). The construction of such a framework began with the advent of Chiral Perturbation Theory (ChPT), proposed by Weinberg \cite{weinberg1} and successfully implemented in calculations of various low-energy processes for pion-pion ($\pi \pi$) and pion-nucleon ($\pi N$) systems \cite{chpt1,chpt2,chpt3,chpt4,chpt5,chpt6}. As shown by Weinberg in his seminal paper \cite{weinberg1}, starting from the most general lagrangian for pions and nucleons consistent with the underlying symmetries of QCD, in particular the approximate and spontaneously broken $SU(2)_{\rm L} \times SU(2)_{\rm R}$ chiral symmetry, a systematic low-energy expansion can be derived for the $S$-matrix in terms of powers $\nu$ of the ratio ($Q/ \Lambda_{\chi}$), where $Q$ stands for a low-momentum scale associated with external pion and nucleon momenta or the pion mass and $\Lambda_{\chi} \sim 1 \; {\rm GeV}$ is the chiral symmetry breaking scale. For purely-pionic and single-nucleon systems the interactions become weak in the low-energy regime, due to the Goldstone boson nature of pions, ensuring that the chiral expansion of the $S$-matrix converges. Such a perturbative expansion corresponds to an infinite number of Feynman diagrams which can be organized according to a power counting scheme based on naive dimensional analysis. At any given order $\nu$ in the chiral expansion, there is only a finite number of diagrams. This power counting scheme, known as {\it Weinberg's power counting}, is an essential ingredient for systematic and controlled calculations of low-energy observables in the ChPT approach, which allows to select the relevant diagrams contributing to the $S$-matrix up to a given order and to estimate the truncation errors involved in the calculations.

The success of the ChPT approach in the $\pi \pi$ and $\pi N$ sectors led Weinberg \cite{weinberg2,weinberg3,weinberg4} to propose its extension to describe the nuclear forces. However, there is a fundamental difficulty with the direct application of ChPT methods in the few-nucleon sector, where the interactions are strong in the low-energy regime. The existence of shallow nuclear bound states (e.g. the deuteron) and large scattering lengths in $S$-wave channels clearly indicates the non-perturbative nature of the nucleon-nucleon ($NN$) interaction at low-energies. Thus, a purely perturbative treatment of few-nucleon systems is expected to fail.

Weinberg pointed out that the non-perturbative dynamics of few-nucleon systems is due to the strong enhancement of the $S$-matrix arising from purely nucleonic intermediate states (infrared enhancement) and then outlined a two-step strategy to perform non-perturbative EFT calculations \cite{weinberg2,weinberg3}. In the first step, the power counting scheme of ChPT (Weinberg's power counting) is applied to an effective nuclear potential rather than directly to the full $S$-matrix. For a $n$-nucleon process, the effective potential is defined as the sum of all possible $n$-nucleon time-ordered perturbation theory diagrams without purely nucleonic intermediate states, the so called {\it irreducible diagrams}, obtained from the chiral expansion of the lagrangian for $\pi \pi$ and $\pi N$ interactions supplemented by a lagrangian for $n$-nucleon contact interactions constrained only by isospin symmetry. In the second step, the effective potential truncated at a given order in the chiral expansion is inserted into the Lippmann-Schwinger (LS) or the Schr{\"o}dinger equation and then iterated to all orders to obtain the full $S$-matrix, thus generating the non-perturbative effects.

The effective nuclear potential includes long-range contributions from pion exchange interactions and short-range contributions parametrized by nucleon contact interactions with an increasing number of derivatives. Weinberg's power counting, used to organize the chiral expansion, naturally implies in the observed hierarchy of the nuclear forces \cite{weinberg4}. At leading order (${\rm LO}$, $\nu=0$), the effective nuclear potential consists of the well known $NN$ one-pion exchange potential (OPEP) plus two non-derivative $NN$ contact interactions. Contributions from multi-pion exchange interactions start at next-to-leading order (${\rm NLO}$, $\nu=2$) and continue through all higher orders. Derivative contact interactions also start to contribute at ${\rm NLO}$ and due to parity appear only in even orders. According to Weinberg's power counting, contributions from three-nucleon ($3N$) and four-nucleon ($4N$) forces should start at ${\rm NLO}$. However, due to additional suppressions and cancelations (both in energy-dependent and energy-independent formulations) the first non-vanishing contributions from $3N$ and $4N$ interactions occur, respectively, at next-to-next-to-leading order (${\rm NNLO}$, $\nu=3$) and next-to-next-to-next-to-leading order (${\rm N^3LO}$, $\nu=4$) \cite{epelbaum1,epelbaum2}. Both pion exchange and contact interaction terms usually grow with momenta and thus can lead to ultraviolet divergences when the effective potential is iterated to all orders in the LS or the Schr{\"o}dinger equation, requiring the use of a non-perturbative regularization and renormalization procedure in order to obtain well-defined finite solutions. Furthermore, multi-pion exchange interactions involve ultraviolet divergent loop integrals which must be consistently regularized and renormalized.

The standard procedure for the non-perturbative renormalization of the $NN$ interaction in the context of Weinberg's approach to ChEFT consists of two steps \cite{epelbaum1,epelbaum2,lepage}. The first step is to solve the regularized LS equation for the scattering amplitude with the $NN$ potential truncated at a given order in the chiral expansion. The most common scheme used to regularize the LS equation is to introduce a sharp or smooth regularizing function that suppresses the contributions from the potential matrix elements for momenta larger than a given cutoff scale, which separates high-energy/short-distance scales and low-energy/long-distance scales, thus eliminating the ultraviolet divergences in the momentum integrals. The second step is to determine the strengths of the contact interactions, the so called low-energy constants (LEC's), by fitting a set of low-energy scattering data. Once the LEC´s are fixed for a given cutoff, the LS equation can be solved to evaluate other observables.  Such a procedure, motivated by Wilson's renormalization group \cite{wilson1,wilson2,wilson3}, relies on the fundamental premise of EFT's (and essentially all renormalization techniques) that physics at low-energy/long-distance scales is insensitive with respect to the details of the dynamics at high-energy/short-distance scales \cite{lepage}, i.e. the relevant high-energy/short-distance effects for describing the low-energy observables can be captured in the cutoff-dependent LEC's. The $NN$ interaction can be considered properly renormalized when the calculated observables are independent of the cutoff scale within the range of validity of the ChEFT or involves a small residual cutoff dependence due to the truncation of the chiral expansion. In the language of Wilson's renormalization group, this means that the LEC's must run with the cutoff scale in such a way that the scattering amplitude becomes (approximately) renormalization group invariant.

Since its pioneering implementation, by Ord\'o\~nez and van Kolck  \cite{ordonez1}, Weinberg's program has been extensively explored to analyze the nuclear forces. Effective $NN$ and $3N$ potentials have been derived to several orders in the chiral expansion and successfully applied in many calculations of nuclear systems \cite{cheft1,cheft2,cheft3,cheft4,cheft5,cheft6,cheft7,cheft8,cheft9,cheft10,cheft11,cheft12,cheft13,cheft14,cheft15,cheft16,cheft17,cheft18,cheft19,
cheft20,cheft21,cheft22,cheft23,cheft24,cheft25,cheft26,cheft27,cheft28,cheft29,cheft30,cheft31,cheft32,cheft33,cheft34}. At ${\rm N^3LO}$, effective nuclear potentials have been constructed \cite{cheft20,cheft23} which provide a remarkably accurate description of low-energy $NN$ scattering phase-shift data and deuteron properties, comparable to that obtained with phenomenological high-precision potentials like the Argonne V18 \cite{argonne} and the Nijmegen \cite{nijmegen}.

In spite of its phenomenological success, conceptual problems have been raised about Weinberg's approach which are related to the formal inconsistency between the naive power counting scheme (Weinberg's power counting) and the proper renormalization of the $NN$ interaction in the non-perturbative regime. Such inconsistency arises from the non-renormalizability of ChEFT in the usual sense: the iteration of the $NN$ potential truncated at a given order in the chiral expansion (using the LS or the Schr{\"o}dinger equation) generates ultraviolet divergences, and hence regularization scale dependencies, which cannot be absorbed by renormalizing the LEC's of the contact interactions included at that same order. Thus, upon cutoff regularization, the limit of infinite cutoff scale cannot be taken while keeping the results cutoff-independent. It is important to observe that in successful calculations of $NN$ systems following Weinberg's approach, such as those described in Refs. \cite{cheft2,cheft3,cheft14,cheft15,cheft16,cheft17,cheft18,cheft19,cheft20,cheft21,cheft22,cheft23}, cutoff scales $\sim 400 - 800 \; {\rm GeV}$ are typically chosen and nearly cutoff-independent stable results are obtained provided the cutoff is varied only over a narrow range of values, in agreement with the prescription advocated by Lepage \cite{lepage} that the cutoff scale should be taken below $\Lambda_{\chi} \sim 1 \; {\rm GeV}$.

The non-perturbative renormalization of the $NN$ interaction in ChEFT has been intensively investigated by many authors \cite{kaplan1,kaplan2,kaplan3,cohen1,cohen2,cohen3,cohen4,cohen5,cohen6,cohen7,cohen8,phillips1,gegelia1,gegelia2,gegelia3,gegelia4,gegelia5,gegelia6,mehen1,mehen2,mehen3,mehen4,bira1,bira2,bira3,bira4,beane1,beane2,
beane3,eiras,nieves,birse1,birse2,birse3,birse4,birse5,birse6,birse7,birse8,birse9,birse10,birse11,birse12,birse13,valderrama1,valderrama2,
valderrama3,valderrama4,valderrama5,valderrama6,valderrama7,valderrama8,valderrama9,valderrama10,valderrama11,epelbaum3,epelbaum4,machleidt2,machleidt3}, with alternative approaches and power counting schemes being proposed, generating a great deal of discussion and controversy. In particular, methods based on Wilson's renormalization group have been successfully applied to analyze the scale dependence of effective $NN$ potentials, both in momentum \cite{birse4,birse5,birse6,birse7,birse8,birse9,birse10,birse11,birse12,birse13} and in coordinate space \cite{valderrama2,valderrama6}, providing a better understanding of the interplay between the power counting and the renormalization of the $NN$ interaction in the non-perturbative regime. Although much progress has been made, the construction of a consistent scheme for the non-perturbative renormalization of the $NN$ interaction still remains an open problem \cite{machleidt3}.

Renormalization group methods have also been used to consistently integrate out the high-momentum components of both ChEFT and high-precision $NN$ potentials in order to derive phase-shift equivalent softer forms, the so called $V_{low-k}$ potentials \cite{vlow1,vlow2,vlow3,vlow4,vlow5,vlow6,vlow7,vlow8,vlow9,vlow10}. An alternative renormalization group approach that has been applied in this context is the Similarity Renormalization Group (SRG), developed by Glazek and Wilson \cite{wilgla1,wilgla2} and independently by Wegner \cite{wegner}. The SRG is a renormalization method based on a series of continuous unitary transformations that evolve hamiltonians with a cutoff on energy differences. Such transformations are the group elements that give the method its name. Viewing the hamiltonian as a matrix in a given basis, the similarity transformations suppress off-diagonal matrix elements as the cutoff is lowered, forcing the hamiltonian towards a band-diagonal form. An important feature of the SRG is that all other operators are consistently evolved by the same unitary transformation. Recently, the SRG has been applied to evolve several $NN$ potentials to equivalent softer forms, effectively decoupling low-energy observables from high-energy degrees of freedom \cite{srg1,srg2,srg3,srg4,srg5}. As shown in these references, such a decoupling is universal and leads to more perturbative $NN$ potentials, improving variational convergence and greatly simplifying calculations in nuclear few- and many-body problems. A recent review on the application of renormalization group methods to nuclear forces and nuclear structure calculations can be found in Ref. \cite{vlowsrg}.

In this work we study the SRG evolution of chiral effective $NN$ potentials derived within the framework of the Subtracted Kernel Method (SKM) \cite{skm1,skm2,skm3,skm4,skm5,skm6}. The SKM is a renormalization scheme designed to treat interactions containing regular and/or singular contact terms which is based on a subtracted scattering equation. Instead of using a cutoff regularizing function, in the SKM the scattering equation is regularized by performing subtractions in the kernel at a given energy scale, while keeping the original interaction intact. An advantage of the SKM approach is that it can be recursively extended to any derivative order of the contact interactions through an iterative process involving multiple subtractions. A similar approach based on subtractive renormalization of the LS equation is described in Refs. \cite{yang1,yang2,yang3}, although there a sharp momentum cutoff is also introduced to regularize the momentum integrals.

This paper is organized as follows. In Section \ref{SRG} we review the basics of the SRG formalism. In Section \ref{DELTA2D} we discuss a simple example in order to illustrate the application of the SRG approach: the Schr{\"o}dinger equation with a two-dimensional Dirac-delta contact interaction. In Section \ref{SRGNN} we review the main aspects involved in the application of the SRG approach to the $NN$ interaction and present the results obtained for the SRG evolution of the Nijmegen potential in the $^1 S_0$ channel. From the SRG evolved potentials we calculate the corresponding phase-shifts and use an exponential regularizing function \cite{srg2,srg3} to verify the explicit decoupling of low-energy observables from the dynamics of the high-energy degrees of freedom. In Section \ref{SKM} we describe the SKM formalism and discuss its recursive extension to any derivative order of the contact interactions. We illustrate the application of the SKM formalism by presenting two simple examples, namely the $NN$ interaction in ${\rm LO}$ ``pionless" EFT, which consists of just a pure Dirac-delta contact interaction, and in ${\rm NLO}$ ``pionless" EFT, which consists of a Dirac-delta contact interaction plus a second order derivative contact interaction. In Section \ref{SRGSKM} we apply the SRG transformation to evolve an effective $NN$ potential obtained by implementing the SKM approach to renormalize the $NN$ interaction. First, we present a detailed and systematic analysis of the SKM approach applied to the $NN$ interaction in ${\rm LO}$ ChEFT. Then, we evolve the SKM-LO ChEFT potential in the $^1 S_0$ channel through the SRG transformation, calculate the corresponding phase-shifts and analyze the decoupling pattern between low- and high-momentum components. Finally, in Section \ref{CONCL} we summarize the results and present our main conclusions.

\section{Similarity Renormalization Group Formalism}
\label{SRG}

\subsection{Glazek-Wilson Formulation}
\label{GWSRGFORM}

The general formulation of the SRG was developed by Glazek and Wilson \cite{wilgla1,wilgla2} in the context of light-front hamiltonian field theory (HLFFT), aiming to obtain effective hamiltonians in which the couplings between high- and low-energy states are eliminated, while avoiding the appearance of artificial divergences in the form of small energy denominators in perturbative expansions. The method is implemented by continuous unitary transformations designed to replace the effects of the removed couplings by scale dependent effective interactions with no small energy denominators, ensuring that the effective hamiltonians produce no ultraviolet divergences. It is important to note that the transformations do not remove any degrees of freedom, but only integrate out couplings between states with energy differences larger than a given scale.

Consider a system described by a canonical hamiltonian written in the form $H=h+V$, where $h$ is the free hamiltonian and $V$ is an interaction. In general, the hamiltonian contains direct couplings between states from all energy scales which can be the source of ultraviolet divergences. In the following discussion, we use the basis of eigenstates of the free hamiltonian, $h |\; i >=\epsilon_i|\; i >$.

Starting from the canonical hamiltonian $H$ we define a {\it bare hamiltonian} $H_{\lambda_0}$ regularized by a large initial cutoff $\lambda_0$ on energy differences at the interaction vertices. Such a cutoff is called {\it similarity cutoff} and here has dimensions of energy. The bare hamiltonian can then be written in the form
\begin{eqnarray}
H_{\lambda_0}&\equiv&h+V_{\lambda_0}\equiv h+f_{\lambda_0} (V+H_{\lambda_0}^{ct}) \;,
\label{hbare}
\end{eqnarray}
\noindent
where $f_{\lambda_0}$ is called {\it similarity function} and $H_{\lambda_0}^{ct}$ are counterterms to be determined through the renormalization process in order to remove dependence in observables upon the cutoff $\lambda_0$. The similarity function $f_{\lambda_0}$ is defined to regularize the hamiltonian by suppressing matrix elements between states with large energy differences. Usually, the similarity function is chosen to be a smooth function of the cutoff, such that
\begin{eqnarray}
&&(i) f_{_{\lambda_0 }} \rightarrow 1, \; {\rm for} \;
|\epsilon_i-\epsilon_j|<< \lambda_0 \; ,\nonumber\\
&&(ii)f_{_{\lambda_0 }}  \rightarrow 0, \; {\rm for} \;
|\epsilon_i-\epsilon_j| >>  \lambda_0 \; .
\label{simfunc}
\end{eqnarray}
\noindent
A simpler choice for the similarity function is a step function $\theta(\lambda_0-|\epsilon_i-\epsilon_j|)$. Although useful in analytic calculations, such a choice can lead to pathologies \cite{billy}.

The next step is to define a unitary transformation $U(\lambda ,\lambda_0)$ that acts on the bare hamiltonian $H_{\lambda_0}$ and changes the similarity cutoff down to a scale $\lambda$, producing a renormalized hamiltonian,
\begin{eqnarray}
H_\lambda&\equiv& U(\lambda ,\lambda_0)\; H_{\lambda_0} \; U^\dagger(\lambda,\lambda_0) \; .
\label{strans}
\end{eqnarray}
\noindent
The unitarity condition satisfied by $U(\lambda ,\lambda_0)$ is given by
\begin{eqnarray}
U(\lambda, \lambda_0)\;  U^\dagger(\lambda, \lambda_0)&\equiv&U^\dagger(\lambda,\lambda_0)\; U(\lambda, \lambda_0)  \equiv 1\;.
\label{unitarity}
\end{eqnarray}

The renormalized hamiltonian $H_\lambda$ is driven towards a band-diagonal form as the similarity cutoff $\lambda$ is lowered and can be written in the form
\begin{eqnarray}
H_\lambda&\equiv& h+f_{\lambda}\;{\overline V}_{\lambda}\;,
\label{hren}
\end{eqnarray}
\noindent
where $f_{\lambda}$ is the similarity function at the cutoff scale $\lambda$ and ${\overline V}_{\lambda}$ is defined as the reduced interaction (i.e., the effective interaction at the cutoff scale $\lambda$ with the similarity function factored out).

The similarity transformation $U(\lambda ,\lambda_0)$ can be defined in terms of an anti-hermitian operator $\eta_{\lambda}$ which generates infinitesimal changes of the cutoff scale $\lambda$,
\begin{eqnarray}
U(\lambda ,\lambda_0)&\equiv&{\cal T} \exp \left(\int_\lambda^{\lambda_0}
\eta_{\lambda^{\prime}}\; d \lambda^{\prime}\right)\; ,
\label{udef}
\end{eqnarray}
\noindent
where ${\cal T}$ puts the operators in order of increasing cutoff scale $\lambda^{\prime}$ from left to right.

Taking the derivative of Eq. (\ref{udef}) and using both the unitarity condition given in Eq. (\ref{unitarity}) and its derivative we get the relation
\begin{equation}
\eta_\lambda=U(\lambda ,\lambda_0)\; \frac{dU^\dagger(\lambda ,\lambda_0)}{d \lambda}
=-\frac{dU(\lambda ,\lambda_0)}{d \lambda}\; U^\dagger(\lambda ,\lambda_0)=-\eta_{\lambda}^\dagger\;.
\label{etadef}
\end{equation}
\noindent
Thus, taking the derivative of Eq. (\ref{strans}) and using the unitarity condition combined with Eq. (\ref{etadef}) we obtain a first-order differential equation for the evolution of the hamiltonian,
\begin{eqnarray}
\frac{d H_{\lambda}}{d \lambda}= \left [ H_\lambda,\eta_\lambda \right]
\label{difstrans}\; ,
\end{eqnarray}
\noindent
which is to be solved with the boundary condition $H_\lambda |_{_{\lambda \rightarrow \lambda_0}} \equiv H_{\lambda_0}$.

The generator $\eta_{\lambda}$ is defined by specifying constraints on the change of the operators $h$ and ${\overline V}_{\lambda}$ with the similarity cutoff $\lambda$. One possible choice is to demand that $h$ is independent of $\lambda$ and that the renormalized hamiltonian does not contain any small energy denominators. Such constraints are defined by the conditions
\begin{eqnarray}
\frac{d h}{d \lambda}\equiv 0 \; \;, \; \; \;
\frac{d {\overline V}_\lambda}{d \lambda}\equiv [V_\lambda,\eta_\lambda]\;.
\label{constraints}
\end{eqnarray}

Since the similarity transformation is unitary, both $H_{\lambda_0}$ and $H_\lambda$ produce the same spectra for the observables. Moreover, when the transformation is implemented in an exact form (i.e., not using a perturbative expansion) the observables calculated using the renormalized hamiltonian $H_\lambda$ are independent of the similarity cutoff $\lambda$. As pointed before, the counterterms $H_{\lambda_0}^{ct}$ are adjusted such that the observables also become independent of $\lambda_0$ in the limit $\lambda_0 \rightarrow \infty$.

An important feature of the SRG is that all operators are consistently renormalized by using the same unitary transformation, i.e. any given operator ${\cal O}$ evolves with the similarity cutoff scale $\lambda$ according to
\begin{eqnarray}
\frac{d {\cal O}_{\lambda}}{d \lambda}= \left [{\cal O}_\lambda,\eta_\lambda \right]
\label{opstrans}\; .
\end{eqnarray}

\subsection{Wegner Formulation}

In the applications described in this work, we employ the formulation for the SRG developed by Wegner \cite{wegner} and applied within the context of many-body problems in condensed matter physics \cite{kehrein}. Wegner's formulation is based on a non-perturbative flow equation that governs the unitary evolution of a hamiltonian with a flow parameter $s$ that ranges from $0$ to $\infty$,
\begin{equation}
\frac{d H_s}{ds}=[\eta_s,H_s]
\label{wegner1}\; ,
\end{equation}
\noindent
which is to be solved with the boundary condition $H_s |_{_{s \rightarrow s_0}} \equiv H_{s_0}$.

The flow equation Eq. (\ref{wegner1}) is analogous to Eq.(\ref{difstrans}), but the specific form $\eta_s=[G_s,H_s]$ is chosen for the anti-hermitian operator that generates the unitary transformation, which gives
\begin{equation}
\frac{d H_s}{ds}=[[G_s,H_s],H_s]\; .
\label{wegner2}
\end{equation}

The operator $G_s$ defines the generator $\eta_s$ and so specifies the flow of the hamiltonian. Wegner's choice in the original formulation is the full diagonal part of the hamiltonian in a given basis, $G_s={\rm diag}(H_s)$. A simpler choice is to use the free hamiltonian, $G_s=h$. Such choices for the generator $\eta_s$ result in a flow parameter $s$ which has dimensions of $({\rm energy})^{-2}$ and correspond in the Glazek-Wilson formulation to the choice of a gaussian similarity function with uniform width $\lambda$. In terms of the similarity cutoff $\lambda$, the flow parameter is then given by the relation $s=\lambda^{-2}$. Although both these choices for $G_s$ are able to force the hamiltonian towards a band-diagonal form as the flow parameter $s$ increases (or as the cutoff $\lambda$ decreases), it has been shown by Glazek and Perry \cite{srgbound} that using $G_s=h$ can produce divergences in theories with bound-states. Such an effect is related to the renormalization group limit cycle behavior \cite{glazek1,glazek2} and happens when $\lambda=1/\sqrt{s}$ approaches a scale at which a bound-state emerges, thus limiting how far the SRG transformation can be implemented. On the other hand, by using $G_s={\rm diag}(H_s)$ the limit cycle behavior can be properly handled and the transformation is guaranteed to converge.

This is an important issue to be considered in the applications of the SRG to nuclear physics. As has been recently shown in several papers \cite{srg1,srg2,srg3,srg4}, the SRG transformation using $G_s=h$ can be applied to evolve the $NN$ interaction with no convergence problems as long as $\lambda$ is kept larger than the bound-state scales. On the other hand, in cases with limit cycle behavior such as the three-nucleon problem \cite{limit1,limit2,limit3,limit4}, the SRG transformation must be implemented by including interactions in $G_s$ \cite{srgbound}.

The consistent renormalization group evolution of many-body nuclear forces is a crucial and challenging issue for nuclear structure calculations using low-momentum interactions \cite{vlowsrg}. As shown in recent works, the SRG flow equations provide an approach to evolve three-body nuclear forces directly, i.e. without having to solve the full three-body problem as required in the $V_{low-k}$ approach. Preliminary studies using simple models \cite{srgmany1,srgmany2} have been carried out  aiming to set up a practical method to consistently evolve three- and many-body interactions with the SRG transformation. In Ref. \cite{srgmany3}, the first application of such a method to evolve three-body nuclear forces has been demonstrated, in calculations of $A \leq 4$ nuclei using a harmonic oscillator basis.

\section{Example: Two-dimensional Dirac-delta Contact Interaction}
\label{DELTA2D}

In order to illustrate the general features of the SRG approach, in this section we discuss a simple example from quantum mechanics. We consider the SRG evolution of the hamiltonian for a system of two non-relativistic particles in two dimensions interacting via a Dirac-delta contact potential \cite{jackiw}-\cite{henderson}. A discretized version of the two-dimensional Dirac-delta contact potential was considered previously by Glazek and Wilson \cite{wilgla3} as a model to study the application of the SRG to an asymptotically free theory. Here we consider the continuous version.

The Schr{\"o}dinger equation in configuration space for the relative motion of two particles in two dimensions with an attractive Dirac-delta contact potential (in units such that $\hbar=1$) is given by
\begin{equation}
-{\bf \nabla}_{\bf r}^2 ~\Psi({\bf r})-\alpha_0 \;\delta^{(2)}({\bf r}) \;
\Psi({\bf r})=E \; \Psi({\bf r})\; .
\label{scheq2d}
\end{equation}
\noindent
The coupling constant $\alpha_0$ is dimensionless, since both the kinetic energy operator and the two-dimensional Dirac-delta contact potential scale as $1/r^2$, and so Eq. (\ref{scheq2d}) is scale invariant (i.e. there is no intrinsic energy or length scale). Although this equation can be easily solved in configuration space by regularizing the Dirac-delta contact potential with a distribution function, we choose to work in momentum space in order to establish a more transparent connection with calculations implemented in the framework of SRG and EFT.

The Schr{\"o}dinger equation in momentum space is given by
\begin{equation}
p^2 \;  \Phi({\bf p})-\frac{\alpha_0}{(2\pi)^2}\; \int \; d^{2}q
\;\;\Phi({\bf q})=E \; \Phi({\bf p}) \; ,
\label{semom2}
\end{equation}
\noindent
where $p$ is the relative momentum and $\Phi({\bf p})$ is the Fourier transform of the configuration space wave-function $\Psi({\bf r})$.

Scale invariance implies that if a bound-state solution exists (with $E<0$) then Eq. (\ref{semom2}) will admit solutions for any negative energy, which correspond to a continuum of bound-states with energies extending down to $-\infty$. The eigenvalue condition for the binding energy $E_0 =-E$ is given by
\begin{equation}
1=\frac{\alpha_0}{2\pi}\;  \int_{0}^{\infty} \; dp \; p \;
\frac{1}{(p^2+E_0)} \; .
\label{bsi2}
\end{equation}
\noindent
The integral in the r.h.s. of Eq. (\ref{bsi2}) diverges logarithmically and so the problem is ill-defined. Nevertheless, using standard regularization and renormalization schemes we can obtain an exact analytic solution \cite{jackiw}.

\subsection{Standard Solution}
\label{STDDELTA2D}

We start by regularizing the integral in Eq. (\ref{bsi2}) using an ultraviolet momentum cutoff $\Lambda$,
\begin{equation}
1=\frac{\alpha_0}{2\pi}\;  \int_{0}^{\Lambda} \; dp \; p \;
\frac{1}{(p^2+E_0)}=\frac{\alpha_0}{4\pi}\; {\rm
ln}\left(1+\frac{\Lambda^2}{E_0}\right) \; .
\end{equation}
\noindent
Solving this equaion for the binding energy $E_0$ we obtain
\begin{equation}
E_0= \Lambda^2 \;(e^{4\pi/\alpha_0}-1)^{-1} \; .
\end{equation}

Clearly, if the coupling constant $\alpha_0$ is fixed, then $E_0 \rightarrow \infty$ when $\Lambda \rightarrow \infty$. To eliminate the divergence and produce a well-defined bound-state, we can renormalize the coupling constant by demanding that it depends on the cutoff $\Lambda$ such that the binding energy $E_0$ remains fixed as $\Lambda \rightarrow \infty$,
\begin{equation}
\alpha_0 \rightarrow \alpha_{\Lambda}=\frac{4\pi}{{\rm ln}\left(1+\frac{\Lambda^2}{E_0}\right)} \;.
\label{couprun}
\end{equation}
\noindent
Thus, the dimensionless renormalized running coupling $\alpha_{\Lambda}$ that characterizes the strength of the Dirac-delta contact interaction is replaced by a dimensionful parameter, the binding energy $E_0$, which can be arbitrarily chosen and fixes the energy scale of the renormalized theory. This is a simple example of {\it dimensional transmutation} \cite{coleman}. The original bare hamiltonian is scale invariant, but the renormalization process introduces an energy scale that characterizes the observables of the theory. It is also interesting to note that the renormalized theory is asymptotically free, since the renormalized running coupling $\alpha_{\Lambda}$ vanishes in the limit $\Lambda \rightarrow \infty$.

We can use the renormalized interaction to evaluate other observables. The usual prescription for the calculations is to obtain the solutions with the
cutoff in place and then take the limit $\Lambda \rightarrow \infty$. If an exact calculation is implemented, the final results must be independent of the regularization and renormalization schemes. As an example, we evaluate the scattering phase-shifts from the solution of the LS for the $T$-matrix with the renormalized interaction,
\begin{equation}
T({\bf p},{\bf p'};k^2)=V({\bf p},{\bf p'})+\int \; d^{2}q \; \; \frac{V({\bf
p},{\bf q})}{k^2-q^2+i \; \epsilon} \; T({\bf q},{\bf p'};k^2) \; ,
\end{equation}
\noindent
where $k$ is the on-shell momentum in the center-of-mass frame.

Only S-wave ($l=0$) scattering occurs, since for the higher waves the centrifugal barrier completely screens the Dirac-delta contact potential. So, we can integrate over the angular variable to obtain
\begin{equation}
T^{(\rm l=0)}_{\Lambda}(p,p';k^2)=V^{(\rm l=0)}_{\Lambda}(p,p')+\int_{0}^{\Lambda} \; dq \; q \;
\frac{V^{(\rm l=0)}_{\Lambda}(p,q)}{k^2-q^2+i \; \epsilon} \; T^{(\rm l=0)}_{\Lambda}(q,p';k^2) \; ,
\label{TLSdelta}
\end{equation}
\noindent
where $V^{(\rm l=0)}_{\Lambda}(p,p')=-\alpha_{\Lambda}/2\pi$. The LS equation for the on-shell $T$-matrix ($p=p'=k$) is given by
\begin{equation}
T^{(\rm l=0)}_{\Lambda}(k,k;k^2)=-\frac{\alpha_{\Lambda}}{2\pi}-\frac{\alpha_{\Lambda}}{2\pi}\;
T^{(\rm l=0)}_{\Lambda}(k,k;k^2) \; \int_{0}^{\Lambda} \; dq \; q \;  \frac{1}{k^2-q^2+i \; \epsilon} \;.
\end{equation}
\noindent
Solving this equation, we obtain
\begin{equation}
T^{(\rm l=0)}_{\Lambda}(k,k;k^2)= -2\;\left[{\rm ln}\left(\frac{k^2}{E_0}\right)+ {\rm ln}\left(\frac{\Lambda^2+E_0}{\Lambda^2+k^2}\right)-i \; \pi\right]^{-1}\; .
\label{TMATDELTAD1}
\end{equation}

The exact on-shell $T$-matrix is then obtained by taking the limit $\Lambda \rightarrow \infty$:
\begin{equation}
T^{(\rm l=0)}(k,k;k^2)= -2\;\left[{\rm ln}\left(k^2/E_0\right)-i \; \pi\right]^{-1}\; .
\label{TMATDELTAD2}
\end{equation}

A more convenient way to evaluate scattering observables is to use the $K$-matrix (reactance matrix). The LS equation for the $K$-matrix is similar to the one for the $T$-matrix, except that standing-wave boundary conditions are imposed for the Green's function. In this way, the $i \epsilon$ prescription used in the LS equation for the $T$-matrix is replaced by the principal value, such that the $K$-matrix is real. The S-wave LS equation for the $K$-matrix is given by
\begin{equation}
K^{(\rm l=0)}_{\Lambda}(p,p';k^2)=V^{(\rm l=0)}_{\Lambda}(p,p')+ {\cal P}  \int_{0}^{\Lambda} \; dq \; q \;
\frac{V^{(\rm l=0)}_{\Lambda}(p,q)}{k^2-q^2} \; K^{(\rm l=0)}_{\Lambda}(q,p';k^2) \; ,
\label{KLSdelta}
\end{equation}
\noindent
where ${\cal P}$ denotes the principal value. The relation between the exact on-shell $K$-matrix and $T$-matrix is given by,
\begin{equation}
K^{-1}(k,k;k^2)= T^{-1}(k,k;k^2) - i\; \pi/2 \; .
\end{equation}
\noindent
For simplicity, here and in what follows we drop the superscript ${\rm l=0}$.

Using the relation between the on-shell $T$-matrix and the phase-shifts,
\begin{equation}
k \;{\rm cot}\; {\delta}(k)-i \; k=-\frac{2 k}{\pi} \; T^{-1}(k,k;k^2) \; ,
\end{equation}
\noindent
we obtain the exact phase-shifts,
\begin{equation}
{\rm cot} \; \delta(k)=\frac{1}{\pi}\; {\rm ln}\left( k^2/E_0 \right)\;.
\label{exactpsdeltaD2}
\end{equation}

It is important to note that keeping a finite momentum cutoff $\Lambda$ in place leads to cutoff-dependent inverse power-law errors in the inverse on-shell $T$-matrix and hence in the phase-shifts. For $\Lambda >> \sqrt{E_0}, k$, such errors can be evaluated perturbatively by double-expanding the difference between the $\Lambda$-dependent inverse on-shell $T$-matrix and the exact inverse on-shell $T$-matrix in powers of $\sqrt{E_0}/{\Lambda}$ and $k/{\Lambda}$,
\begin{equation}
\Delta T^{-1}=T^{-1}_{\Lambda}(k,k;k^2)-T^{-1}(k,k;k^2)= -\frac{(E_0+k^2)}{2 \Lambda^2}+\frac{(E_0^2-k^4)}{4 \Lambda^4} + {\cal O}\left[\frac{E_0^3}{\Lambda^6};\frac{k^6}{\Lambda^6}\right].
\label{DINVTMATDELTAD2}
\end{equation}

\subsection{SRG Evolution}
\label{SRGDELTA2D}

The hamiltonian in momentum space for the relative motion of two non-relativistic particles in two dimensions with an attractive Dirac-delta contact potential can be written in the form
\begin{equation}
H({\bf p},{\bf p'})=T_{\rm rel}({\bf p},{\bf p'})+V({\bf p},{\bf p'}) \; ,
\end{equation}
\noindent
where $T_{\rm rel}({\bf p},{\bf p'})= p^2 \delta^{(2)}({\bf p}-{\bf p'})$ is the free hamiltonian (kinetic energy) and $V({\bf p},{\bf p'})=-{\alpha_0}/(2\pi)^2$ corresponds to the Fourier transform of the Dirac-delta contact potential in configuration space. One should note that the matrix elements of the Dirac-delta contact potential in momentum space are constants, clearly showing that the hamiltonian directly couples an infinite number of scales. Such couplings are precisely the source of the logarithmic divergences that appear in the Schr{\"o}dinger equation, as shown in the previous subsection.

We apply the SRG approach to renormalize this hamiltonian using Wegner's formulation. Choosing the generator $\eta_s=[T_{\rm rel},H_s]$ and integrating out the angular variable, we obtain the flow equation for the relative momentum space matrix elements of the potential in the basis of the free hamiltonian eigenstates, $T_{\rm rel} |p >=p^2|p >$, given by
\begin{eqnarray}
\frac{dV_s(p,p')}{ds}=-(p^2-p'^2)^2 \; V_{s}(p,p')+\int_{0}^{\infty}dq \; q \;
(p^2+p'^2-2 q^2) V_{s}(p,q) V_{s}(q,p').
\label{flowdelta}
\end{eqnarray}
\noindent
For convenience, we consider here and in the rest of the paper that the flow parameter $s$ is related to a similarity cutoff $\lambda$ with dimensions of momentum, such that $s=\lambda^{-4}$.

In principle, we could impose the boundary condition at $s_0=0$ ($\lambda_0 \rightarrow \infty$),
\begin{equation}
H_{s_0=0}(p,p') = H(p,p') = T_{\rm rel}(p,p')+V(p,p')\;.
\end{equation}
\noindent
where $T_{\rm rel}(p,p')=p^2 \delta(p-p')$ and $V(p,p')=-{\alpha_0}/2\pi$. However, the bare hamiltonian (i.e., with no similarity cutoff) produces logarithmic divergences and so the boundary condition should be imposed at some other value $s_0 \neq 0$, leading to dimensional transmutation. Formally, a renormalization prescription must be specified that allows one to obtain a well-defined hamiltonian at $s=s_0$ and thus fixes the underlying theory.

A simple prescription is to use the renormalized running coupling $\alpha_{\Lambda}$ derived in subsection (\ref{STDDELTA2D}) to define the potential in the bare hamiltonian. We introduce an ultraviolet momentum cutoff $\Lambda$ and set
\begin{equation}
V(p,p') \rightarrow V_{\Lambda}(p,p')=-\frac{\alpha_{\Lambda}}{2\pi}=-\frac{2}{{\rm ln}\left(1+\frac{\Lambda^2}{E_0}\right)} \; ,
\end{equation}
\noindent
Using this prescription, the bare hamiltonian becomes well-defined and the boundary condition can be imposed at $s_0=0$. Note that the underlying theory is fixed at $\lambda_0 \rightarrow \infty$ by fitting the binding energy $E_0$.

In previous works \cite{szpigel1}-\cite{szpigel3} we solved the flow equation Eq. (\ref{flowdelta}) perturbatively, using the idea of ``coupling-coherence" \cite{oehme1}-\cite{coupcoh2}. We assumed a solution in the form of an expansion in powers of $\alpha_{s,\Lambda}/2\pi$,
\begin{equation}
V_{s,\Lambda}(p,p')=\left[-\frac{\alpha_{s,\Lambda}}{2\pi}+\sum_{n=2}^{\infty}\left(\frac{\alpha_{s,\Lambda}}{
2\pi}\right)^{n}\; F^{(n)}_{s,\Lambda}(p,p')\right] \; e^{-s(p^2-p'^2)^2}\; ,
\label{coc2}
\end{equation}
\noindent
where $F^{(n)}_{s,\Lambda}(p,p')$ are irrelevant operators that vanish when $p=p'=0$, and solved Eq. (\ref{flowdelta}) iteratively, obtaining analytic expressions order-by-order for the perturbative approximations of the renormalized potential, such as:
\vspace{0.5cm}

\noindent
(a) marginal operator with initial coupling $\alpha_{\Lambda}$,
\begin{equation}
V_{s,\Lambda}(p,p')=-\frac{\alpha_{\Lambda}}{2\pi}\; e^{-s(p^2-p'^2)^2} \; ;
\label{coc2a}
\end{equation}
\noindent
(b) marginal operator with second-order running coupling $\alpha_{s,\Lambda}^{(2)}$,
\begin{equation}
V_{s,\Lambda}(p,p')=-\frac{\alpha_{s,\Lambda}^{(2)}}{2\pi}\; e^{-s(p^2-p'^2)^2} \; ;
\label{coc2b}
\end{equation}
\noindent
(c) marginal operator plus second-order irrelevant operator $F^{(2)}_{s,\Lambda}$ with second-order running coupling $\alpha_{s,\Lambda}^{(2)}$,
\begin{equation}
V_{s,\Lambda}(p,p')=\left[-\frac{\alpha_{s,\Lambda}^{(2)}}{2\pi}+\left(\frac{\alpha_{s,\Lambda}^{(2)}}{2\pi}\right)^2 \; F^{(2)}_{s,\Lambda}(p,p')\right] \; e^{-s(p^2-p'^2)^2} \; ;
\label{coc2c}
\end{equation}
\noindent
Then, we used these approximations in a sequence of bound-state calculations and analyzed the scaling of the binding energy errors with the similarity cutoff $\lambda = s^{-1/4}$ at each order.

In Refs. \cite{szpigel1,szpigel2} a comparison was made with bound-state calculations based on Lepage's treatment of EFT \cite{lepage}. For simplicity, a separable effective potential was used that in momentum space consists of the Fourier transform of a Dirac-delta potential regularized by a smooth function of a momentum cutoff $\Lambda$ and a series of approximately local effective interactions which correspond to the derivatives of the Dirac-delta potential,
\begin{equation}
V^{\rm eff}_{\Lambda}(p,p')=\left[C_0(\Lambda) + C_2(\Lambda) \;
\frac{(p^2+p'^2)}{2\Lambda^2}+C_4(\Lambda) \;
\frac{(p^4+p'^4)}{4\Lambda^4}+C'_4(\Lambda) \; \frac{p^2 \; p'^2}{2\Lambda^4}+
\dots \right] \;  e^{-p^2/2\Lambda^2}\; e^{-p'^2/2\Lambda^2} \; .
\label{pe}
\end{equation}
\noindent
Following the method described by Steele and Furnstahl \cite{steele1, steele2}, the parameters $C_i(\Lambda)$ were determined order-by-order by fitting the values for the exact inverse on-shell $K$-matrix. The method consists in fitting the difference between the inverse on-shell $K$-matrix obtained from the effective potential at a given order and the exact inverse on-shell $K$-matrix to an interpolating polynomial in $k^2/\Lambda^2$,
\begin{equation}
\Delta K^{-1}=K^{-1}_{\Lambda}(k,k;k^2)-K^{-1}(k,k;k^2)= A_0 + A_2\;\frac{k^2}{\Lambda^2} + A_4\;\frac{k^4}{\Lambda^4} + \cdots
\end{equation}
\noindent
and then minimizing the coefficients $A_i$ with respect to the variations of the parameters $C_i(\Lambda)$.

The results from the error analysis have shown that for the EFT calculations the errors scale like inverse powers of the momentum cutoff $\Lambda$, as expected since the effective interactions are irrelevant operators. A systematic power-law improvement is obtained as each term is added to the effective potential and the corresponding parameter $C_i(\Lambda)$ is adjusted by the fitting procedure. For the SRG calculations, the error analysis displayed two different scaling regimes. When the similarity cutoff $\lambda$ is larger than the momentum cutoff $\Lambda$ (introduced to define the bare hamiltonian) the scaling is similar to that for EFT, i.e. the errors are power-law. When the similarity cutoff $\lambda$ becomes smaller than the momentum cutoff $\Lambda$, in addition to the power-law errors there are inverse logarithmic errors that result from the truncation of the perturbative expansion Eq. (\ref{coc2}) for the renormalized potential at a given order in $\alpha_{s,\Lambda}$.

Although not trivial, the scaling behavior of the errors in the perturbative SRG can be qualitatively understood through fairly simple analytical calculations. By regrouping the terms in the SRG potential given by Eq. (\ref{coc2}) we can formally write it as a momentum expansion similar to the EFT potential given by Eq. (\ref{pe}),
\begin{equation}
V_{s,\Lambda}(p,p')=  \left[g_0(\alpha_{s,\Lambda})+g_2(\alpha_{s,\Lambda})\sqrt{s}\; \;\frac{( p^2+p'^2)}{2}+ \cdots  \right]\; e^{-s(p^2-p'^2)^2}\; ,
\label{mexp}
\end{equation}
\noindent
where, in this case, the expansion parameters $g_i(\alpha_{s,\Lambda})$ are analytic functions of the running coupling $\alpha_{s,\Lambda}$,
\begin{equation}
g_i(\alpha_{s,\Lambda})=a_i(s,\Lambda)\; \alpha_{s,\Lambda}+b_i(s,\Lambda)\; \alpha_{s,\Lambda}^2 + \cdots \; .
\end{equation}

From this expansion, we can identify two interdependent sources of perturbative errors in the calculations of observables using the SRG potential truncated at a given order in the running coupling $\alpha_{s,\Lambda}$. First, there are inverse logarithmic errors introduced by the perturbative approximation for $\alpha_{s,\Lambda}$. Second, there are errors introduced by the perturbative expansion for the functions $g_i(\alpha_{s,\Lambda})$ in powers of $\alpha_{s,\Lambda}$, which are a combination of inverse power-law and inverse logarithmic errors. Consider for instance the approximation (c) given by Eq. (\ref{coc2c}), with $\Lambda$ and $E_0$ fixed ($\Lambda >> E_0$). The second-order running coupling $\alpha_{s,\Lambda}^{(2)}$ and irrelevant operator $F^{(2)}_{s,\Lambda}$ are given respectively by
\begin{equation}
\alpha_{s,\Lambda}^{(2)}=\frac{\alpha_{\Lambda}}{1-\frac{\alpha_{\Lambda}}{8\pi}\; \left[\gamma+{\rm ln}\left(2s \Lambda^4 \right)-{\rm Ei}\left(-2s \Lambda^4 \right)\right]}
\end{equation}
\noindent
and
\begin{eqnarray}
F^{(2)}_{s,\Lambda}(p,p')&=&  \frac{1}{4}\left[\gamma+{\rm ln}(2s \; p^2 \; p'^2)-{\rm Ei}(-2s \; p^2 \; p'^2)\right]+ \frac{1}{4}\left[\gamma+{\rm ln}(2s \Lambda^4)-{\rm Ei}(-2s \Lambda^4)\right]\nonumber\\
&-&\frac{1}{4}\left[\gamma+{\rm ln}\left(s \left[(p^2-\Lambda^2)^2 +(p'^2-\Lambda^2)^2-(p^2-p'^2)^2\right]\right)\right.\nonumber\\
&&\left. \; \; \; \; \; -{\rm Ei}\left(-s \left[(p^2-\Lambda^2)^2 +(p'^2-\Lambda^2)^2-(p^2-p'^2)^2\right]\right)\right]\; ,
\end{eqnarray}
\noindent
where ${\rm Ei}(x)$ is the exponential integral function. Expanding $F^{(2)}_{s,\Lambda}$ in powers of $\sqrt{s}\;p^2 = p^2/\lambda^2$ and $\sqrt{s}\;p'^2 = p'^2/\lambda^2$, we obtain
\begin{equation}
V_{\lambda,\Lambda}^{(2)}(p,p')= \left[{\tilde g}_0(\alpha_{\lambda,\Lambda}^{(2)})+{\tilde g}_2(\alpha_{\lambda,\Lambda}^{(2)}) \;\frac{( p^2+p'^2)}{2 \lambda^2}+{\tilde g'}_4(\alpha_{\lambda,\Lambda}^{(2)}) \;\frac{( p^2 p'^2)}{2 \lambda^4}+ \cdots \right] \; e^{-(p^2-p'^2)^2/\lambda^4}\; ,
\end{equation}
\noindent
where
\begin{eqnarray}
&&{\tilde g}_0(\alpha_{\lambda,\Lambda}^{(2)})=-\frac{\alpha_{\lambda,\Lambda}^{(2)}}{2\pi}\; ; \\
&&{\tilde g}_2(\alpha_{\lambda,\Lambda}^{(2)})=\frac{(\alpha_{\lambda,\Lambda}^{(2)})^2\lambda^2}{8\pi^2 \Lambda^2}\left(1-e^{-2\Lambda^4/\lambda^4}\right)\; ; \;\; \\
&&{\tilde g'}_4(\alpha_{\lambda,\Lambda}^{(2)})=-\frac{2\alpha_{\lambda,\Lambda}^{(2)}}{\pi}+\frac{(\alpha_{\lambda,\Lambda}^{(2)})^2}{4\pi^2}\left(1-e^{-2\Lambda^4/\lambda^4}\right) \; .
\end{eqnarray}

When $\lambda >> \Lambda$ the second-order running coupling $\alpha_{\lambda,\Lambda}^{(2)}$ runs little (approaching its asymptotic value $\alpha_{\Lambda}$) and so the functions ${\tilde g}_2(\alpha_{\lambda,\Lambda}^{(2)})$ and ${\tilde g}_4(\alpha_{\lambda,\Lambda}^{(2)})$. Evaluating the difference between the truncated SRG potential $V_{\lambda,\Lambda}^{(2)}(p,p')$ and the initial potential $V_{\Lambda}(p,p')=-\alpha_{\Lambda}/(2\pi)$ and expanding in powers of $\Lambda/\lambda$, we obtain
\begin{equation}
\delta V_{\lambda,\Lambda}(p,p')\equiv V_{\lambda,\Lambda}^{(2)}(p,p')-V_{\Lambda}(p,p')=-\frac{\alpha_{\Lambda}^{2}\Lambda^4}{8\pi^2 \lambda^4} + \frac{\alpha_{\Lambda}^{2}\Lambda^2}{4\pi^2 \lambda^2}\;\frac{( p^2+p'^2)}{2 \lambda^2}- \frac{4\alpha_{\Lambda}}{\pi} \;\frac{( p^2 p'^2)}{2 \lambda^4}+ \cdots \; ,
\end{equation}
\noindent
One can clearly see that in this region perturbation theory in $\delta V_{\lambda,\Lambda}$ leads to errors that scale like powers of $1/\lambda$, similar to EFT. In the limit $\lambda \rightarrow \infty$ the results become exact.

When $\lambda$ becomes smaller than $\Lambda$ there is a crossover to a different scaling regime. The second-order running coupling $\alpha_{\lambda,\Lambda}^{(2)}$ runs significantly in this region, and so the functions ${\tilde g}_2(\alpha_{\lambda,\Lambda}^{(2)})$ and ${\tilde g}_4(\alpha_{\lambda,\Lambda}^{(2)})$, leading to errors that scale like a combination of powers of $1/\lambda$ and inverse powers of ${\rm ln}(\lambda)$.

In this work we solve Eq.(\ref{flowdelta}) numerically, obtaining an exact (non-perturbative) solution for the SRG evolved potential (apart from numerical errors). The relative momentum space is discretized on a grid of $N$ gaussian integration points (we have used 200 mesh points in our calculations), leading to a system of $N^2$ non-linear first-order coupled differential equations which is solved by using an adaptative fifth-order Runge-Kutta algorithm.

\begin{figure}[h]
\centerline{\includegraphics[width=10.5 cm]{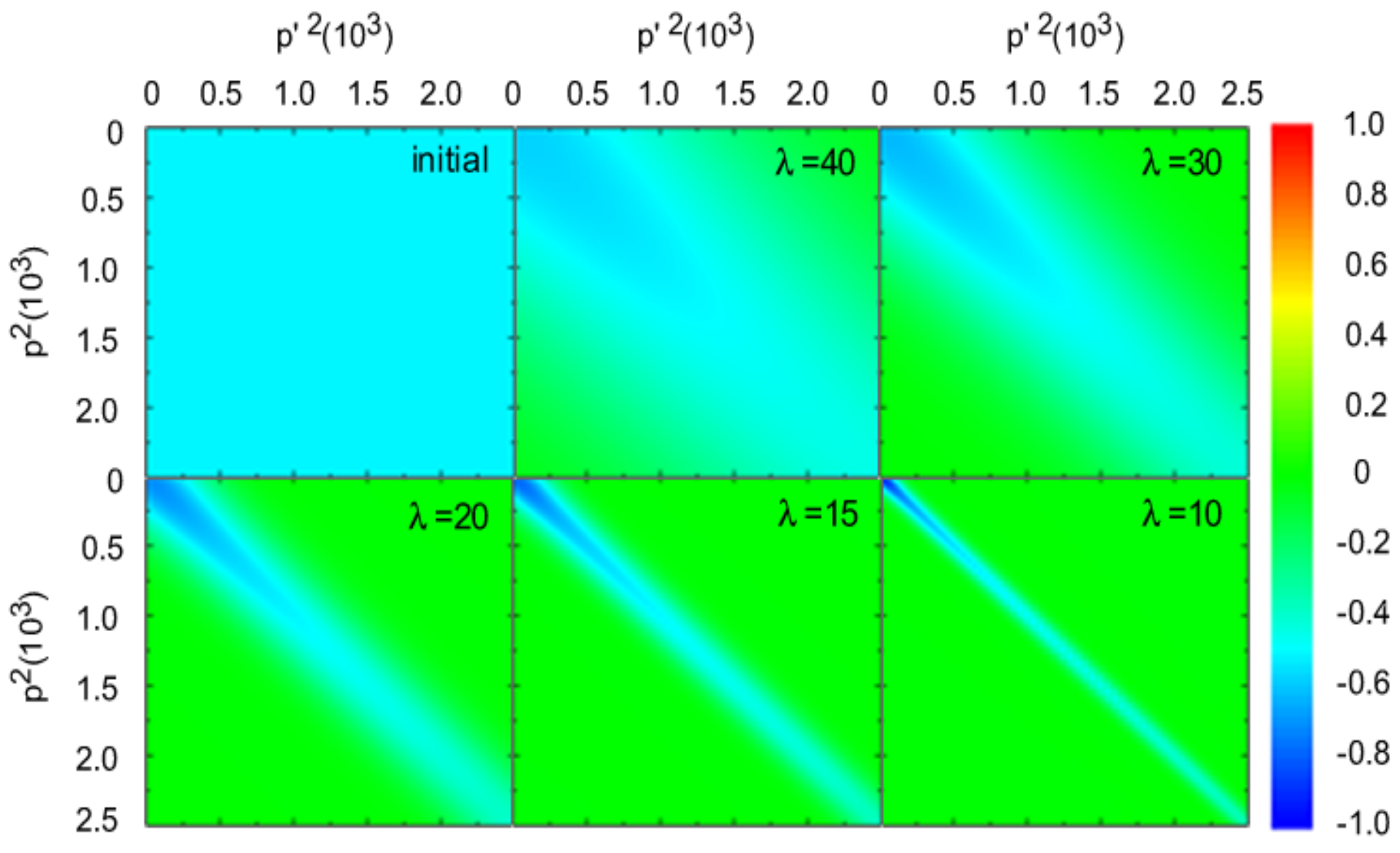}} \caption{(Color online) SRG evolution of the two-dimensional Dirac-delta potential for $\Lambda = 50$ and $E_0=1$ in arbitrary units (contour plots).}
\label{fig1}
\end{figure}

\begin{figure}[h]
\centerline{\includegraphics[width=4.8 cm]{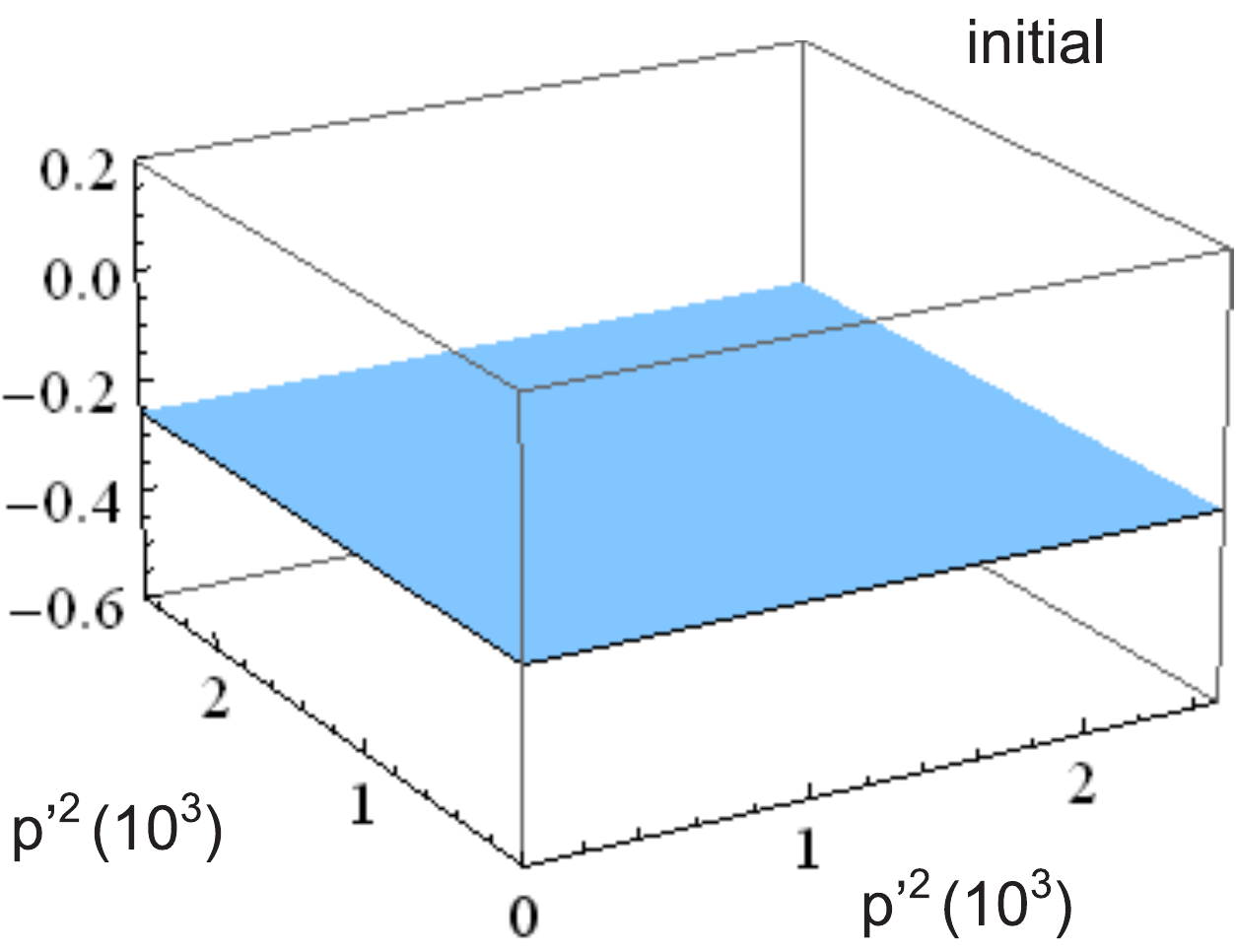}\hspace*{.5cm}\includegraphics[width=4.8 cm]{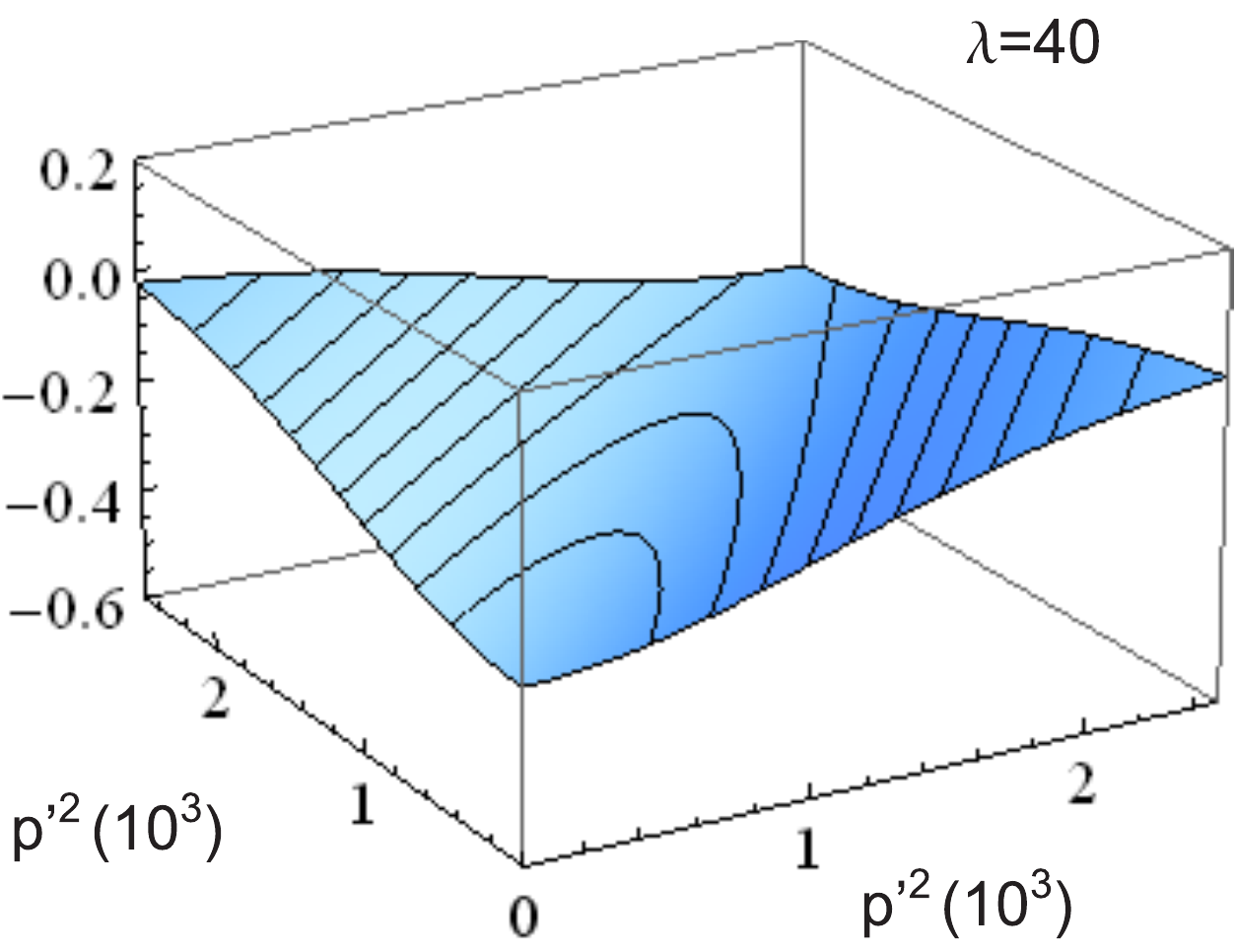}\hspace*{.5cm}\includegraphics[width=4.8 cm]{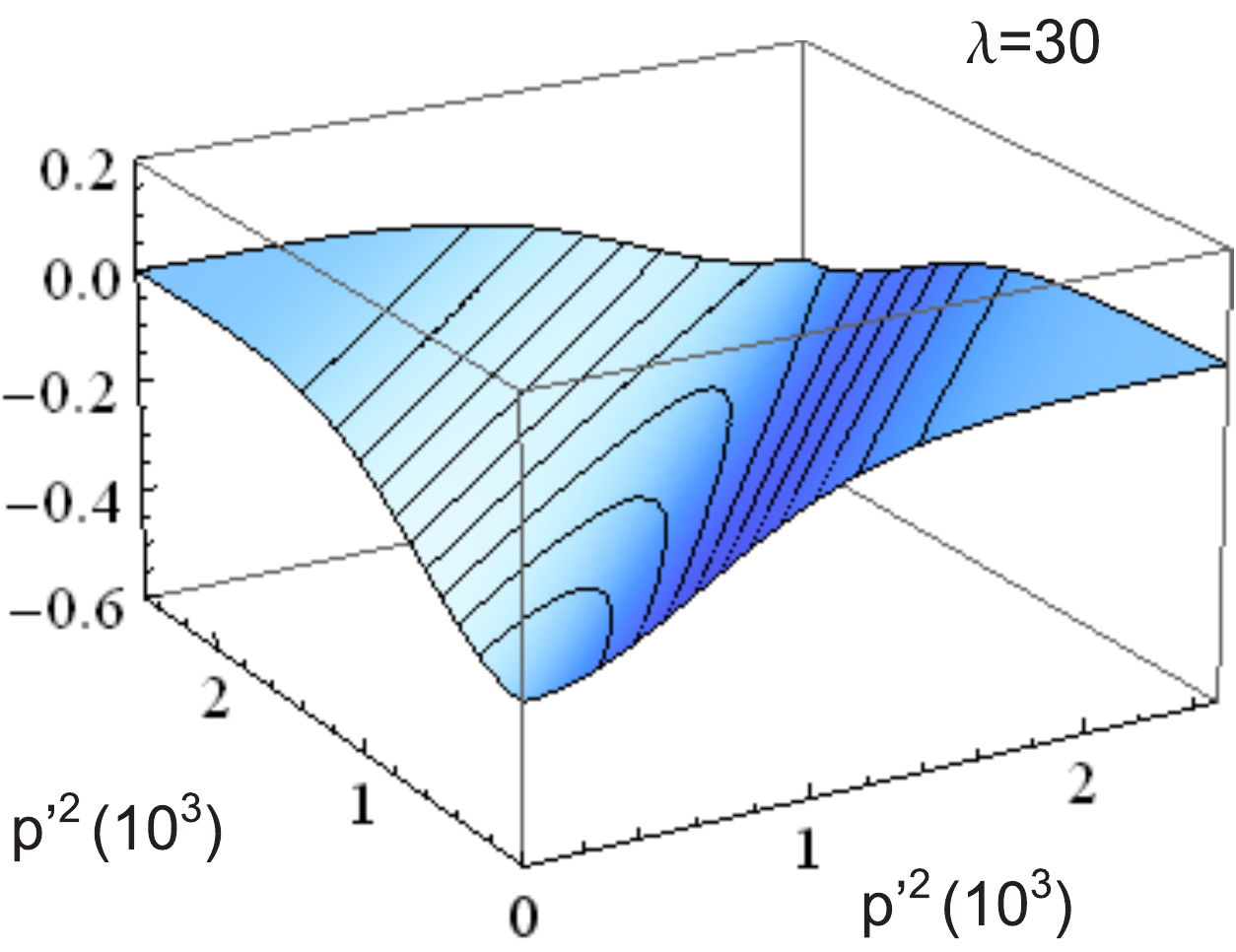}}\vspace*{.5cm}
\centerline{\includegraphics[width=4.8 cm]{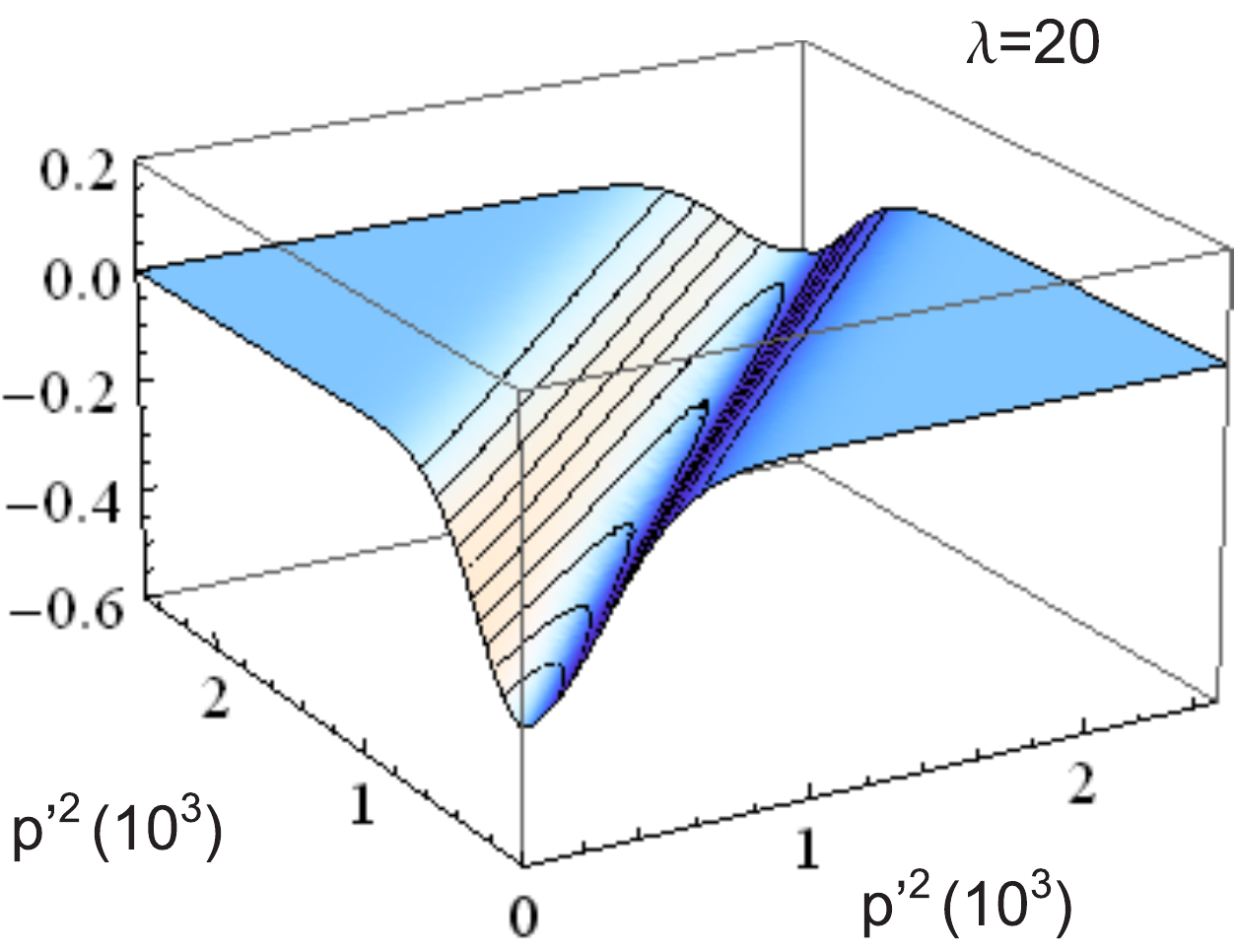}\hspace*{.5cm}\includegraphics[width=4.8 cm]{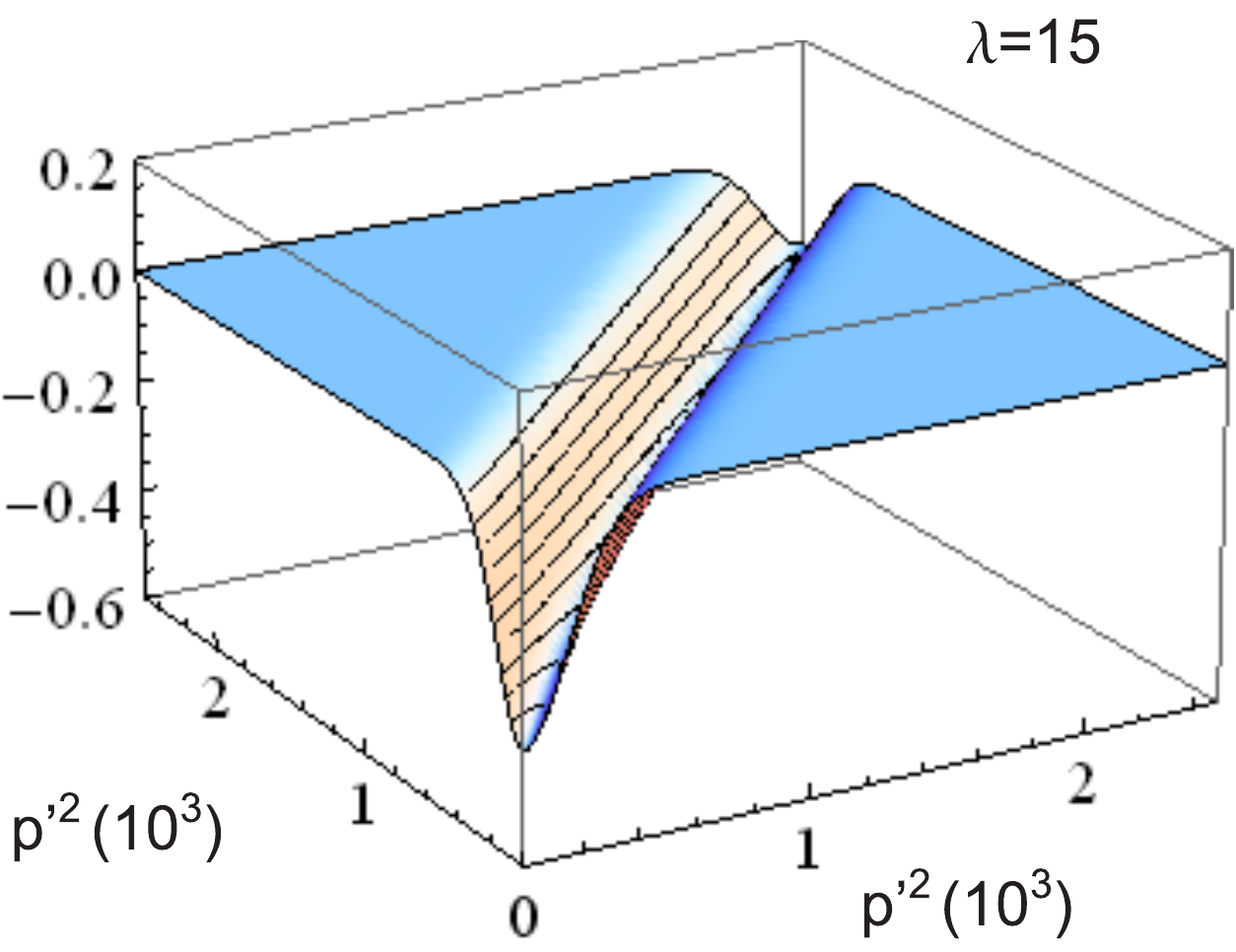}\hspace*{.5cm}\includegraphics[width=4.8 cm]{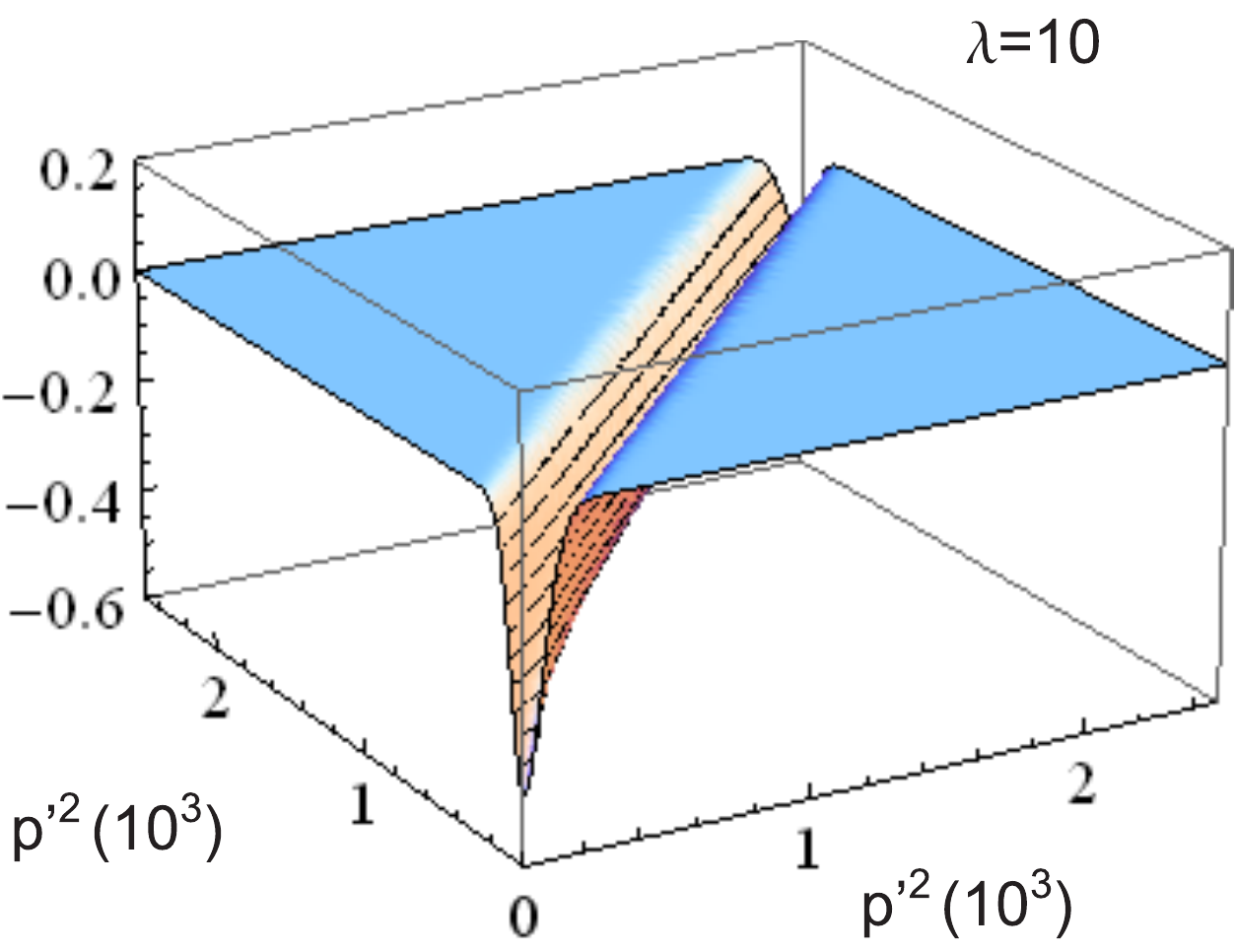}}
\caption{(Color online) SRG evolution of the two-dimensional Dirac-delta potential for $\Lambda = 50$ and $E_0=1$ in arbitrary units (surface plots).}
\label{fig2}
\end{figure}

In Figs. (\ref{fig1}) and (\ref{fig2}) we show respectively the contour and the surface plots for the evolution of the two-dimensional Dirac-delta contact potential with the similarity cutoff $\lambda=s^{-1/4}$, starting from an initial (bare) potential $V_{\Lambda}(p,p')$ defined by using the renormalized running coupling $\alpha_{\Lambda}$ in Eq. (\ref{couprun}) fixed at a momentum cutoff $\Lambda=50$ to give a binding energy $E_0=1$ (in arbitrary units). The matrix elements of the initial potential $V_{\Lambda=50}(p,p')$ are constants and we clearly see a systematic suppression of the off-diagonal matrix elements as the similarity cutoff $\lambda$ is reduced, such that the potential is driven towards a band-diagonal form.
\vspace*{.5cm}

\begin{figure}[h]
\centerline{\includegraphics[width=7.6 cm]{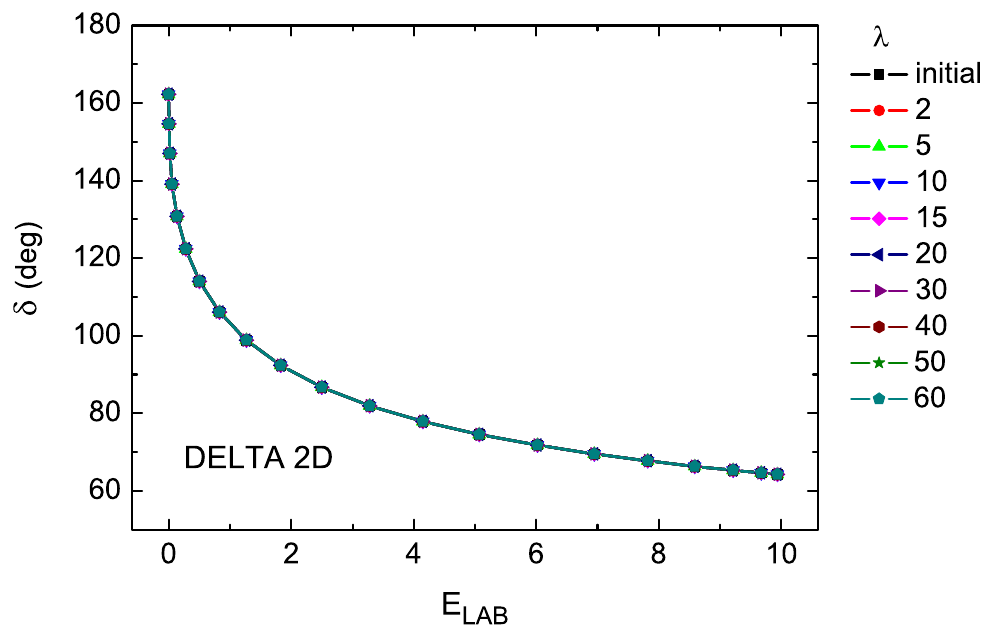}}
\caption{(Color online) Phase-shifts for the initial two-dimensional Dirac-delta potential and the SRG potentials evolved to several values of $\lambda$.}
\label{fig3}
\end{figure}

In Fig. \ref{fig3} we show the results for the phase-shifts as a function of the laboratory energy $E_{\rm LAB}$ calculated from the numerical solution of the LS equation for the $K$-matrix with the initial two-dimensional Dirac-delta contact potential $V_{\Lambda=50}(p,p')$ and with the corresponding SRG potentials evolved to several values of the similarity cutoff $\lambda$.

\begin{figure}[h]
\centerline{\includegraphics[width=7.5 cm]{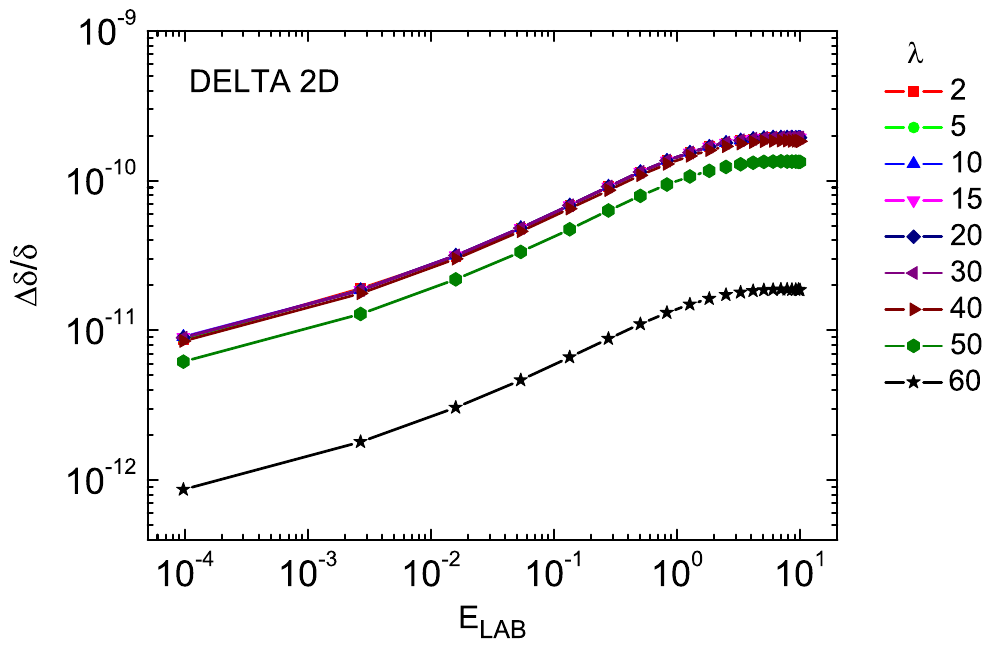}\hspace*{.5cm} \includegraphics[width=7.5 cm]{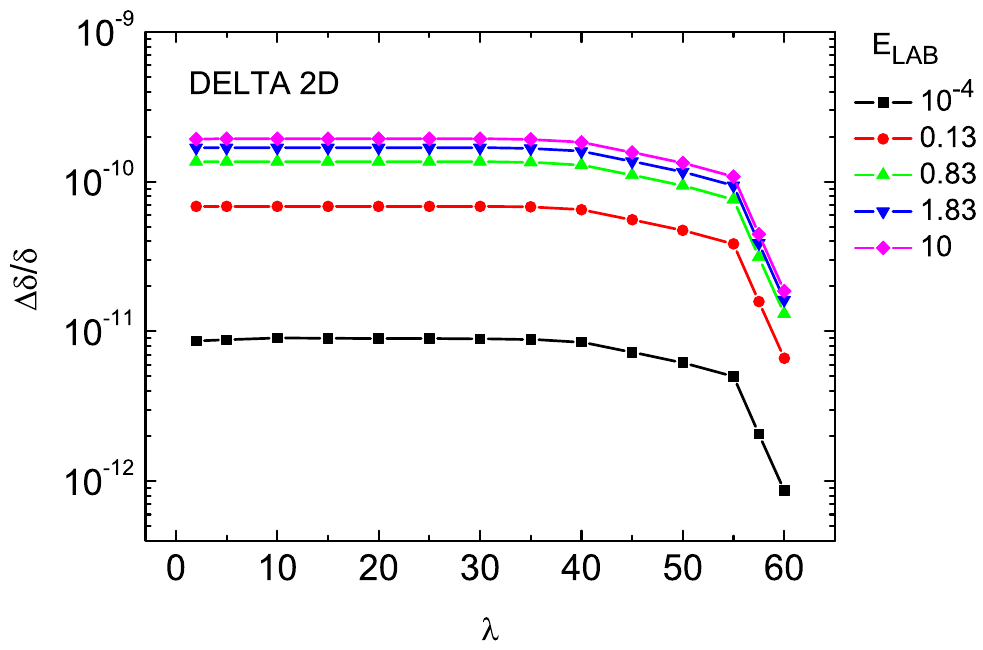}}
\caption{(Color online) Relative errors in the phase-shifts for the SRG potentials evolved from the initial two-dimensional Dirac-delta potential. Left panel: as a function of $E_{\rm LAB}$ for several values of $\lambda$; Right panel: as a function of $\lambda$ for several values of $E_{\rm LAB}$.}
\label{fig4}
\end{figure}

In Fig. \ref{fig4} we show the relative errors in the phase-shifts for the SRG evolved potentials with respect to the results for the initial potential $V_{\Lambda=50}(p,p')$, as a function of $E_{\rm LAB}$ for several values of the similarity cutoff $\lambda$ (left) and as a function of the similarity cutoff $\lambda$ for several values of $E_{\rm LAB}$ (right). Note that, in order to avoid mixing up with the errors from using a momentum cutoff $\Lambda=50$ and with the errors from the numerical solution of the LS equation, we compare the phase-shifts for the SRG evolved potentials with those for the initial potential $V_{\Lambda=50}(p,p')$, rather than with the exact results given by Eq. (\ref{exactpsdeltaD2}). As one can observe, the phase-shifts obtained with the SRG evolved potentials are the same as those obtained with the initial potential, apart from relative differences smaller than $10^{-9}$ due to residual numerical errors (which depend on the discretization and on the accuracy and relative error tolerance set in the Runge-Kutta solver used to evolve the potential). This result is expected, since the SRG transformation is unitary. One can also observe two different scaling regions in the error plots for fixed $E_{\rm LAB}$ (right). When the similarity cutoff $\lambda$ is smaller than the momentum cutoff $\Lambda(=50)$ the errors are practically constant. When $\lambda$ becomes larger than $\Lambda$ there is a crossover to a different scaling regime where the errors significantly decrease.

The change in the scaling behavior of the residual errors is a numerical artifact that can be understood in the following way. When $\lambda >> \Lambda$, the SRG potential evolves slowly and stays very near the initial potential. Only a few steps are required to integrate the differential equations using the Runge-Kutta solver, resulting in very small errors in the matrix-elements of the SRG evolved potential and hence in the phase-shifts. As $\lambda$ approaches $\Lambda$, the SRG potential starts to evolve more rapidly. An increasing number of steps is required to integrate the differential equations, resulting in larger errors in the phase-shifts. When $\lambda << \Lambda$, the errors approach an asymptotic value.

\section{SRG for the $NN$ Interaction}
\label{SRGNN}

In this section we review the main aspects involved in the application of the SRG approach to the $NN$ interaction \cite{srg1,srg2,srg3,srg4,srg5}. We start from the hamiltonian in the center-of-mass frame for a system of two nucleons, which can be written in the form $H=T_{\rm rel}+V$, where $T_{\rm rel}$ corresponds to the relative kinetic energy and $V$ corresponds to the $NN$ potential. Here and in what follows we use units such that $\hbar=c=M=1$, where $M$ is the nucleon mass.

Using the generator $\eta_s=[T_{\rm rel},H_s]$, Wegner's flow equation for the SRG evolution of the $NN$ potential is given by
\begin{eqnarray}
\frac{dV_{s}(p,p')}{ds}=-(p^2-p'^2)^2 \; V_{s}(p,p')+\frac{2}{\pi} \int_{0}^{\infty}dq \; q^2\;
(p^2+p'^2-2 q^2)\; V_{s}(p,q)\; V_{s}(q,p'),
\label{flowNN}
\end{eqnarray}
\noindent
where $V_{s}(p,p')$ is used as a brief notation for the projected $NN$ potential matrix elements,
\begin{equation}
V^{(JLL'S;I)}_{s}(p,p')= \langle\;p{(LS)J;I}|V_{s}|\;p'{(L'S)J;I} \;\rangle \; ,
\end{equation}
\noindent
in a partial-wave relative momentum space basis, $|\;q{(LS)J;I} \;\rangle$, with normalization such that
\begin{equation}
1=\frac{2}{\pi} \int_{0}^{\infty}dq \; q^2 \; |\;q{(LS)J;I} \;\rangle \;\langle \; q{(LS)J;I}\;|.
\label{PWnorm}
\end{equation}
\noindent
The superscripts $J$, $L(L')$, $S$ and $I$ denote respectively the total angular momentum, the orbital angular momentum, the spin and the isospin quantum numbers of the $NN$ state. For non-coupled channels ($L=L'=J$), the matrix elements $V_{s}(p,p')$ are simply given by $V_{s}(p,p')\equiv V^{(JJJS;I)}_{s}(p,p')$. For coupled channels ($L,L'=J \pm 1$), the $V_{s}(p,p')$ represent $2 \times 2$ matrices of matrix elements for the different combinations of $L$ and $L'$:
\begin{eqnarray}
 V_{s}(p,p')
  \equiv \begin{pmatrix}
    V^{(JLLS;I)}_{s}(p,p')  & V^{(JLL'S;I)}_{s}(p,p') \\ \\
    V^{(JL'LS;I)}_{s}(p,p')  & V^{(JL'L'S;I)}_{s}(p,p')
  \end{pmatrix} \;.
\end{eqnarray}
\noindent
As a consequence of the choice of the SRG transformation generator $\eta_s=[T_{\rm rel},H_s]$, each interaction channel
evolves with the cutoff $\lambda=s^{-1/4}$ independently of the other channels \cite{UCOM}.

The scattering observables for each $NN$ interaction channel can be calculated by iterating the corresponding SRG evolved potential through the LS equation for the partial-wave $T$-matrix (for simplicity, we drop the subscript $s$ denoting the flow parameter and the superscripts denoting the quantum numbers of the $NN$ state):
\begin{equation}
T(p,p';k^2)=V(p,p')+\frac{2}{\pi}\; \int_{0}^{\infty} \; dq \; q^2 \;
\frac{V(p,q)}{k^2-q^2+i \; \epsilon} \; T(q,p';k^2) \; .
\end{equation}
\noindent
The phase-shifts are then obtained from the relation $S(k^2) = 1 - 2ik~T(k,k;k^2)$. For the non-coupled channels,
\begin{equation}
S(k^2) ={\rm exp}(2\;i\;\delta^{s,t}_J) ,
\end{equation}
\noindent
and for the coupled channels (using the Stapp parametrization \cite{stapp}),
\begin{equation}
S(k^2) =
\left( \begin{array}{cc}
\cos{(2\epsilon^t)}\;{\rm exp}(2\;i\;\delta^t_{J-1}) & i\;\sin{(2\epsilon^t)}\;{\rm exp}(i\;\delta^t_{J-1}+i\;\delta^t_{J+1})  \\ \\
i\;\sin{(2\epsilon^t)}\;{\rm exp}(i\;\delta^t_{J+1}+i\;\delta^t_{J-1}) & \cos{(2\epsilon^t)}\;{\rm exp}(2\;i\;\delta^t_{J+1})
\end{array} \right) \; ,
\end{equation}
\noindent
where the superscripts $s$ and $t$ stand respectively for the spin singlet states and spin triplet states, $J=L+S$ is the total angular momentum of the $NN$ state and $\epsilon^t$ is the mixing parameter.

For convenience, in our calculations we iterate the potential through the LS equation for the partial-wave $K$-matrix
\begin{equation}
K(p,p';k^2)=V(p,p')+\frac{2}{\pi}\; {\cal P}\int_{0}^{\infty} \; dq \; q^2 \;
\frac{V(p,q)}{k^2-q^2} \; K(q,p';k^2) \; ,
\label{LSKNN}
\end{equation}

The relation between the $T$-matrix and the $K$-matrix on-shell is given by:
\begin{equation}
K^{-1}(k,k;k^2)=T^{-1}(k,k;k^2)-i \; k \; .
\end{equation}

\begin{figure}[h]
\centerline{\includegraphics[width=11.5 cm]{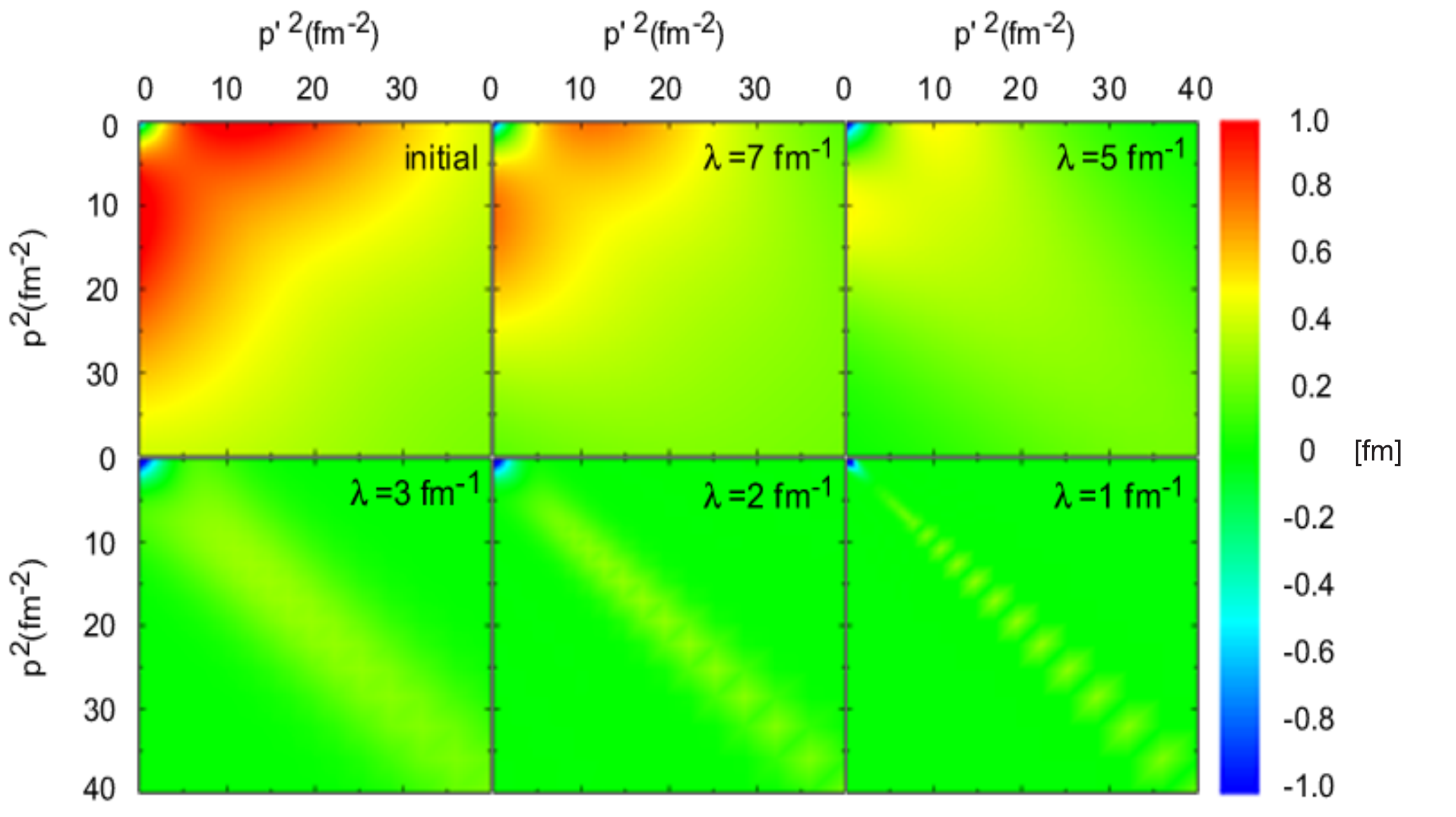}}
\caption{(Color online) SRG evolution of the Nijmegen potential in the $^1 S_0$ channel (contour plot).}
\label{fig5}
\end{figure}

\begin{figure}[h]
\centerline{\includegraphics[width=4.8 cm]{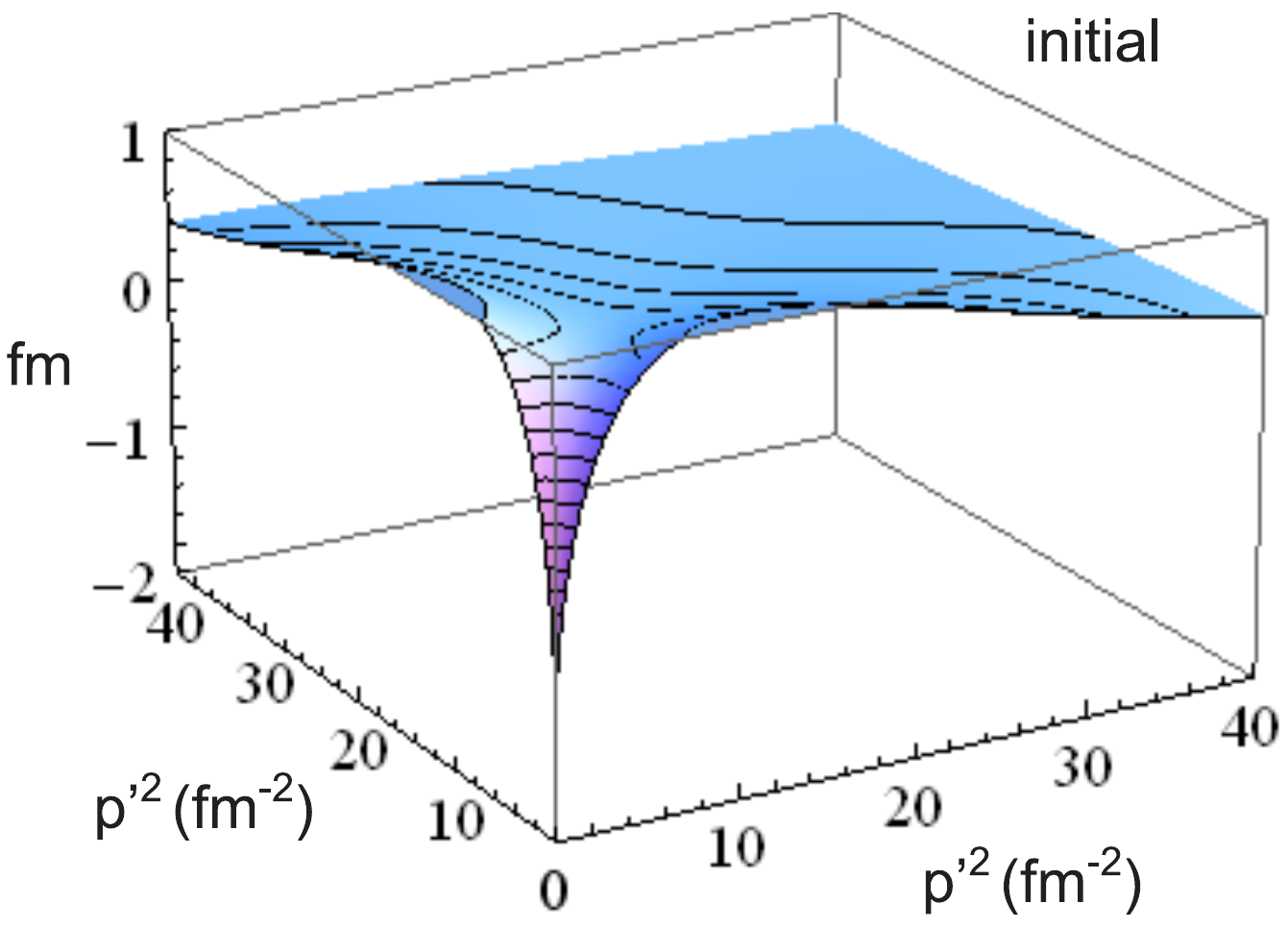}\hspace*{.5cm}\includegraphics[width=4.8 cm]{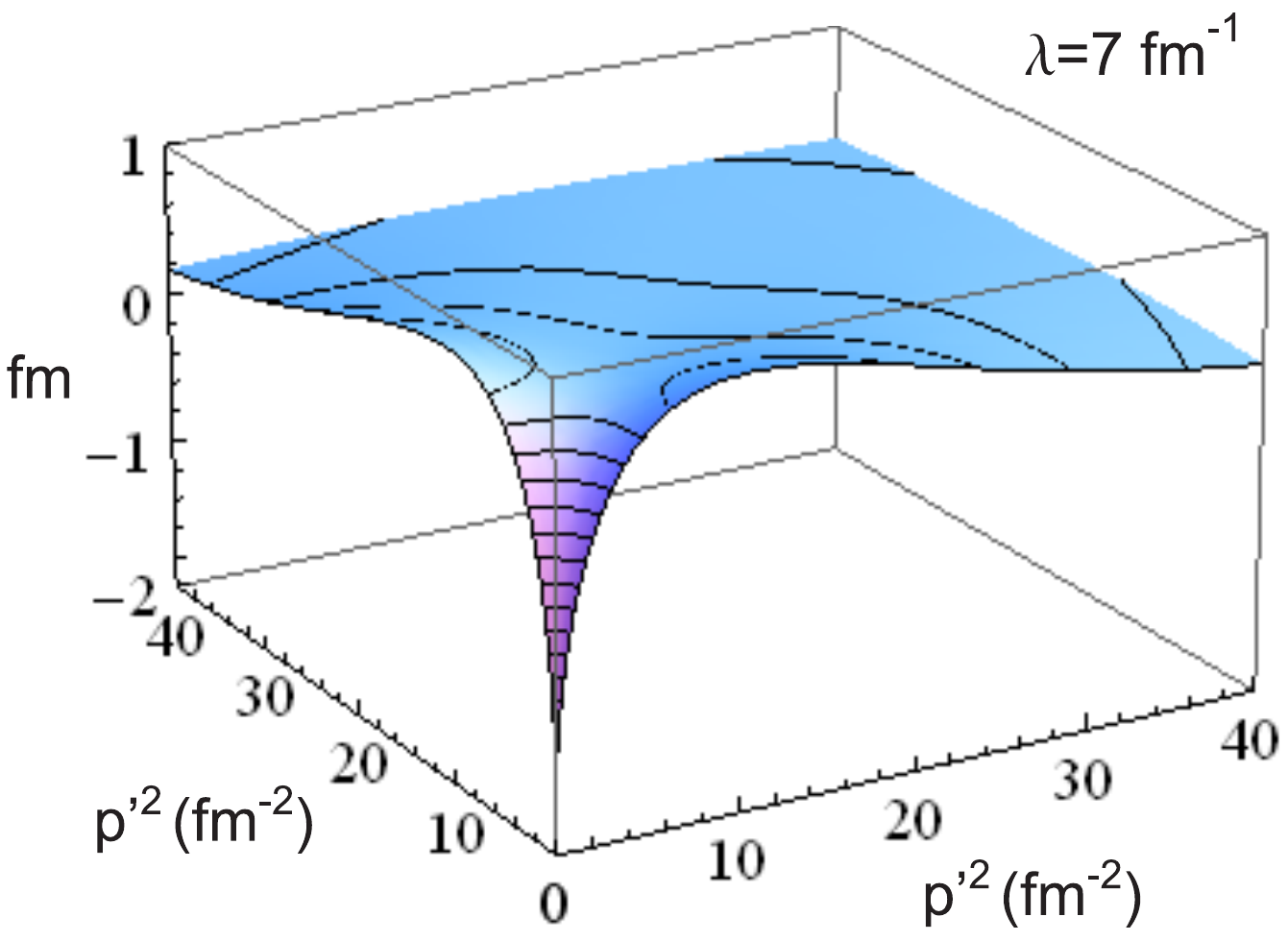}\hspace*{.5cm}\includegraphics[width=4.8 cm]{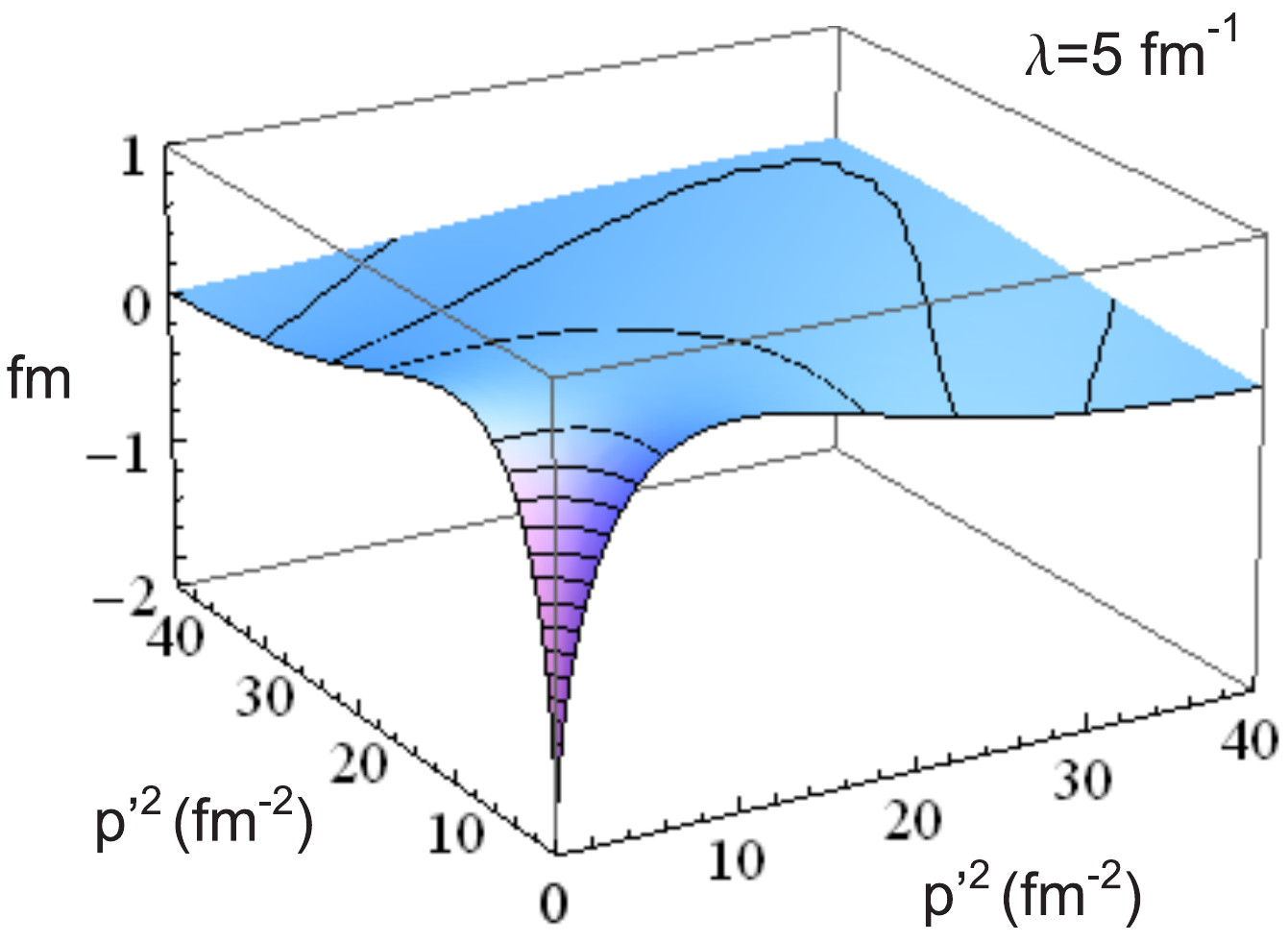}}\vspace*{0.8cm}
\centerline{\includegraphics[width=4.8 cm]{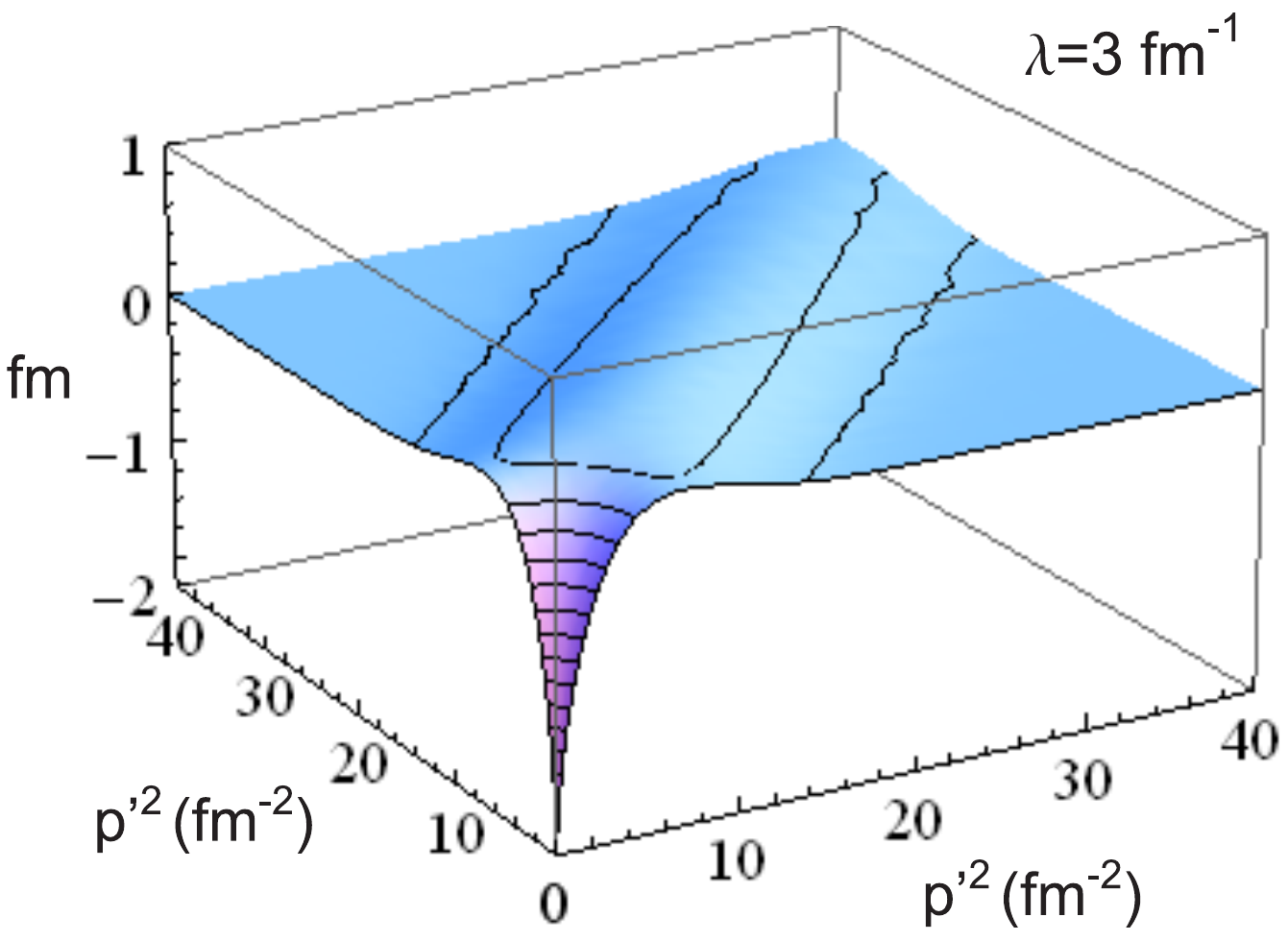}\hspace*{.5cm}\includegraphics[width=4.8 cm]{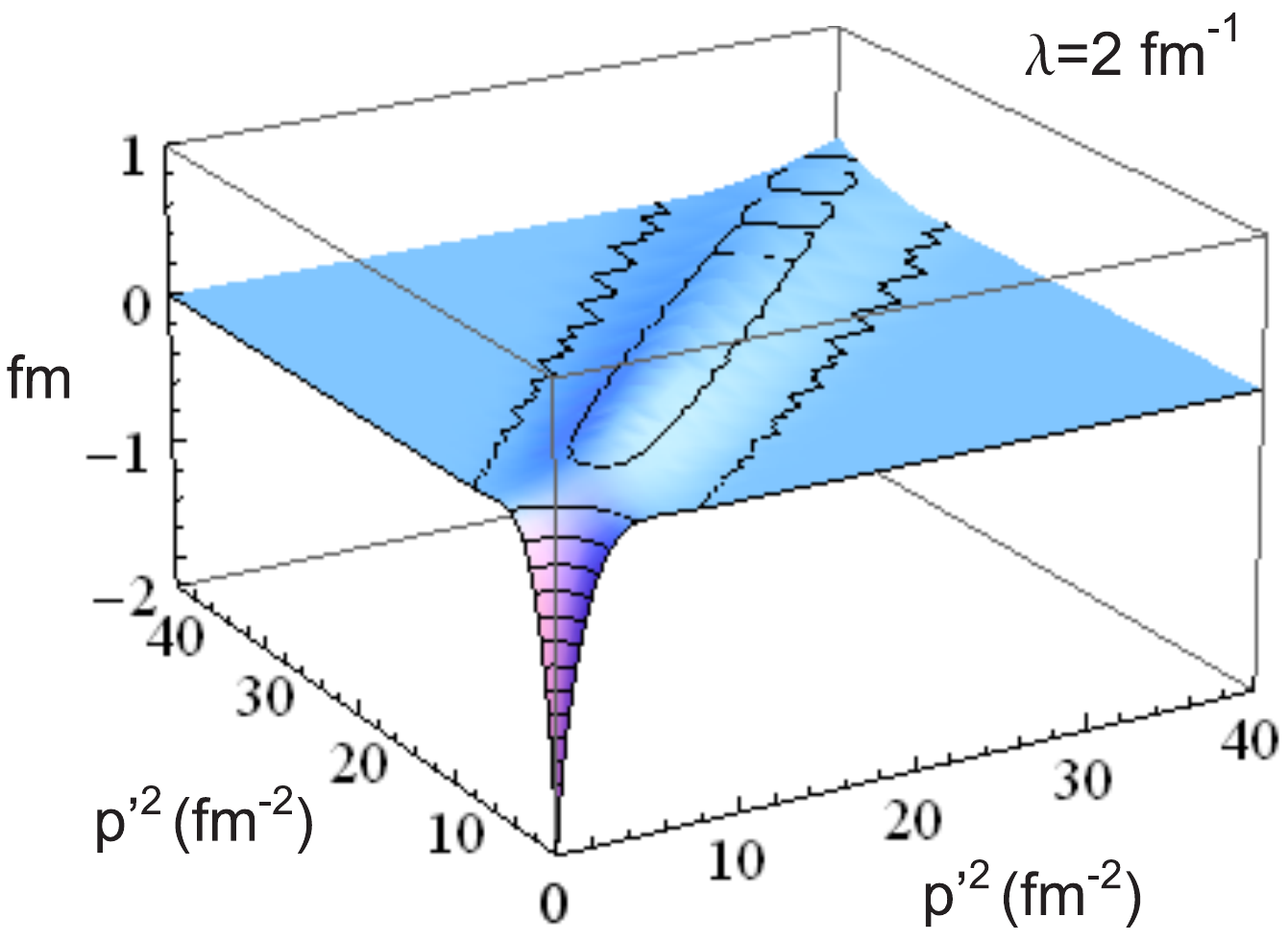}\hspace*{.5cm}\includegraphics[width=4.8 cm]{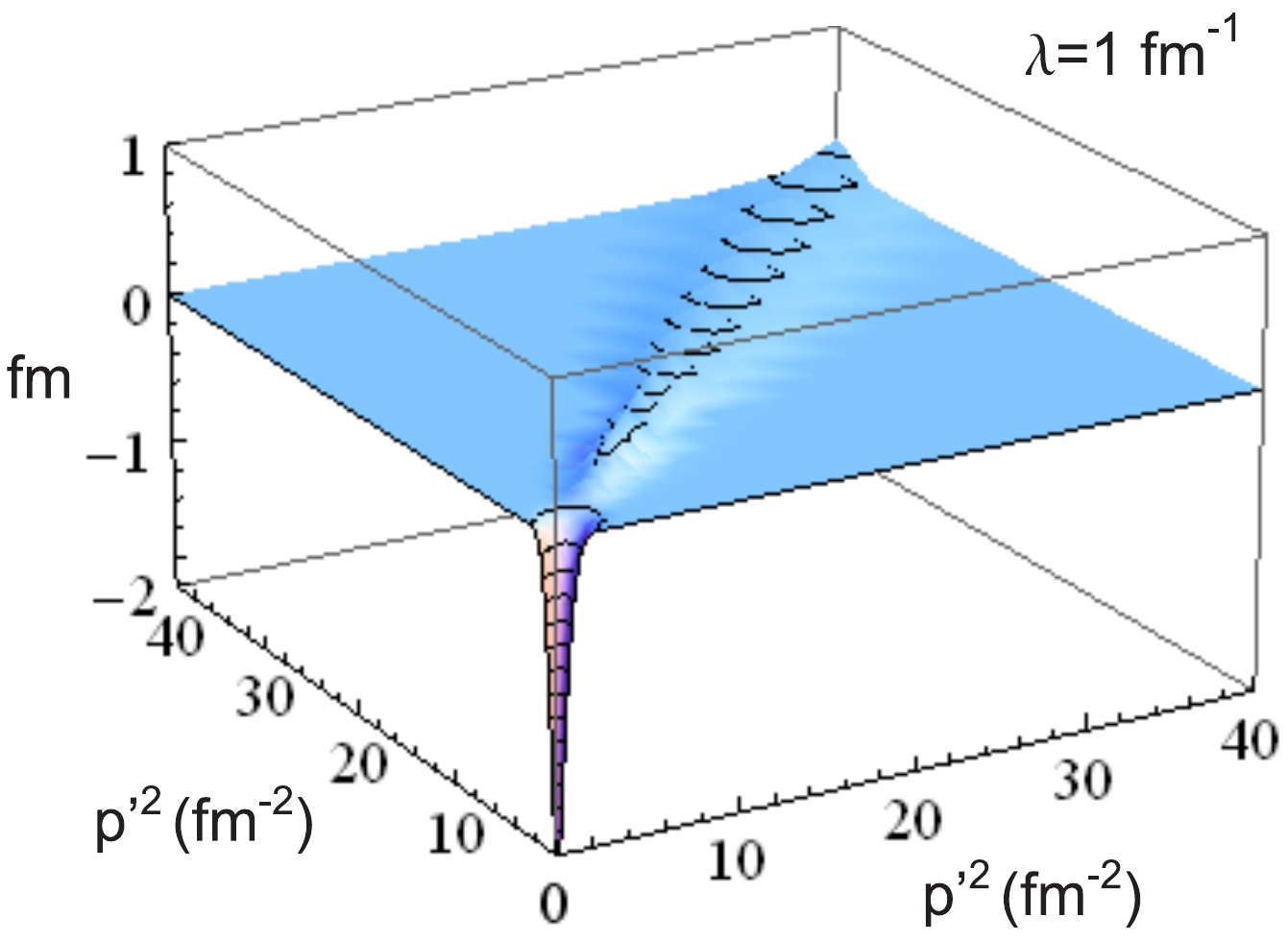}}
\caption{(Color online) SRG evolution of the Nijmegen potential in the $^1 S_0$ channel (surface plots).}
\label{fig6}
\end{figure}

In order to illustrate the application of the SRG to the $NN$ interaction, we solve Eq. (\ref{flowNN}) numerically to evolve the
Nijmegen potential \cite{nijmegen} in the $^1 S_0$ channel. In Figs. (\ref{fig5}) and (\ref{fig6}) we show respectively the contour and the surface plots for the SRG evolution of the Nijmegen potential in the $^1 S_0$ channel. As one can observe, the initial potential ($\lambda \rightarrow \infty$) exhibits strong matrix elements outside the diagonal, extending up to high-momenta (large energy differences). As the potential evolves with the similarity cutoff $\lambda$, the off-diagonal matrix elements are systematically suppressed while the high-momentum components are gradually concentrated near the diagonal and the low-momentum components (attractive) are enhanced. Therefore, the Nijmegen potential evolved through the SRG tends to a band-diagonal form.
\vspace*{1.2cm}

\begin{figure}[h]
\centerline{\includegraphics[width=8.0 cm]{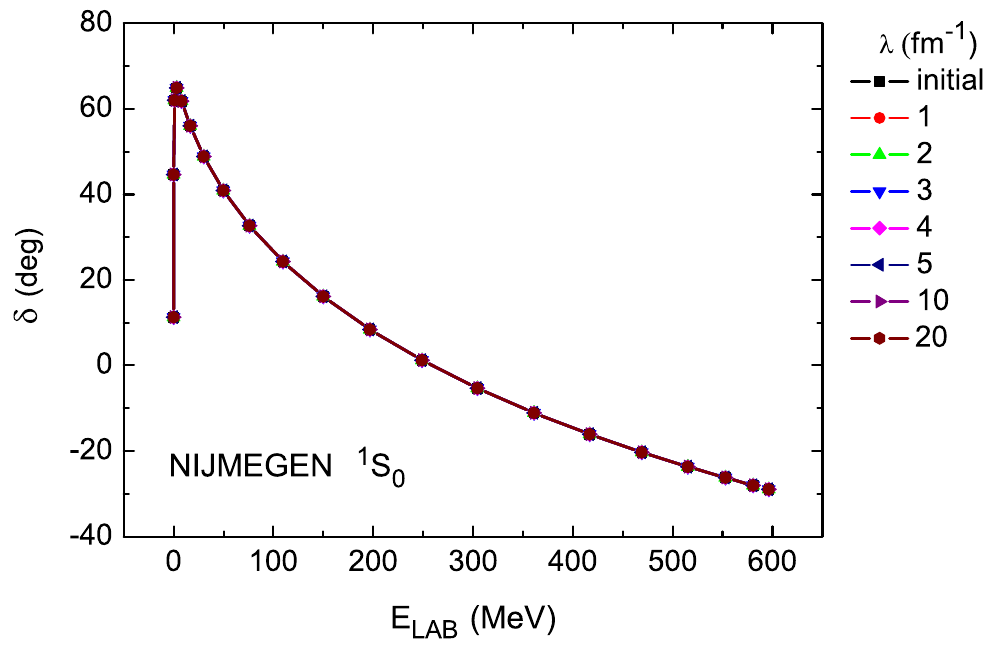}}
\caption{(Color online) Phase-shifts in the $^1 S_0$ channel for the initial Nijmegen potential and the SRG potentials evolved to several values of $\lambda$.}
\label{fig7}
\end{figure}

In Fig. \ref{fig7} we show the $NN$ phase-shifts in the $^1 S_0$ channel as a function of the laboratory energy $E_{\rm LAB}$ calculated from the numerical solution of the LS equation for the partial-wave $K$-matrix with the initial Nijmegen potential and with the corresponding SRG potentials evolved to several values of the similarity cutoff $\lambda$. As expected, the phase-shifts obtained with the SRG evolved potentials are the same as those obtained with the initial potential, apart from relative differences smaller than $10^{-9}$ due to numerical errors. This result is obtained for all energies, even those not constrained to scattering data (larger than $E_{\rm LAB}\sim 350 MeV$).

Now, we test the decoupling of low-energy observables from high-energy degrees of freedom for the Nijmegen potential evolved through the SRG, following the analysis introduced by Bogner et al. \cite{srg2,srg3}. Such an analysis consists in applying an exponential regularizing function to the potential, which suppresses the contributions to the observables from matrix elements $V_{s}(p,p')$ with $p,p'$ larger than a given momentum cut $k_{\rm max}$,
\begin{equation}
V^{(k_{\rm max}, n)}_{s}(p,p')={\rm exp}[-(p^2/k_{\rm max}^2)^n]\; V_{s}(p,p') \;{\rm exp}[-(p'^2/k_{\rm max}^2)^n] \;,
\label{smooth}
\end{equation}
\noindent
where $n$ is an integer number that specifies the smoothness of the regularizing function.

\begin{figure}[ht]
\centerline{\includegraphics[width=7.0 cm]{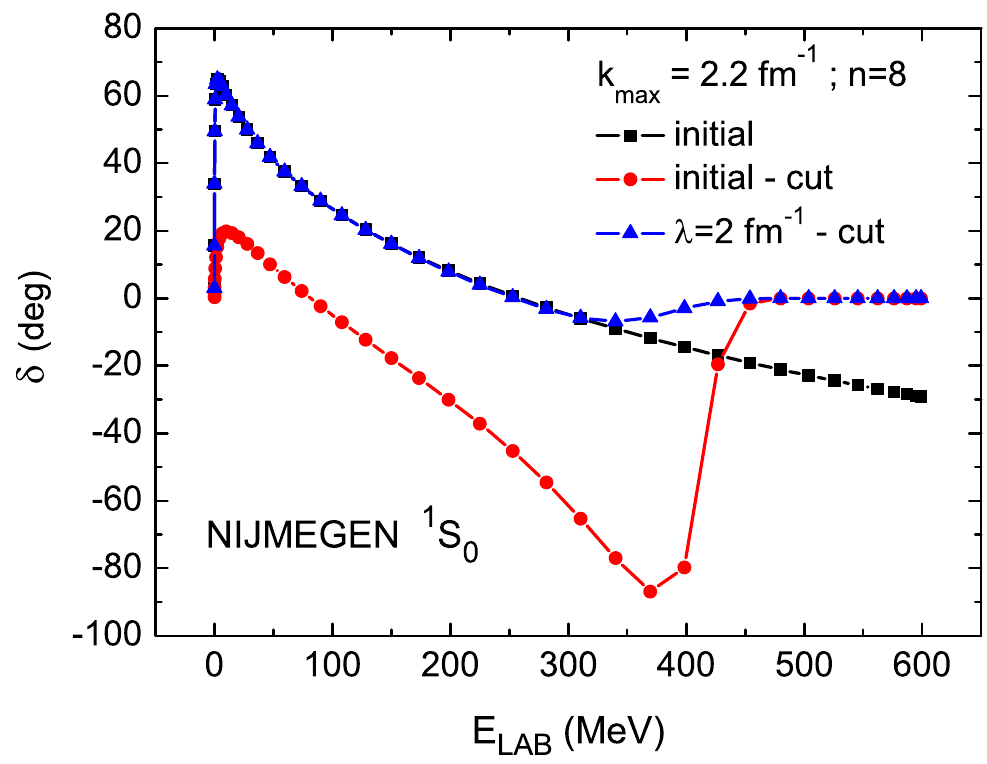}\hspace{0.8cm}\includegraphics[width=7.0 cm]{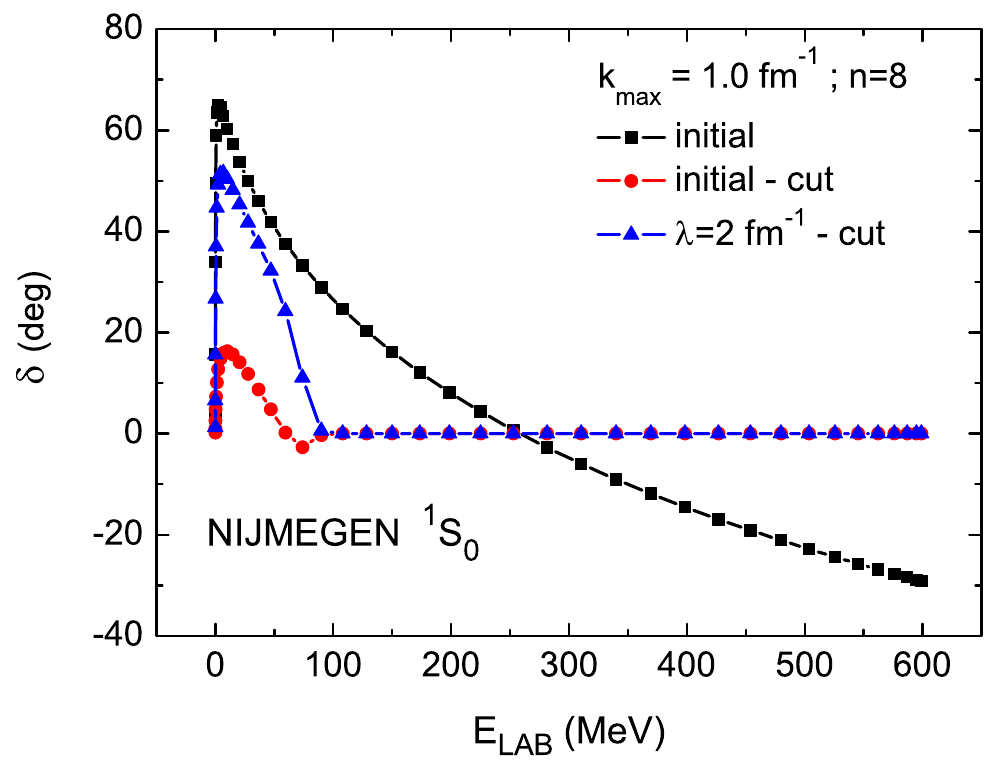}}
\caption{(Color online) Phase-shifts in the $^1 S_0$ channel for the initial Nijmegen potential and the corresponding SRG potential evolved to a similarity cutoff $\lambda=2.0 \; {\rm fm}^{-1}$ with the application of an exponential regularizing function. Left panel: using n=8 and $k_{max}=2.2 ~{\rm fm^{-1}}$; Right panel: using n=8 and $k_{max}=1.0 ~{\rm fm^{-1}}$.}
\label{fig8}
\end{figure}

We apply the exponential regularizing function to cut the initial Nijmegen potential ($\lambda \rightarrow \infty$) in the $^1 S_0$ channel and the corresponding SRG potential evolved to a similarity cutoff $\lambda=2.0 \; {\rm fm}^{-1}$ and then calculate the phase-shifts  as a function of the laboratory energy $E_{\rm LAB}$. In Fig. \ref{fig8} we show the results obtained by cutting the potentials at $k_{\rm max}=2.2 \; {\rm fm}^{-1}$ (left) and $k_{\rm max}=1.0 \; {\rm fm}^{-1}$ (right), with $n=8$.

As one can observe, the phase-shifts obtained for the cut initial potential are completely modified when compared to those obtained for the uncut potential. For the SRG evolved potential cut at $k_{\rm max}=2.2 \; {\rm fm}^{-1}$ (above $\lambda$), the phase-shifts at low energies agree with those obtained for the uncut SRG evolved potential and deviations start to occur near $E_{\rm LAB}\sim 2 \; k_{\rm max}^2/M$. When the SRG evolved potential is cut at $k_{\rm max}=1.0 \; {\rm fm}^{-1}$ (below $\lambda$), there is no agreement between the phase-shifts practically at all energies. One should note that the phase-shifts obtained for the cut potentials (both the initial and the SRG evolved) tend to zero for energies $E_{\rm LAB} > 2 \; k_{\rm max}^2/M $. Therefore, an explicit decoupling between the low- and high-momentum components is verified for the SRG evolved potential, such that the high-momentum matrix elements above the similarity cutoff $\lambda$ can be removed without causing a significant distortion in the calculations of low-energy observables. This result is similar to those described in Refs. \cite{srg2,srg3} for other $NN$ potentials. It is important to emphasize that different decoupling patterns are expected by using the generator $\eta_s=[G_s,H_s]$ with other choices for the operator $G_s$, such as $G_s={\rm diag}(H_s)$.

\section{The Subtracted Kernel Method}
\label{SKM}

\subsection{Formalism}
\label{SKMFORM}

We begin the description of the SKM approach \cite{skm1,skm2,skm3,skm4,skm5,skm6} by considering the formal LS equation for the $T$-matrix of a two-body system, which can be written in operator form as
\begin{eqnarray}
T(E) &=& V + V~G_{0}^{+}(E)~T(E)=V \left[ 1+ G_{0}^{+}(E)~T(E) \right]\; \nonumber \\
&=& V + T(E)~G_{0}^{+}(E)~V= \left[ 1+ T(E)~G_{0}^{+}(E) \right]~V \; ,
\label{LS}
\end{eqnarray}
where $V$ is the interaction potential and $G_{0}^{+}(E)$ is the free Green's function for the two-body system with outgoing-wave boundary conditions, given in terms of the free hamiltonian $H_0$ by
\begin{equation}
G_{0}^{+}(E) = (E - H_{0} + i \epsilon)^{-1} \; .
\end{equation}
\noindent
For singular potentials, like those containing Dirac-delta contact interactions and their derivatives, Eq. (\ref{LS}) becomes ill-defined due to the ultraviolet divergencies that appear in the implicit momentum integration. In the SKM approach a regularized and renormalized scattering equation is derived by performing subtractions in propagator at a certain energy scale.

We first consider a singular potential containing a regular term and a Dirac-delta contact interaction, in which case only one subtraction is enough to get a finite scattering amplitude. In momentum space, this potential can be written in the  form $V = V_{\rm reg} + C_{0}$, where $C_{0}$ is the strength of the contact interaction. Using Eq. (\ref{LS}), the potential $V$ can be formally written in terms of the $T$-matrix at a given energy scale $-\mu^2$ (in units such that $\hbar=c=2m=1$, where $m$ is the reduced mass of the two-body system):
\begin{equation}
V = T(-\mu^2)~\left[1 + G_{0}^{+}(-\mu^2)~T(-\mu^2) \right]^{-1}= \left[1 + T(-\mu^2)~G_{0}^{+}(-\mu^2) \right]^{-1}~T(-\mu^2) \; .
\label{veqtmu}
\end{equation}
\noindent
For convenience we choose a negative energy for the subtraction scale, such that the free Green's function $G_{0}^{+}(-\mu^2)$ is real (this is not a restriction, since the method works as well for a positive energy subtraction scale).

Replacing the potential $V$ in Eq. (\ref{LS}) by its expression in terms of $T(-\mu^2)$ given in Eq. (\ref{veqtmu}), we obtain
\begin{equation}
T(E) = T(-\mu^2)~\left[1 + G_{0}^{+}(-\mu^2)~T(-\mu^2) \right]^{-1} + T(E)~G_{0}^{+}(E)~T(-\mu^2)~\left[1 + G_{0}^{+}(-\mu^2)~T(-\mu^2) \right]^{-1}   \; .
\label{LS2}
\end{equation}
\noindent
Now, multiplying both sides of Eq. (\ref{LS2}) from the right by $\left[1 + G_{0}^{+}(-\mu^2)~T(-\mu^2) \right]$ and re-arranging the terms, we obtain the subtracted kernel LS equation for the $T$-matrix,
\begin{equation}
T(E) = T(-\mu^2) + T(-\mu^2)~\left[ G_{0}^{+}(E) - G_{0}^{+}(-\mu^2) \right]~T(E) \; ,
\label{SKLS}
\end{equation}
\noindent
which has the same operator structure as the Eq. (\ref{LS}) for the formal LS equation, but with the potential $V$ replaced according to Eq. (\ref{veqtmu}) by the $T$-matrix at the subtraction scale $-\mu^2$ and the original free propagator replaced by a propagator with one subtraction at the same scale.

The subtracted kernel LS equation provides a finite solution for the $T$-matrix at any given energy $E$, once its value at the subtraction scale $-\mu^2$ is known. Thus, the input for the solution of Eq. (\ref{SKLS}) is  $T(-\mu^2)$, which is called ``driving term" and contains the physical information apparently lost due to the removal of the propagation through intermediate states at the scale $-\mu^2$. A simple ansatz consists in considering that the driving term is given by
\begin{equation}
T(-\mu^2)=V(-\mu^2)= V_{\rm reg} + C_{0}(-\mu^2) \; ,
\label{driving1}
\end{equation}
\noindent
where $C_{0}(-\mu^2)$ is the renormalized strength of the contact interaction at the subtraction scale $-\mu^2$, which is fixed by fitting experimental data for scattering observables and, therefore, encodes the physical information.

In order to clarify the application of the SKM scheme through a simple example, we consider the scattering of two nucleons with the LO ``pionless" EFT potential, which consists of just a pure Dirac-delta contact interaction, $V = C_0$, and gives an analytic solution for the $T$-matrix. Using the ansatz $T(-\mu^2)=V(-\mu^2)=C_{0}(-\mu^2)$, the subtracted kernel LS equation for the $T$-matrix is given in operator form by
\begin{equation}
T(E) = C_{0}(-\mu^2) + C_{0}(-\mu^2)~\left[ G_{0}^{+}(E) - G_{0}^{+}(-\mu^2) \right]~T(E) \;.
\end{equation}

As pointed out in subsection (\ref{STDDELTA2D}), due to the centrifugal barrier only S-wave scattering can occur for a pure Dirac-delta contact interaction. Using a partial-wave relative momentum space basis with normalization given by Eq. (\ref{PWnorm}), we obtain the subtracted kernel LS equation for the matrix elements of the $T$-matrix in the S-wave channel
\begin{equation}
T^{s,t}(p,p';k^2) = C_{0}^{s,t}(-\mu^2) +\frac{2}{\pi} \int dq~q^2~C_{0}^{s,t}(-\mu^2)~\left[ \frac{1}{k^2 - q^2 + i \epsilon} - \frac{1}{-\mu^2 - q^2 } \right]~T^{s,t}(q,p';k^2) \;,
\label{SKLSCI}
\end{equation}
\noindent
where $k=\sqrt{E}$ is the on-shell momentum in the center-of-mass frame and the superscripts $s$ and $t$ stand respectively for the spin-singlet channel $^1 S_0$ and the spin-triplet channel $^3 S_1$.

Solving Eq. (\ref{SKLSCI}) by matrix inversion, we obtain
\begin{equation}
T^{s,t}(p,p';k^2)= \left[ \frac{1}{C_{0}^{s,t}(-\mu^2)} + (\mu+i \; k) \right]^{-1} \; .
\label{TCI}
\end{equation}
\noindent
One should note that in the case of a pure Dirac-delta contact interaction the $T$-matrix is momentum independent.

The renormalized strength of the contact interaction, $C_{0}^{s,t}(-\mu^2)$, can then be fixed by using the scattering length $a_{s,t}$ as a physical input, which is defined by the relation $a_{s,t}=T^{s,t}(0,0,k^2=0)$. From Eq. (\ref{TCI}) we obtain
\begin{equation}
C_{0}^{s,t}(-\mu^2) = \frac{a_{s,t}}{1-\mu \; a_{s,t}} \; ,
\label{C0LO}
\end{equation}
\noindent
which leads to a $\mu$ independent $T$-matrix, given by
\begin{equation}
T^{s,t}(p,p';k^2) = \left[ \frac{1}{a_{s,t}} + i \; k \right]^{-1} \; .
\label{TLO}
\end{equation}
\noindent
Such a result is equivalent to the one obtained from LO ``pionless" EFT calculations using a momentum cutoff $\Lambda$ to regularize the divergent integral in the formal LS equation and taking the limit $\Lambda \rightarrow \infty$ \cite{cohen5,cohen6,phillips1,gegelia2,gegelia3}. Other regularization schemes, such as dimensional regularization with power divergence subtraction (PDS) \cite{kaplan1,kaplan1,kaplan3} and momentum subtractions \cite{gegelia1,gegelia2,mehen1,mehen2}, also lead to the same result in LO. Thus, this example clearly demonstrates that, in the case of a potential containing just a pure Dirac-delta contact interaction, a finite $T$-matrix can be obtained by solving the subtracted kernel LS equation with only one subtraction.

For a general singular potential containing Dirac-delta contact interactions plus regular momentum dependent terms and/or derivative contact interactions, like the $NN$ potential to a given order in ChEFT, the subtracted kernel LS equation must be solved numerically. In this case, a finite $T$-matrix can still be obtained, but using the fitting procedure described before to fix the renormalized strengths in the driving term $T(-\mu^2)$ generates a residual dependence with respect to $\mu$ in the $T$-matrix, and hence in the scattering observables. As we will show in subsection (\ref{FP}), a more rigorous procedure can be used in order to obtain scattering observables that are strictly invariant under the change of the subtraction point, which is based on a renormalization group equation for the running of the driving term $T(-\mu^2)$ with the subtraction scale $-\mu^2$. Furthermore, as we will see in subsection (\ref{SKMREC}), when the original potential specifically contains derivative contact interactions the SKM scheme must be generalized to allow for multiple subtractions, leading to a highly non-trivial dependence of the driving term on the subtraction scale $-\mu^2$.

\subsection{Renormalized Potential and Fixed-point Hamiltonian}
\label{FP}

Once the renormalized strength of the contact interaction $C_{0}(-\mu^2)$ in Eq. (\ref{driving1}) is fixed, and so the driving term $T(-\mu^2)$ is known, a renormalized potential $V_{\cal R}$ can be formally defined from Eq. (\ref{veqtmu}) \cite{skm3}:
\begin{equation}
V \rightarrow V_{\cal R} \equiv \left[1 + T(-\mu^2)~G_{0}^{+}(-\mu^2) \right]^{-1}~T(-\mu^2) \; .
\label{vren1}
\end{equation}

Multiplying both sides of Eq. (\ref{vren1}) from the left by $\left[1 + T(-\mu^2)~G_{0}^{+}(-\mu^2)\right]$ and re-arranging the terms, we obtain the integral equation (in operator form)
\begin{equation}
V_{\cal R} = T(-\mu^2) - T(-\mu^2)~G_{0}^{+}(-\mu^2)~ V_{\cal R} \; .
\label{vreneq}
\end{equation}
\noindent
which formally relates the renormalized potential $V_{\cal R}$ to the driving term $T(-\mu^2)$.

Replacing $V$ by $V_{\cal R}$ in Eq. (\ref{LS}), we obtain the LS equation for the renormalized $T$-matrix:
\begin{equation}
T_{\cal R}(E) = V_{\cal R} + V_{\cal R}~G_{0}^{+}(E)~ T_{\cal R} \; .
\label{treneq}
\end{equation}

The renormalized potential $V_{\cal R}$ is not well-defined for singular interactions. Nevertheless, for a driving term $T(-\mu^2)$ like the one defined by Eq. (\ref{driving1}), containing a regular term plus a Dirac-delta contact interaction, Eq. (\ref{treneq}) gives a finite solution for the renormalized $T$-matrix, $T_{\cal R}(E)$, which is equivalent to the one obtained from the subtracted kernel LS equation Eq.(\ref{SKLS}), i.e. $T_{\cal R}(E)=T(E)$. This result provides {\it a posteriori} justification for the formal manipulations of $V_{\cal R}$ and $V$ used to obtain Eqs. (\ref{SKLS}) and (\ref{vreneq}). In numerical calculations it is more convenient to regularize the integral implicit in Eq. (\ref{vreneq}) by introducing an ultraviolet momentum cutoff $\Lambda$, such that we can work with a well-defined renormalized potential ${\tilde V}_{\cal R}(\Lambda)$. Using this potential, we construct the LS equation for a cutoff regularized $T$-matrix,
\begin{equation}
{\tilde T}_{\cal R}(E;\Lambda) = {\tilde V}_{\cal R}(\Lambda) + {\tilde V}_{\cal R}(\Lambda)~{\tilde G}_{0}^{+}(E;\Lambda)~ {\tilde T}_{\cal R}(E;\Lambda) \; ,
\label{regtreneq}
\end{equation}
\noindent
where ${\tilde G}_{0}^{+}(E;\Lambda)$ denote the Green's function in the regularized integral.

In the limit $\Lambda \rightarrow \infty$ the result obtained by solving Eq. (\ref{regtreneq}) with the cutoff regularized potential ${\tilde V}_{\cal R}(\Lambda)$ should be the same as the one obtained from the direct solution of Eq. (\ref{treneq}) with $V_{\cal R}$. The same procedure could be applied for the numerical solution of the subtracted kernel LS equation Eq. (\ref{SKLS}). It is important to emphasize that the momentum cutoff $\Lambda$ used in this context just plays the role of an instrumental regulator for the numerical integration. The relevant scale parameter is the subtraction scale $-\mu^2$, at which the physical information is introduced through the renormalized strength of the contact interaction $C_{0}(-\mu^2)$ in the driving term $T(-\mu^2)$.

The subtraction scale $-\mu^2$ is arbitrary, and so all scattering observables should not depend on its choice. In order to fulfill this condition, the $T$-matrix must be independent of the subtraction scale $-\mu^2$:
\begin{equation}
\frac{\partial T_{\cal R}}{\partial\mu^2} = \frac{\partial T}{\partial\mu^2}=0 \; .
\label{INVt}
\end{equation}
\noindent
From Eqs. (\ref{treneq}) and (\ref{INVt}) we find that the renormalized potential $V_{\cal R}$ is also independent of $-\mu^2$,
\begin{equation}
\frac{\partial V_{\cal R}}{\partial \mu^2}=0 \; ,
\label{INVvren}
\end{equation}
\noindent
which implies that the renormalized hamiltonian $H_{\cal R}=H_{0}+V_{\cal R}$ corresponds to a renormalization group fixed-point, i.e. it is stationary with respect to the change of the subtraction scale $-\mu^2$.

Using Eqs. (\ref{vreneq}) and (\ref{INVvren}) or Eqs. (\ref{SKLS}) and (\ref{INVt}), a renormalization group equation can be obtained for the driving term $T(-\mu^2)$ \cite{skm3} in the form of a non-relativistic Callan-Symanzik equation (NRCS) \cite{CS1,CS2,CS3},
\begin{equation}
\frac{\partial T(-\mu^2)}{\partial\mu^2} = T(-\mu^2)~\frac{\partial G_{0}^{+}(-\mu^2)}{\partial\mu^2}~T(-\mu^2) \; ,
\label{CSE}
\end{equation}
\noindent
with the boundary condition, imposed at some reference scale ${\bar \mu}$, given by $T(-\mu^2)|_{\mu \rightarrow {\bar \mu}}= V(-{\bar \mu}^2)= V_{\rm reg} + C_{0}(-{\bar \mu}^2)$. The NRCS equation provides a definite rule for the evolution of the driving term $T(-\mu^2)$ with the sliding scale $\mu$, from the reference scale ${\bar \mu}$, such that the $T$-matrix remains invariant under the dislocation of the subtraction point.

\subsection{Recursiveness of the SKM for Multiple Subtractions}
\label{SKMREC}

Using a notation that will be convenient when considering multiple subtractions, the LS equation for the $T$-matrix with one subtraction, Eq. (\ref{SKLS}), can be written as
\begin{equation}
T(E) = V^{(1)}(-\mu^2) + V^{(1)}(-\mu^2)~G_{1}^{+}(E;-\mu^2)~T(E) \; ,
\label{SKLS2}
\end{equation}
\noindent
where $V^{(1)}(-\mu^2) \equiv T(-\mu^2)$ is the driving term and $G_{1}^{+}(E;-\mu^2)\equiv G_{0}^{+}(E) - G_{0}^{+}(-\mu^2)$ denote the subtracted Green's function with one subtraction, which can also be written in the form
\begin{equation}
G_{1}^{+}(E;-\mu^2) = (-\mu^2-E)~G_{0}^{+}(-\mu^2)~G_{0}^{+}(E)= F_{1}(E;-\mu^2)~G_{0}^{+}(E)\; ,
\end{equation}
\noindent
The function $F_{1}(E;-\mu^2)$ can be regarded as a form factor that modifies the free Green's function $G_{0}^{+}(E)$ by introducing a factor proportional to $q^{-2}$ in the momentum integration. Such a form factor is enough to regularize the divergence generated in the LS equation by a singular potential containing just a pure Dirac-delta contact interaction. However, for singular potentials containing terms that generate higher-order power divergences in the LS equation, like derivative contact interactions, the SKM scheme must be generalized. As shown in Refs. \cite{skm2, skm4}, in this case it is necessary to perform multiple subtractions in the kernel of the LS equation by using an iterative procedure.

For a general number of subtractions $n$, we define a $n$-fold subtracted kernel LS equation, given by
\begin{equation}
T(E) = V^{(n)}(E;-\mu^2) + V^{(n)}(E;-\mu^2)~G_{n}^{+}(E;-\mu^2)~T(E) \; .
\label{LSn}
\end{equation}
\noindent
The $n$-fold subtracted Green's function $G_{n}^{+}(E;-\mu^2)$ is defined by
\begin{equation}
G_{n}^{+}(E;-\mu^2) \equiv \left[(-\mu^2-E)~ G_{0}^{+}(-\mu^2) \right]^{n}~G_{0}^{+}(E)= F_{n}(E;-\mu^2)~G_{0}^{+}(E)  \; .
\label{Gn}
\end{equation}
\noindent
Note that the $n$-fold form factor $F_{n}(E;-\mu^2)$ introduces a factor proportional to $q^{-2n}$ in the momentum integration, thus regularizing divergences to order $q^{2n-1}$. The driving term $V^{(n)}(E;-\mu^2)$ for the $n$-fold subtracted LS equation is recursively constructed through an iterative procedure, starting from $V^{(1)}(-\mu^2)$. The recursion formula is given by

\begin{equation}
V^{(m)}(E;-\mu^2) = {\bar V}^{(m)}(E;-\mu^2) + V^{(m)}_{\rm sing}(-\mu^2)  \; ,
\label{Vm}
\end{equation}
\noindent
where
\begin{equation}
{\bar V}^{(m)}(E;-\mu^2) = \left[1 - (-\mu^2-E)^{m-1}~V^{(m-1)}(E;-\mu^2)~ G_{0}^{+}(-\mu^2)^m \right]^{-1} ~ V^{(m-1)}(E;-\mu^2)  \; ,
\label{Vbarm}
\end{equation}
\noindent
and the term $V^{(m)}_{\rm sing}(-\mu^2)$ contains the higher-order singular interactions that generate divergent integrals which can be regularized by performing $m$ subtractions. One should note from Eq. (\ref{Vm}) the that the driving term $V^{(m)}(E;-\mu^2)$ at each iteration is derived in two steps. First, we calculate ${\bar V}^{(m)}(E;-\mu^2)$ from $V^{(m-1)}(E;-\mu^2)$, solving the integral equation obtained by manipulating Eq. (\ref{Vbarm}):
\begin{equation}
{\bar V}^{(m)}(E;-\mu^2) =V^{(m-1)}(E;-\mu^2)+(-\mu^2-E)^{m-1}~V^{(m-1)}(E;-\mu^2)~ G_{0}^{+}(-\mu^2)^m ~ {\bar V}^{(m)}(E;-\mu^2)  \; .
\label{Vbarmint}
\end{equation}
\noindent
Then, we introduce the corresponding higher-order singular interactions in the driving term by adding $V^{(m)}_{\rm sing}(-\mu^2)$.

As in the case of one subtraction, once the driving term $V^{(n)}(E;-\mu^2)$ with $n$ subtractions is known we can formally manipulate Eq. (\ref{LSn}) to define a renormalized potential,
\begin{equation}
V_{\cal R} \equiv \left[1 + V^{(n)}(E;-\mu^2)\left(G_{0}^{+}(E)- G_{n}^{+}(E;-\mu^2) \right) \right]^{-1}~V^{(n)}(E;-\mu^2) \; ,
\label{vrenn}
\end{equation}
\noindent
which is iterated in the LS equation Eq. (\ref{treneq}) for $T_{\cal R}(E)$ and leads to a renormalization group fixed-point hamiltonian. In the same way, a NRCS equation for the driving term $V^{(n)}(E;-\mu^2)$ can be obtained from the invariance of the n-fold subtracted $T$-matrix with respect to the subtraction scale $-\mu^2$,
\begin{equation}
\frac{\partial V^{(n)}(E;-\mu^2)}{\partial\mu^2} = -V^{(n)}(E;-\mu^2)~\frac{\partial G_{n}^{+}(E;-\mu^2)}{\partial\mu^2}~V^{(n)}(E;-\mu^2) \; ,
\label{CSEn}
\end{equation}
\noindent
with the boundary condition given by $V^{(n)}(E;-\mu^2)|_{\mu \rightarrow {\bar \mu}}= V^{(n)}(E;-{\bar \mu}^2)$ imposed at some reference scale $\bar \mu$.

As an example to illustrate how the SKM scheme works when multiple subtractions are performed, we consider the scattering of two nucleons in the $S$-wave channel with the NLO ``pionless" EFT potential, which consists of a Dirac-delta contact interaction plus a second-order derivative contact interaction. In a partial-wave relative momentum space basis, the matrix elements of such a potential are given by:
\begin{equation}
V(p,p') = C_0 + C_2~(p^2+{p'}^2) \; .
\label{vc0c2}
\end{equation}
\noindent
Even thought the NLO potential contains only terms of order $q^2$, when iterated in the LS equation it generates integrals that diverge as much as $q^5$. Therefore, we have to construct a 3-fold subtracted kernel LS equation in order to get a finite result for the $T$-matrix. The iterative procedure starts from the $T$-matrix with one subtraction at $-\mu^2$, such that:
\begin{equation}
V^{(1)}(p,p';-\mu^2)=C_0(-\mu^2) \; .
\label{v1}
\end{equation}

The next step is to derive the driving term $V^{(2)}$. We calculate ${\bar V}^{(2)}$ from $V^{(1)}$ through Eq. (\ref{Vbarmint}):
\begin{equation}
{\bar V}^{(2)}(p,p';k^2;-\mu^2) = V^{(1)}(p,p';-\mu^2) + \frac{2}{\pi} ~ \int_{0}^{\infty} dq ~ q^2 ~ \left(\frac{\mu^2+k^2}{\mu^2+q^2}\right)~ \frac{V^{(1)}(p,q;-\mu^2)}{(-\mu^2-q^2)}~{\bar V}^{(2)}(q,p';k^2;-\mu^2) \; .
\label{v2bareq}
\end{equation}
\noindent
Solving this equation by matrix inversion, we obtain
\begin{equation}
{\bar V}^{(2)}(p,p';k^2;-\mu^2)=\left[\frac{1}{C_0(-\mu^2)}+ I_0 \right]^{-1} \; ,
\label{v2bar}
\end{equation}
\noindent
with
\begin{equation}
I_0 \equiv I_0(k^2;\mu^2)= \frac{2}{\pi} ~\int_{0}^{\infty} dq ~q^2~\frac{(\mu^2+k^2)}{(\mu^2+q^2)^2} = \frac{(\mu^2+k^2)}{2\mu} \; .
\label{I0}
\end{equation}
\noindent
Since there are no higher-order singular terms to be added at this iteration, we simply have
\begin{equation}
V^{(2)}(p,p';k^2;-\mu^2)={\bar V}^{(2)}(p,p';k^2;-\mu^2)=\left[\frac{1}{C_0(-\mu^2)}+ \frac{(\mu^2+k^2)}{2\mu} \right]^{-1} \; .
\label{v2}
\end{equation}

The last step of the iterative procedure is to derive the driving term $V^{(3)}$. First, we calculate ${\bar V}^{(3)}$ from $V^{(2)}$ through the integral equation
\begin{equation}
{\bar V}^{(3)}(p,p';k^2;-\mu^2) = V^{(2)}(p,p';k^2;-\mu^2) + \frac{2}{\pi} ~ \int_{0}^{\infty} dq ~ q^2 ~ \left(\frac{\mu^2+k^2}{\mu^2+q^2}\right)^{2}~ \frac{V^{(2)}(p,q;k^2;-\mu^2)}{(-\mu^2-q^2)}~{\bar V}^{(3)}(q,p';k^2;-\mu^2) \; ,
\label{v3bareq}
\end{equation}
\noindent
which is solved by matrix inversion giving
\begin{equation}
{\bar V}^{(3)}(p,p';k^2;-\mu^2)= {\tilde C}_0(k^2;-\mu^2)=\left[\frac{1}{C_0(-\mu^2)}+ I_0 + I_1 \right]^{-1}\; ,
\label{v3bar}
\end{equation}
\noindent
with
\begin{equation}
I_1 \equiv I_1(k^2;\mu^2)= \frac{2}{\pi} ~\int_{0}^{\infty} dq ~q^2~\frac{(\mu^2+k^2)^2}{(\mu^2+q^2)^3} = \frac{(\mu^2+k^2)^2}{(2\mu)^3} \; .
\label{I1}
\end{equation}
\noindent
Then, we add the term $V^{(3)}_{\rm sing}(p,p';-\mu^2)=C_2(-\mu^2)~(p^2+{p'}^2)$ to obtain
\begin{equation}
V^{(3)}(p,p';k^2;-\mu^2)=\left[\frac{1}{C_0(-\mu^2)}+ \frac{(\mu^2+k^2)}{2\mu} + \frac{(\mu^2+k^2)^2}{(2\mu)^3} \right]^{-1}+C_2(-\mu^2)~(p^2+{p'}^2)\; .
\label{v3}
\end{equation}
\noindent
The resulting 3-fold subtracted LS equation for the $T$-matrix is then given by:
\begin{eqnarray}
T(p,p';k^2) &=& V^{(3)}(p,p';k^2;-\mu^2) +\frac{2}{\pi} \int_{0}^{\infty} dq~q^2~\left(\frac{\mu^2+k^2}{\mu^2+q^2}\right)^3~ \frac{V^{(3)}(p,q;k^2;-\mu^2)}{k^2 - q^2 + i \epsilon}~T(q,p';k^2) \; .
\label{SKLSn}
\end{eqnarray}

Using the method described in Refs. \cite{birse1,birse2,cohen6,gegelia2}, suitable for separable potentials, Eq. (\ref{SKLSn}) can be solved analytically. The driving term $V^{(3)}$ given in Eq. (\ref{v3}) can be written in the form of a two-term separable potential,
\begin{equation}
V^{(3)}(p,p';k^2;-\mu^2)=\sum_{i,j=0}^{1} p^{2i}\; \Lambda_{ij}(k^2;\mu^2)\; p'^{2j} \;,
\label{v3sep}
\end{equation}
\noindent
where $\Lambda_{ij}(k^2;\mu^2)$ are the matrix elements of
\begin{equation}
{\bf \Lambda}(k^2;\mu^2)=\left( \begin{array}{ccc}
{\tilde C}_0 & C_2\\
C_2 & 0
\end{array} \right) \; .
\end{equation}
\noindent
with ${\tilde C}_0 \equiv {\tilde C}_0(k^2;-\mu^2)$ and $C_2 \equiv C_2(-\mu^2)$.

The solution of Eq. (\ref{SKLSn}) can then be written in the form
\begin{equation}
T(p,p';k^2)=\sum_{i,j=0}^1 p^{2i}\; {\cal T}_{ij}(k^2;\mu^2)\; p'^{2j} \; .
\label{Tmat}
\end{equation}
\noindent
The unknown matrix ${\bf {\cal T}}(k^2;\mu^2)$ satisfies the equation
\begin{equation}
{\bf {\cal T}}(k^2;\mu^2)={\bf \Lambda}(k^2;\mu^2) + {\bf \Lambda}(k^2;\mu^2) \;{\bf {\cal J}}(k^2;\mu^2) \;{\bf {\cal T}}(k^2;\mu^2)\; ,
\label{tau}
\end{equation}
\noindent
where
\begin{equation}
{\bf {\cal J}}(k^2;\mu^2)=\left( \begin{array}{ccc}
J_0 & J_1\\
J_1 & J_2
\end{array} \right) \; ,
\end{equation}
\noindent
with
\begin{equation}
J_{n}\equiv J_{n}(k^2;\mu^2)=
\frac{2}{\pi}~\int_{0}^{\infty} dq~\left(\frac{\mu^2+k^2}{\mu^2+q^2}\right)^3~\frac{q^{2n+2}}{k^2 - q^2 + i \epsilon}\; \; \; (n=0,1,2).
\end{equation}
\noindent
Note that the factor $F_{3}(k^2;-\mu^2)=(\mu^2+k^2)^{3}/({\mu^2+q^2})^{3}$, introduced by the subtraction procedure, regularizes the otherwise-divergent integrals $J_{n}(k^2;\mu^2)$ for $n=0,1,2$.

Solving Eq. (\ref{tau}) for ${\bf {\cal T}}(k^2;\mu^2)$ and substituting in Eq. (\ref{Tmat}) yields
\begin{eqnarray}
T(p,p';k^2)=\frac{{\tilde C}_0+C_2~(p^2 + p'^2)+C_2^{2}~[J_2-(p^2 + p'^2)~J_1+p^2 ~ p'^2~ J_0]}
{1-{\tilde C}_0 ~ J_0 - 2~C_2~J_1 -C_2^2~[J_2 \; J_0-J_1^2 ]}\; .
\end{eqnarray}

Using the identity
\begin{equation}
\frac{2}{\pi}~\int_{0}^{\infty} dq~f(q)~\frac{q^{2n+2}}{k^2 - q^2 + i \epsilon}=
\frac{2}{\pi}~{\cal P} \int_{0}^{\infty} dq~f(q)~\frac{q^{2n+2}}{k^2 - q^2}\; - i~k^{2n+1}.
\end{equation}
\noindent
and the relation (valid for $n \geq 0$)
\begin{equation}
\frac{2}{\pi}~{\cal P} \int_{0}^{\infty} dq~f(q)~\frac{q^{2n+2}}{k^2 - q^2}=-\frac{2}{\pi}~ \int_{0}^{\infty} dq~f(q)~q^{2n}+k^2~\frac{2}{\pi}~{\cal P} \int_{0}^{\infty} dq~f(q)~\frac{q^{2n}}{k^2 - q^2} \; ,
\end{equation}
\noindent
the integrals $J_{n}$ can be written in the form
\begin{eqnarray}
&&J_{0}=W_{0}+k^2~W-i~k \; ;\\
&&J_{1}=W_{1}+k^2~W_{0}+k^4~W-i~k^3\; ;\\
&&J_{2}=W_{2}+k^2~W_{1}+k^4~W_{0}+k^6~W-i~k^5 \; ,
\end{eqnarray}
\noindent
with
\begin{equation}
W_{n} \equiv W_{n}(k^2;\mu^2)=-\frac{2}{\pi}~\int_{0}^{\infty} dq~\left(\frac{\mu^2+k^2}{\mu^2+q^2}\right)^3~q^{2n}\; ; \;\;\; W \equiv W(k^2;\mu^2)=\frac{2}{\pi}~{\cal P} \int_{0}^{\infty} dq~\left(\frac{\mu^2+k^2}{\mu^2+q^2}\right)^3~\frac{1}{k^2 - q^2}\; .
\end{equation}

The inverse on-shell 3-fold subtracted $T$-matrix ($p=p'=k$) can then be written in the form
\begin{eqnarray}
\frac{1}{T(k,k;k^2)}=\frac{[C_2~W_{1}-1]^2}{{\tilde C}_0 + C_2^2~W_{2}+k^2~C_2~[2-C_2~W_{1}]}-(W_{0}+k^2~W)+i~k\; ,
\label{TonNLO}
\end{eqnarray}
\noindent
One should note that the unitarity of the 3-fold subtracted $T$-matrix is ensured by the term $i~k$.

\begin{figure}[t]
\centerline{\includegraphics[width=7.2 cm]{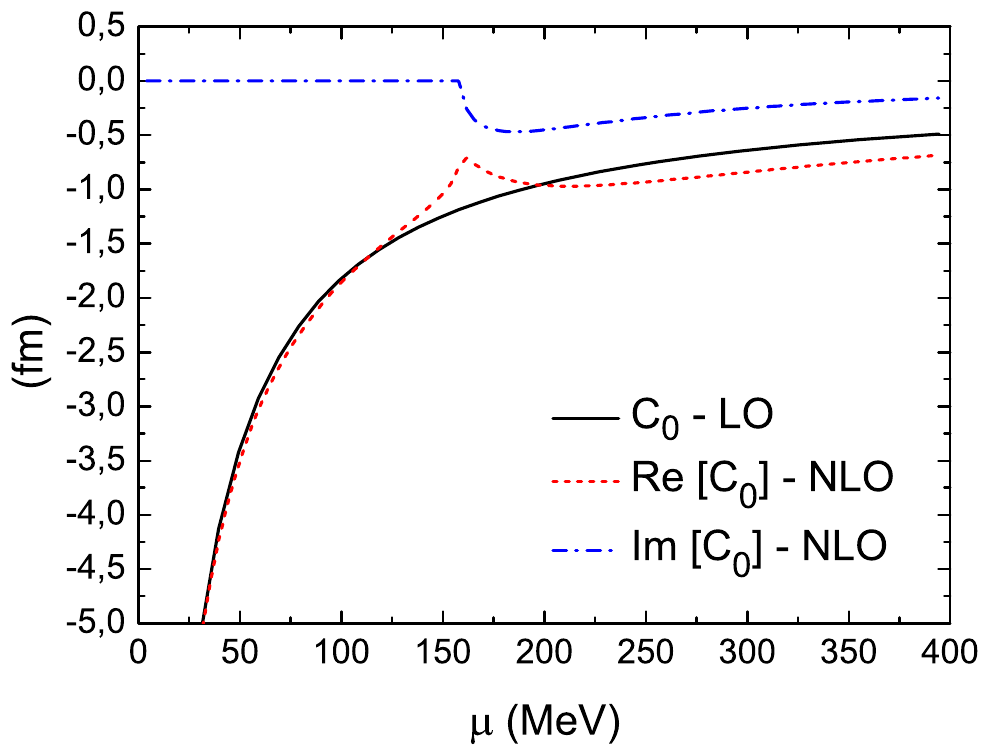}\hspace*{.8cm}\includegraphics[width= 7.2 cm]{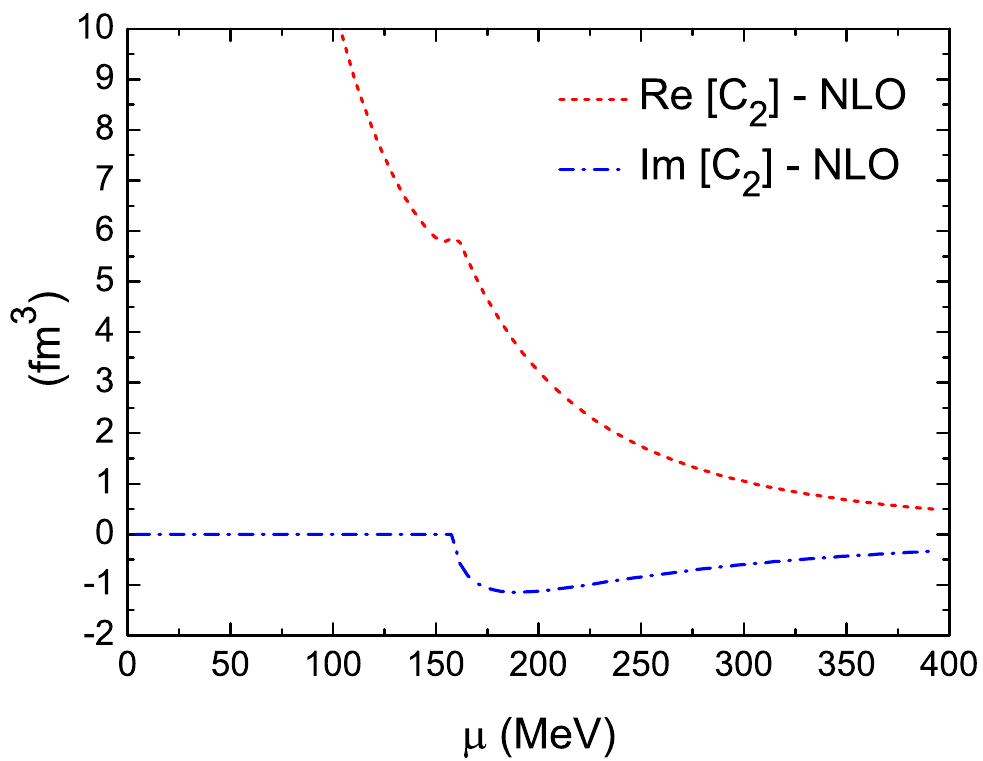}}
\caption{(Color online) Renormalized strengths $C_0(-\mu^2)$ and $C_2(-\mu^2)$ adjusted to fit the $^1 S_0$ channel scattering length and effective range.}
\label{fig9}
\end{figure}

The renormalized strengths $C_0(-\mu^2)$ and $C_2(-\mu^2)$ can be fixed by fitting the experimental values of the scattering length $a$ and the effective range $r_e$. Expanding Eq. (\ref{TonNLO}) in powers of $k$ yields
\begin{eqnarray}
\frac{1}{T(k,k;k^2)}=A_0[C_0,C_2;-\mu^2]+A_1[C_0,C_2;-\mu^2]~k^2 + {\cal O}(k^4)+i~k \; ,
\label{TonNLOexp}
\end{eqnarray}
\noindent
with the coefficients $A_0[C_0,C_2;-\mu^2]$ and $A_2[C_0,C_2;-\mu^2]$ given by
\begin{eqnarray}
A_0[C_0,C_2;-\mu^2]=\frac{64 + 64 ~C_0~ \mu + 16 ~C_2 ~\mu^3 + 10 ~C_0~ C_2~ \mu^4 - 8~ C_2^2~ \mu^6 - 5 ~C_0~ C_2^2~ \mu^7 }{64~ C_0 - 24 ~C_2^2 ~\mu^5 - 15 ~C_0~ C_2^2~ \mu^6 } \; ;
\end{eqnarray}
\begin{eqnarray}
A_1[C_0,C_2;-\mu^2]= &-&\frac{256 \left[32~C_2 - 8~ C_2^2~ \mu^3 + C_2^3 ~\mu^6 + 2 C_2^4 ~\mu^9 \right]}{\left[64 ~C_0 - 24~ C_2^2~ \mu^5 - 15~ C_0~ C_2^2 ~\mu^6\right]^2} \\ \nonumber \\
&-& \; \frac{64 ~C_0 \left[112~C_2~\mu - 82~C_2^2~ \mu^4 + 5~ C_2^3~ \mu^7 + 10~ C_2^4~ \mu^{10} \right]}{\left[64 ~C_0 - 24~ C_2^2~ \mu^5 - 15~ C_0~ C_2^2 ~\mu^6\right]^2}\nonumber \\ \nonumber \\
&-&\frac{4~ C_0^2 \left[128~C_2 ~\mu^2 - 632~ C_2^2 ~\mu^5 + 25 ~C_2^3~ \mu^8 + 50~ C_2^4 ~\mu^{11} \right]}{\left[64 ~C_0 - 24~ C_2^2~ \mu^5 - 15~ C_0~ C_2^2 ~\mu^6\right]^2} \; .
\end{eqnarray}
\noindent
Matching Eq. (\ref{TonNLOexp}) to the first two terms in the effective range expansion (ERE),
\begin{eqnarray}
\frac{1}{T_{\rm er}(k,k;k^2)}=-\left[k~{\rm cot}~\delta(k)-i~k \right]=-\left[-\frac{1}{a}+\frac{1}{2}~r_e~k^2 + {\cal O}(k^4)-i~k \right] \; ,
\end{eqnarray}
\noindent
we obtain a system of two coupled non-linear equations for $C_0(-\mu^2)$ and $C_2(-\mu^2)$,
\begin{eqnarray}
A_0[C_0,C_2;-\mu^2]= \frac{1}{a} \; ; \; \; \; A_1[C_0,C_2;-\mu^2]= -\frac{1}{2}~r_e \; .
\label{A0A1}
\end{eqnarray}
\noindent
The solutions of these equations for the renormalized strengths $C_0(-\mu^2)$ and $C_2(-\mu^2)$ are highly non-trivial functions of the scale $\mu$. In Fig. (\ref{fig9}) we show one of the solutions obtained by using parameters for the $NN$ scattering in the $^1 S_{0}$ channel as physical input, i.e. an unnaturally large scattering length $a=-23.7~{\rm fm}$ and a positive effective range $r_e=2.7~{\rm fm}$. For comparison, we also show the result obtained for $C_0(-\mu^2)$ in the calculation with just a pure Dirac-delta contact interaction, given by Eq. (\ref{C0LO}). As one can observe, for values of the scale $\mu$ up to $\mu_{\rm max}\sim 0.8~{\rm fm}^{-1}$ the solutions for $C_0(-\mu^2)$ and $C_2(-\mu^2)$ are real. For $\mu > \mu_{\rm max}$ the solutions become complex, corresponding to a non-Hermitian driving term, and thus lead to the violation of unitarity. This result implies that in order to fit a positive effective range while preserving the unitarity of the subtracted $T$-matrix we cannot use an arbitrarily large value for the scale $\mu$. Such a behavior is similar to what is found in NLO ``pionless" EFT calculations using the cutoff regularization scheme \cite{cohen1,cohen3,cohen4,cohen5,cohen6,phillips1,gegelia2,gegelia3} and can be related to the Wigner bound on the effective range \cite{wigner}, a general result based on the physical principles of causality and unitarity which states that for a Hermitian potential of range $R$ the effective range $r_e$ is constrained by
\begin{equation}
r_e \leq 2~\left[R - \frac{R^2}{a}+\frac{R^2}{a^2}\right]\; .
\label{wigner}
\end{equation}
\noindent
If we take $ R \sim 1/\mu$, then for $a=-23.7~{\rm fm}$ and $r_e=2.7~{\rm fm}$ we obtain the condition $\mu < 0.78~{\rm fm}^{-1}$ which is consistent with the results of our calculations.

\begin{figure}[t]
\centerline{\includegraphics[width=7.2 cm]{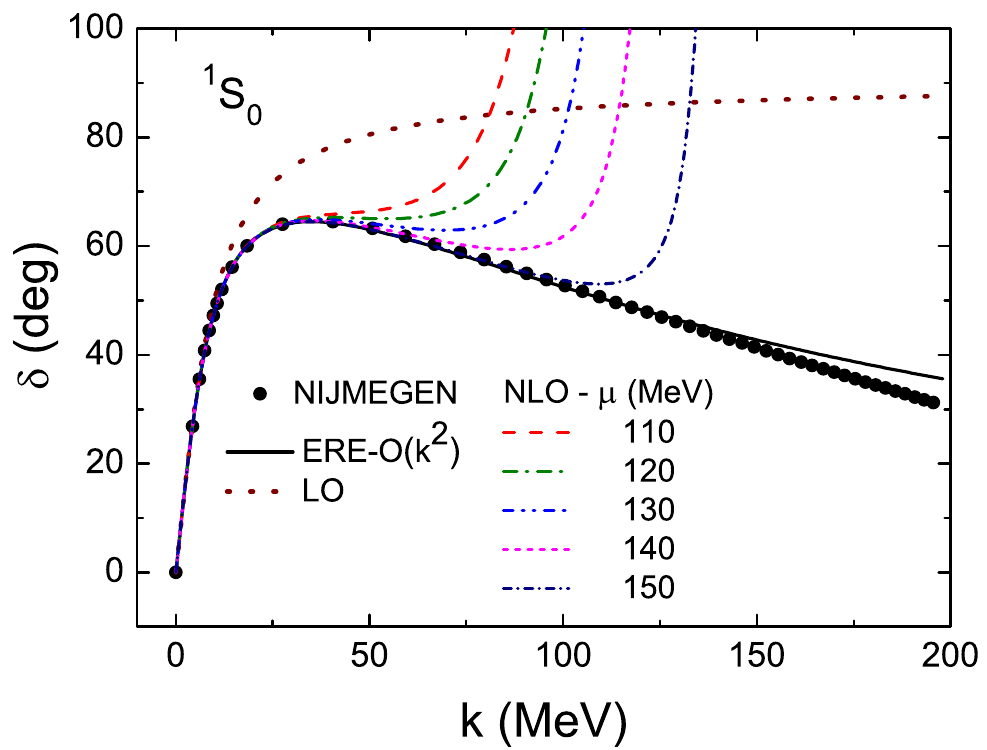}\hspace*{.8cm}\includegraphics[width= 7.2 cm]{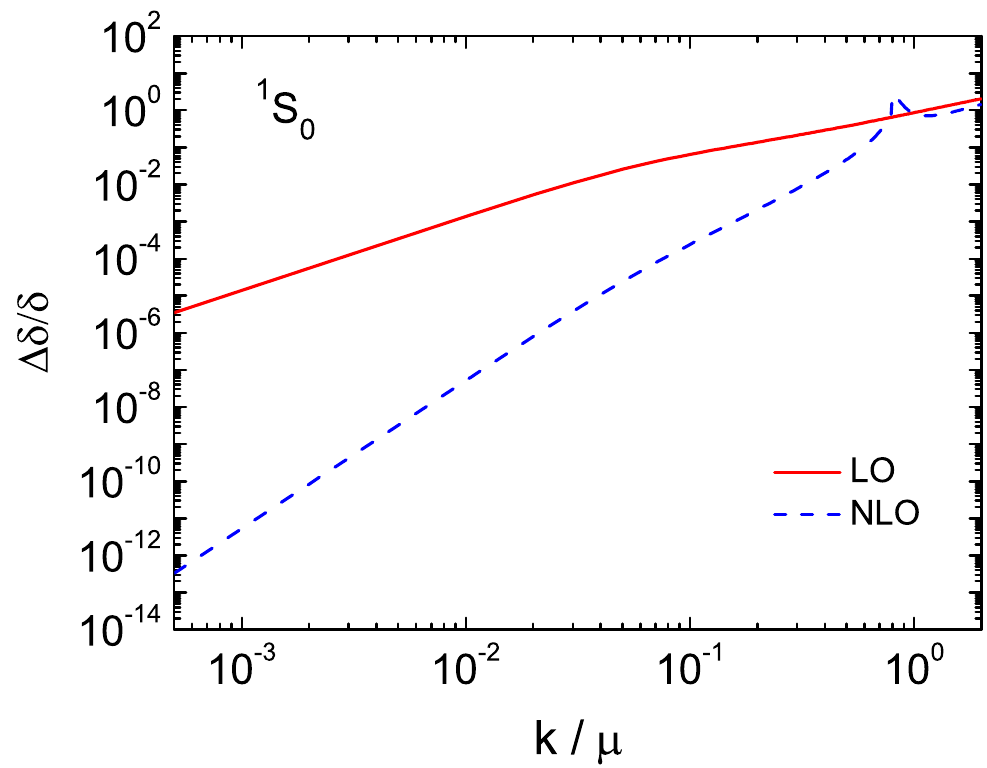}}
\caption{Left panel: (Color online) Left panel: Phase-shifts in the $^1 S_{0}$ channel as a function of the on-shell momentum for the LO and NLO potentials renormalized through the SKM procedure, compared to the results from the Nijmegen PWA and from the ERE to order $k^2$; Right panel: relative errors in the phase-shifts (with respect to the results from the ERE) as a function of $(k/\mu)$ for the LO and NLO potentials ($\mu=130~{\rm MeV}$). }
\label{fig10}
\end{figure}

In the left panel of Fig. (\ref{fig10}) we show the $NN$ phase-shifts in the $^1 S_{0}$ channel obtained for the LO potential from Eq. (\ref{TLO}) and those obtained for the NLO potential from Eq. (\ref{TonNLO}) with several values of the scale $\mu$, compared to the results from the Nijmegen partial-wave analysis (PWA). We also show the $NN$ phase-shifts obtained from the ERE to order $k^2$, which provides a very good fit to the results from the Nijmegen PWA up to on-shell momenta $k$ of order $m_{\pi} \sim 140$ MeV. As one can observe, the phase-shifts for the NLO potential renormalized through the SKM procedure are in good agreement with those obtained from the ERE up to a given value of the on-shell momentum that becomes larger as the scale $\mu$ increases. Note that for very low on-shell momenta (up to $\sim 40$ MeV) the phase-shifts are nearly independent of the scale $\mu$. Such results would be expected, since the renormalized strengths $C_0(-\mu^2)$ and $C_2(-\mu^2)$ in the NLO potential are fixed by matching the ERE to order $k^2$ and so dependence on the scale $\mu$ in the inverse on-shell 3-fold subtracted $T$-matrix should start at order $(k/\mu)^4$. This can be clearly verified from the log-log plots for the absolute value of the relative errors in the phase-shifts with respect to the results from the ERE, shown in the right panel of Fig. (\ref{fig10}), whose slopes are given by the dominant power of $k/\mu$ in the errors.

It is important to emphasize that once a reference scale ${\bar \mu}$ is specified and the renormalized strengths are fixed to fit the observables used as physical input (e.g., the scattering length and the effective range), the subtraction point $-\mu^2$ can be changed without modifying the results for the calculated phase-shifts, as long as the driving term $V^{(3)}(p,p';k^2;-\mu^2)$ in the 3-fold subtracted LS equation for the $T$-matrix is evolved through the NRCS equation Eq. (\ref{CSEn}) with the boundary condition given by $V^{(3)}(p,p';k^2;-\mu^2)|_{\mu \rightarrow {\bar \mu}}=V^{(3)}(p,p';k^2;-{\bar \mu^2})$.

We have just shown two simple analytical examples of how to renormalize $NN$ singular interactions using the SKM approach with one and multiple subtractions. We close this section with some remarks about the advantages and disadvantages of the SKM formalism and its possible extension to renormalize $3N$ interactions.

As pointed out in the introduction, an advantage of the SKM formalism is that it can be extended to any derivative order of the contact interactions, through an interative process involving multiple subtractions. Such an iterative process allows for a systematic treatment of ChEFT $NN$ potentials up to higher-orders, which include interactions that generate higher divergences. This also makes the SKM convenient to test alternative power-counting schemes in which higher-order contact interactions are promoted to lower-order \cite{bira3}. Furthermore, the renormalization group invariance of the SKM, ensured by the evolution of the driving term through a NRCS-type equation, allows one to arbitrarily move the renormalization scale (i.e. the subtraction point $\mu$) while maintaining the observables invariant. One can even slide the renormalization scale beyond the limits imposed by the Wigner bound, which is not possible within the cutoff regularization scheme. A disadvantage of the SKM is that the computational load increases as more subtractions are performed, since the recursive process requires more matrix inversions. So, while the cutoff regularization scheme demands only one inversion to compute the $T$-matrix, the SKM with $n$ subtractions requires $n-1$ inversions to compute the driving term $V^{(n)}$ plus one inversion to compute the $n$-fold subtracted $T$-matrix.

The SKM formalism presented here provides a powerful method to renormalize two-body interactions. Similar subtractive renormalization methods \cite{subren1,subren2,subren3,subren4}, developed in the context of EFT for three-body systems with contact interactions, suggest that the SKM formalism can be generalized to treat three-body interactions. As in the SKM approach with one subtraction, in the methods described in Refs. \cite{subren1,subren2,subren3,subren4} renormalized amplitudes are obtained by performing a subtraction in the three-body scattering equations, so we believe that an extension of the SKM approach to derive renormalized ChEFT potentials including $3N$ interactions might work as well. Once obtained, such SKM renormalized $3N$ potentials can then be used as the starting point for the SRG evolution. Investigating the possibility of applying the SKM formalism to the $3N$ system is certainly a project we intend to pursue in forthcoming works.

\section{SRG Evolution of $NN$ Potentials in the SKM Approach}
\label{SRGSKM}

\subsection{SKM for the $NN$ Interaction in LO ChEFT}
\label{SKMNN}

In this subsection we present a detailed and systematic analysis of the SKM approach applied to the $NN$ interaction in LO ChEFT, which consists of the OPEP plus a Dirac-delta contact interaction. This is a simple and convenient example to perform a detailed investigation, since only one subtraction is required to renormalize the LO potential. Such an application has already been described in Ref. \cite{skm1} for the singlet $^1S_0$ channel and the triplet $^3S_1-^3D_1$ coupled channel. Here, we restrict our calculations to the $^1S_0$ channel only. In Ref. \cite{skm4} the SKM approach was applied to the $NN$ interaction in the $^1S_0$ channel with the inclusion of a second-order derivative contact interaction in addition to the LO ChEFT interaction, requiring the use of a 3-fold subtracted kernel LS equation in order to regularize the $q^5$ divergences. The implementation of the method with the full $NN$ interaction in NLO ChEFT, which includes the NLO two-pion exchange potential (TPEP), was considered in Ref. \cite{skm5} in the case of the $^1S_0$ channel. In Ref. \cite{skm6} the method was applied to describe peripheral waves in uncoupled channels with the inclusion of the NNLO TPEP, requiring the use of a 4-fold subtracted kernel LS equation.

The $NN$ interaction in LO ChEFT is given by
\begin{eqnarray}
V(\vec p, \vec {p'})=-\frac{g_a^2}{4(2\pi)^3f_\pi^2} \vec\tau_1\cdot\vec\tau_2
\frac{\vec\sigma_1\cdot(\vec {p'}-\vec {p})\;
\vec\sigma_2\cdot(\vec {p'}-\vec {p})}
{ (\vec{p'}-\vec{p})^2+m_\pi^2}+ \frac{1}{2\pi^2}
\left[
C_0^{s} \; \left( \frac{1 - \vec\tau_1\cdot\vec\tau_2}{4}  \right)
+  C_0^{t} \left(\frac{3 +\vec\tau_1\cdot\vec\tau_2}{4} \right)
\right], \label{VLO}
\end{eqnarray}
\noindent
where $\vec\tau_i$ and $\vec\sigma_i$ ($i=1,2$) are the usual isospin and spin Pauli operators, $g_a=1.25$ is the axial coupling constant, $f_\pi =93$ MeV is the pion weak-decay constant and $m_\pi=138$ MeV is the pion mass. The coefficients $C_0^{s}$ and $C_0^{t}$ correspond to the strengths of the contact interactions respectively for the $^1 S_0$ and $^3S_1-^3D_1$ channels.

In a partial-wave relative momentum space basis with normalization given by Eq. (\ref{PWnorm}), the matrix elements of the LO ChEFT potential projected in the $^1 S_0$ channel are given by
\begin{eqnarray}
V(p,p')=V_{1\pi}(p,p') + C_0^{s} \; ,
\label{VLO1S0}
\end{eqnarray}
\noindent
where $V_{1\pi}(p,p')$ are the corresponding OPEP matrix elements, given by
\begin{eqnarray}
V_{1\pi}(p, p^\prime)
&=&
\frac{g_a^2}{32\pi f_\pi^2}
\left(2-\int^1_{-1}dx \frac{m_\pi^2}{p^2+{p'}^2-2 p {p'}x+m_\pi^2}
\right).
\label{VOPE1S0}
\end{eqnarray}

We implement the SKM procedure by using the $K$-matrix instead of the $T$-matrix. We also introduce an ultraviolet momentum cutoff $\Lambda$ which is convenient for the numerical calculations when we further consider the evolution of the SKM renormalized potential through the SRG transformation. Such a cutoff allows one to work with a finite-range gaussian grid of momentum integration points, which is required for an efficient numerical solution of Wegner's flow equation using a Runge-Kutta solver. As pointed before, the cutoff $\Lambda$ must be regarded as an instrumental regulator for the numerical integrations, whose effects on the calculated quantities should vanish in the limit $\Lambda \rightarrow \infty$.

The subtracted kernel LS equation for the cutoff regularized $K$-matrix in the $^1 S_0$ channel, with the momentum cutoff $\Lambda$ included through a step function $\theta(\Lambda-q)$, is given by
\begin{eqnarray}
{\tilde K}(p,p';k^2;\Lambda) =  {\tilde V}^{(1)}(p,p';-\mu^2;\Lambda) +
\frac{2}{\pi}~{\cal P}\int_0^\infty dq \; q^2~\theta(\Lambda-q)~
\left(\frac{\mu^2+k^2}{\mu^2+q^2}\right)
\frac{ {\tilde V}^{(1)}(p,q;-\mu^2;\Lambda)}{k^2-q^2}
{\tilde K}(q,p'; k^2;\Lambda) \; .
\label{KLS1SO}
\end{eqnarray}
\noindent
Note that in the limit $\Lambda \rightarrow \infty$ the result obtained by solving this equation should be the same as the one obtained by solving the equation for the unregularized $K$-matrix, i.e. ${\tilde K}(p,p';k^2;\Lambda \rightarrow \infty)\rightarrow K(p,p';k^2)$.

The driving term ${\tilde V}^{(1)}(p,p';-\mu^2;\Lambda)$ at the subtraction scale $-\mu^2$ in Eq. (\ref{KLS1SO}) is defined by the ansatz
\begin{eqnarray}
{\tilde V}^{(1)}(p,p';-\mu^2;\Lambda) \equiv {\tilde K}(p,p';-\mu^2;\Lambda)=V_{1\pi}(p,p') + {\tilde C}_0^{s}(-\mu^2;\Lambda) \ .
\label{KLO1SOMU}
\end{eqnarray}
\noindent
The renormalized strength of the contact interaction ${\tilde C}_0^{s}(-\mu^2;\Lambda)$ is fixed at the scale $\mu$ for a given momentum cutoff $\Lambda$ by fitting the scattering length $a_s$ in the $^1 S_0$ channel. We solve Eq. (\ref{KLS1SO}) numerically and evaluate the scattering length from the on-shell $K$-matrix ($p=p'=k$) at $E=k^2=0$ through the relation $a_s={\tilde K}(0,0;k^2=0;\Lambda)$, adjusting the value of ${\tilde C}_0^{s}(-\mu^2;\Lambda)$ so as to reproduce the experimental value $a_s=-23.7~{\rm fm}$ with a given accuracy.

\begin{figure}[p]
\centerline{\includegraphics[width=7.0 cm]{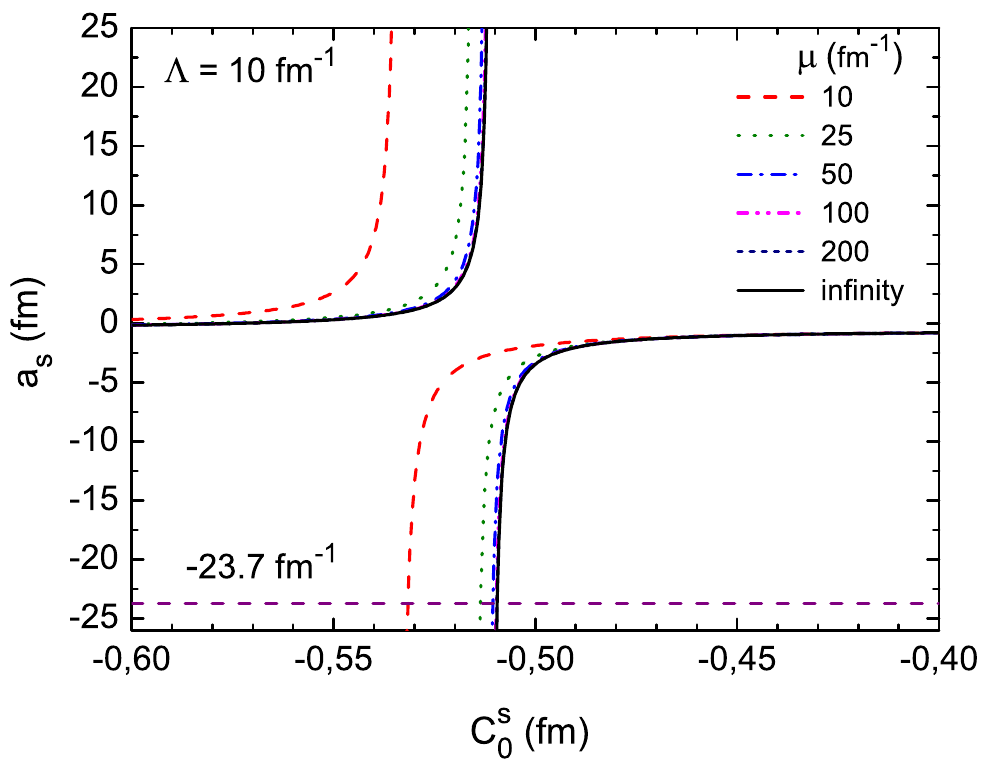}\hspace*{.5cm}\includegraphics[width=7.0 cm]{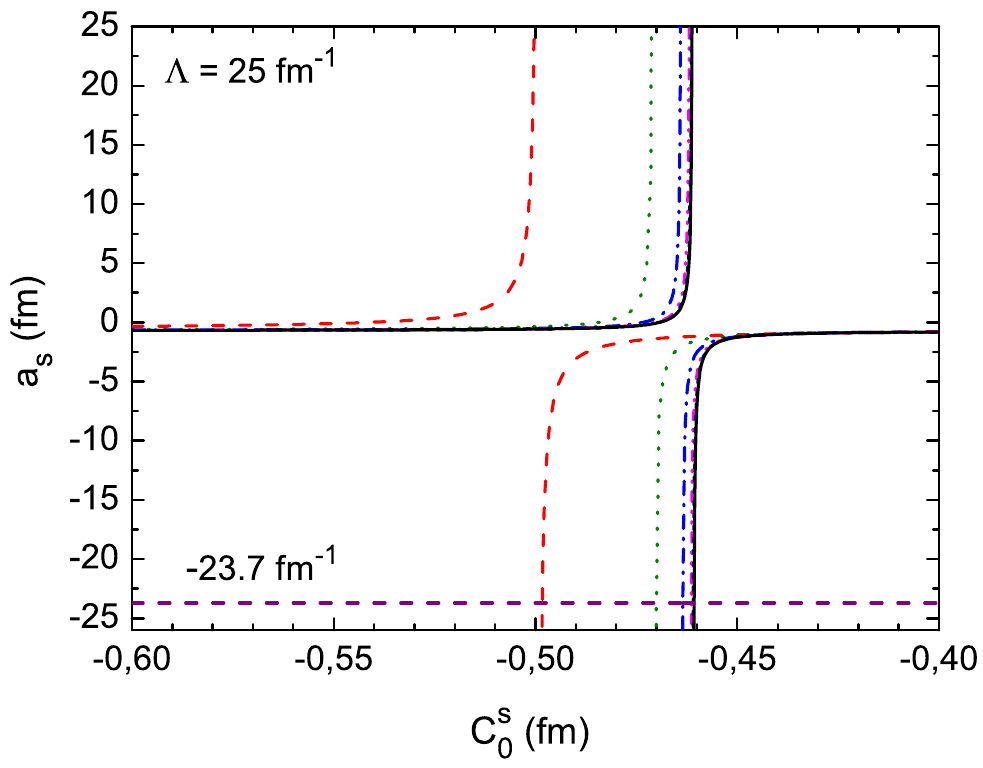}}\vspace*{.8cm}
\centerline{\includegraphics[width=7.0 cm]{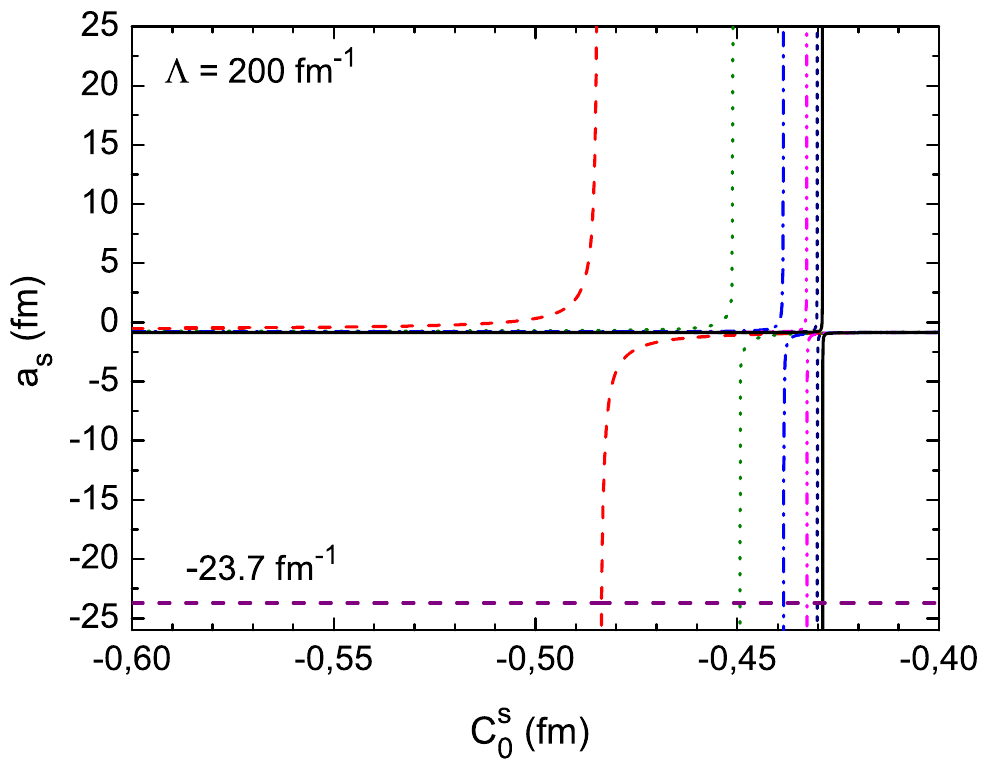}\hspace*{.5cm}\includegraphics[width=7.0 cm]{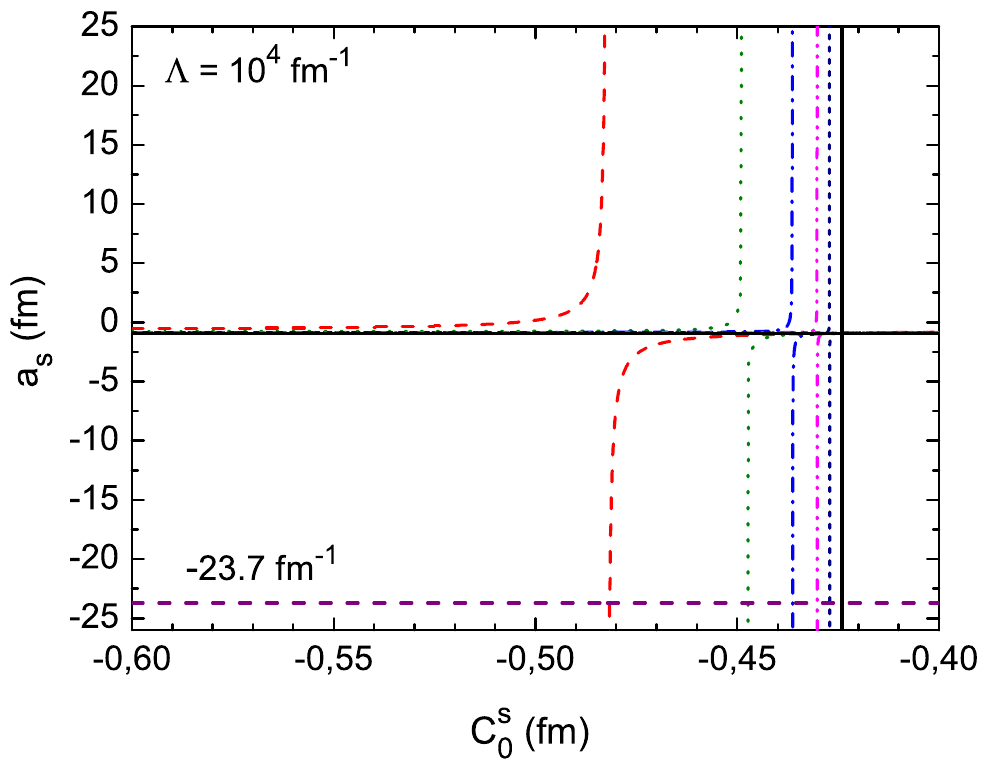}}
\caption{(Color online) Scattering length $a_s$ as a function of ${\tilde C}_0^{s}$ for several values of the scale $\mu$ with the momentum cutoff $\Lambda$ fixed.}
\label{fig11}
\end{figure}

\begin{figure}[p]
\centerline{\includegraphics[width=7.0 cm]{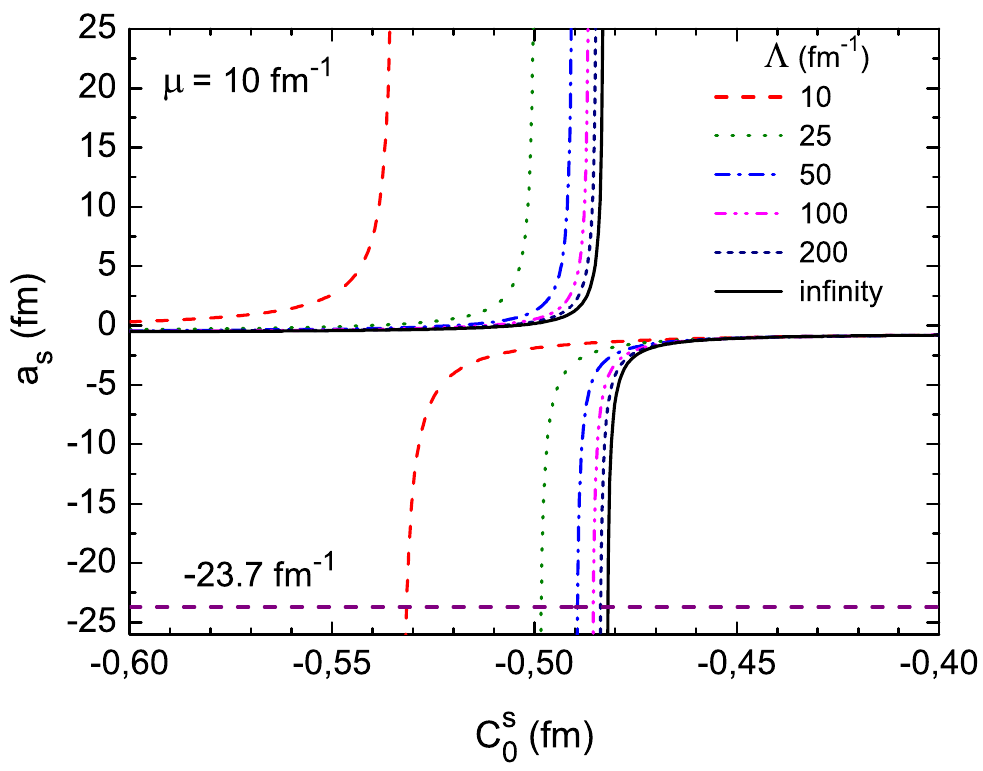}\hspace*{.5cm}\includegraphics[width=7.0 cm]{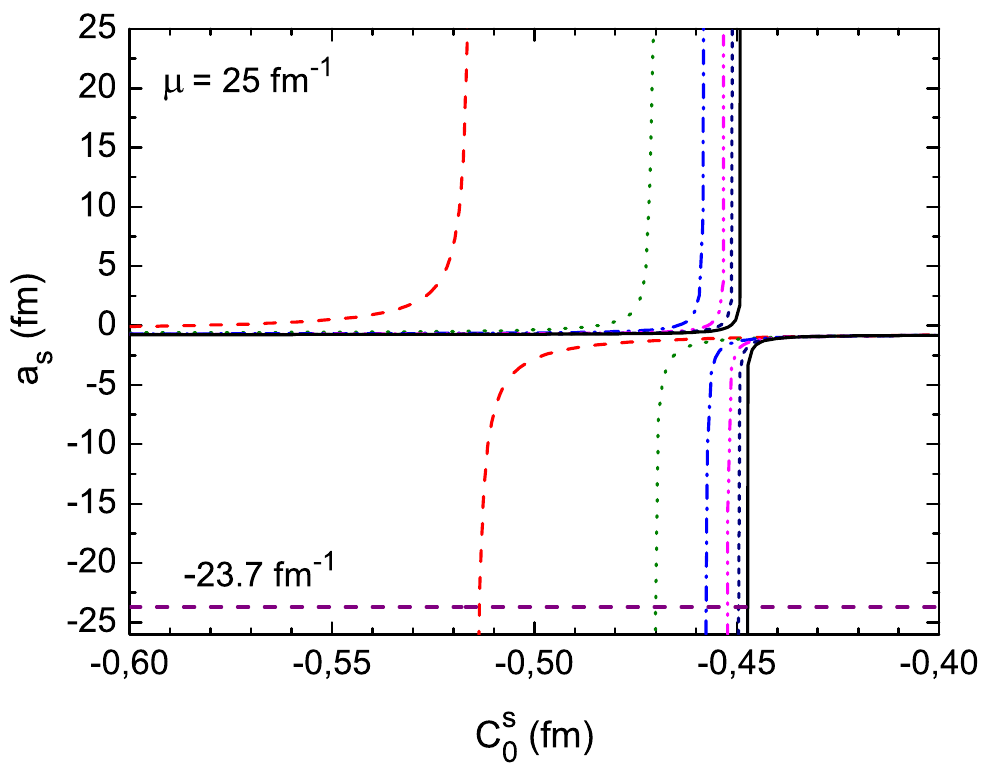}}\vspace*{.8cm}
\centerline{\includegraphics[width=7.0 cm]{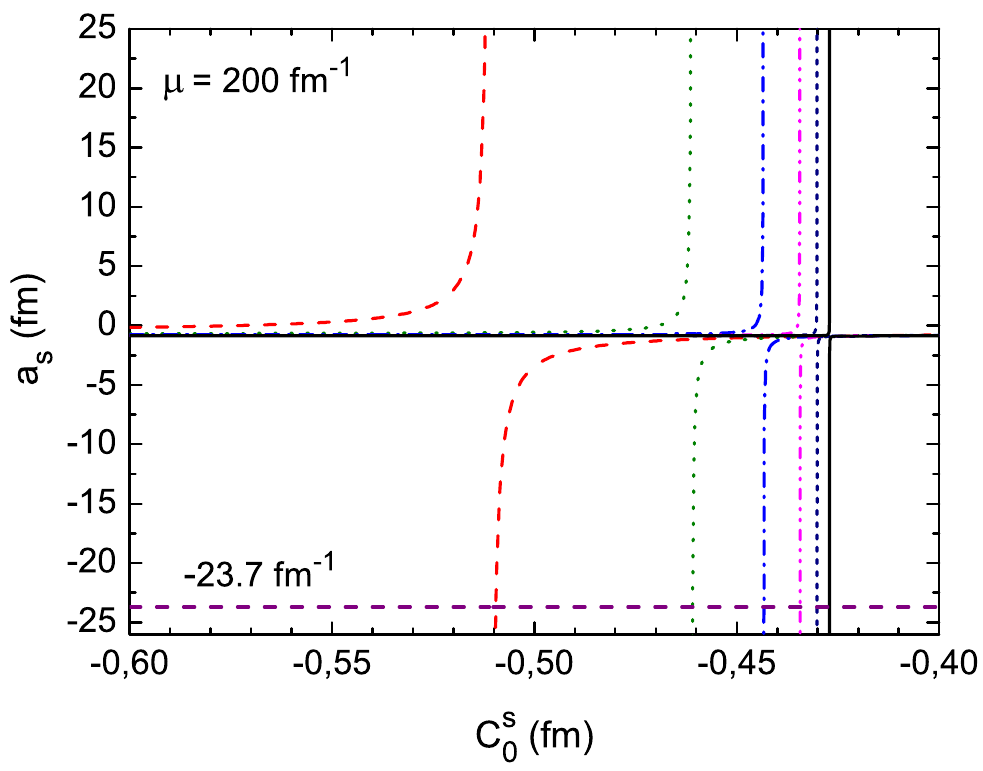}\hspace*{.5cm}\includegraphics[width=7.0 cm]{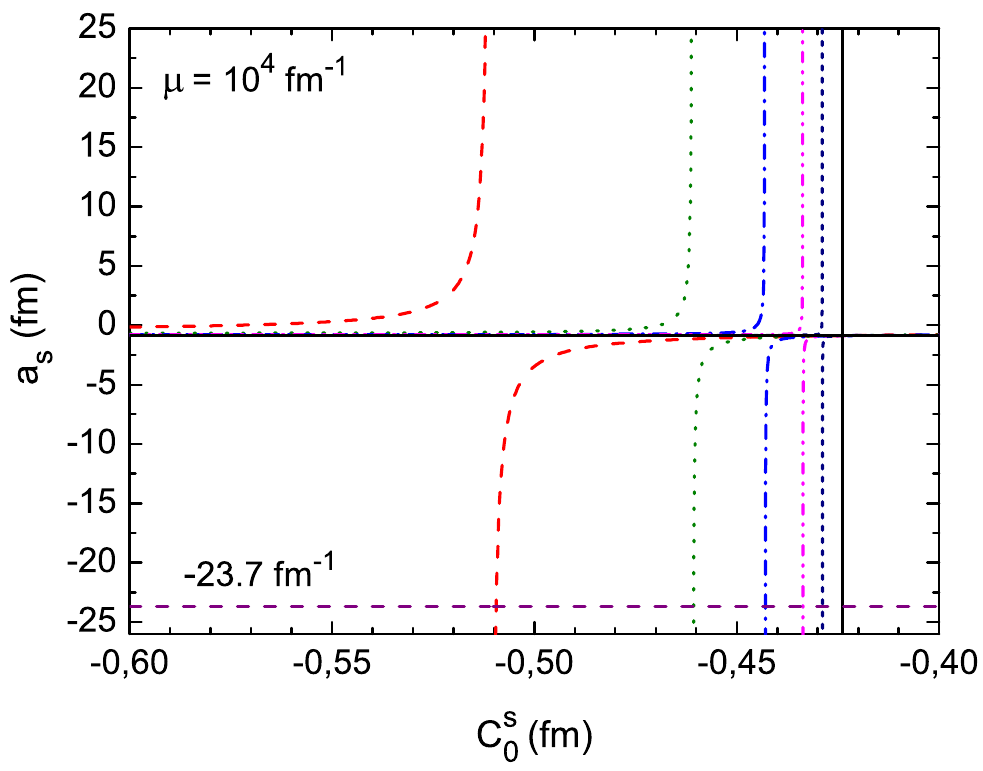}}
\caption{(Color online) Scattering length $a_s$ as a function of ${\tilde C}_0^{s}$ for several values of the momentum cutoff $\Lambda$ with the scale $\mu$ fixed.}
\label{fig12}
\end{figure}

In Figs. (\ref{fig11}) and (\ref{fig12}) we show the scattering length $a_s$ as a function of the strength ${\tilde C}_0^{s}(-\mu^2;\Lambda)$, respectively for several values of the scale $\mu$ with the momentum cutoff $\Lambda$ fixed and for several values of the momentum cutoff $\Lambda$ with the scale $\mu$ fixed. As one can observe from both plots, the position of the resonance pole (associated with the Feshbach resonance \cite{feshbach} due to presence of a shallow virtual bound-state in the $^1 S_0$ channel) converges to a limit value as the scale $\mu$ and the momentum cutoff $\Lambda$ increase. The same happens with the renormalized strength of the contact interaction (the value adjusted to reproduce $a_s=-23.7~{\rm fm}$), which is very close to the position of the pole.

\begin{figure}[p]
\centerline{\includegraphics[width=8.0 cm]{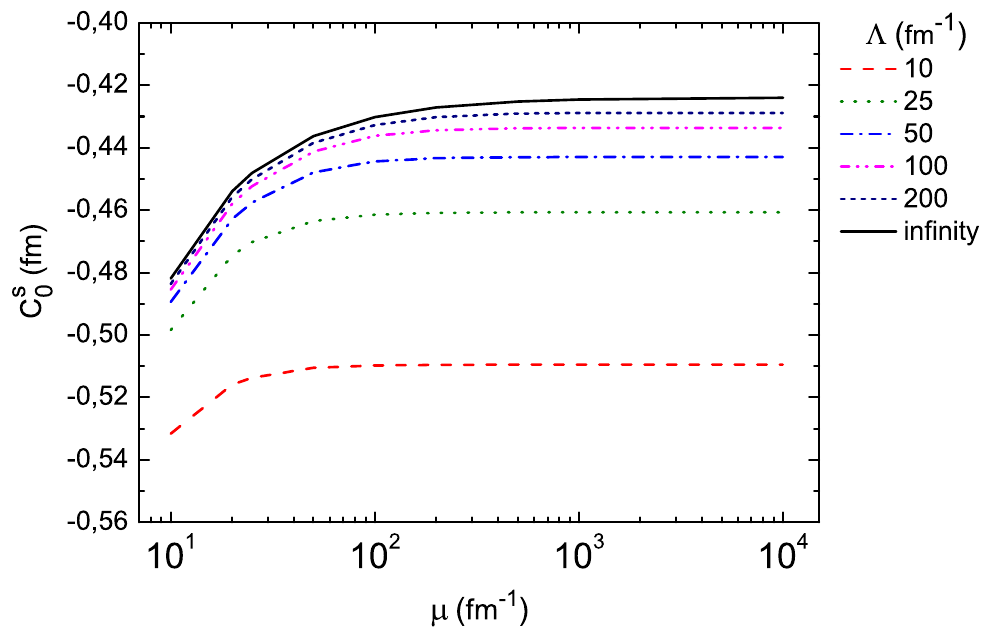}\hspace*{.3cm}\includegraphics[width= 8.0 cm]{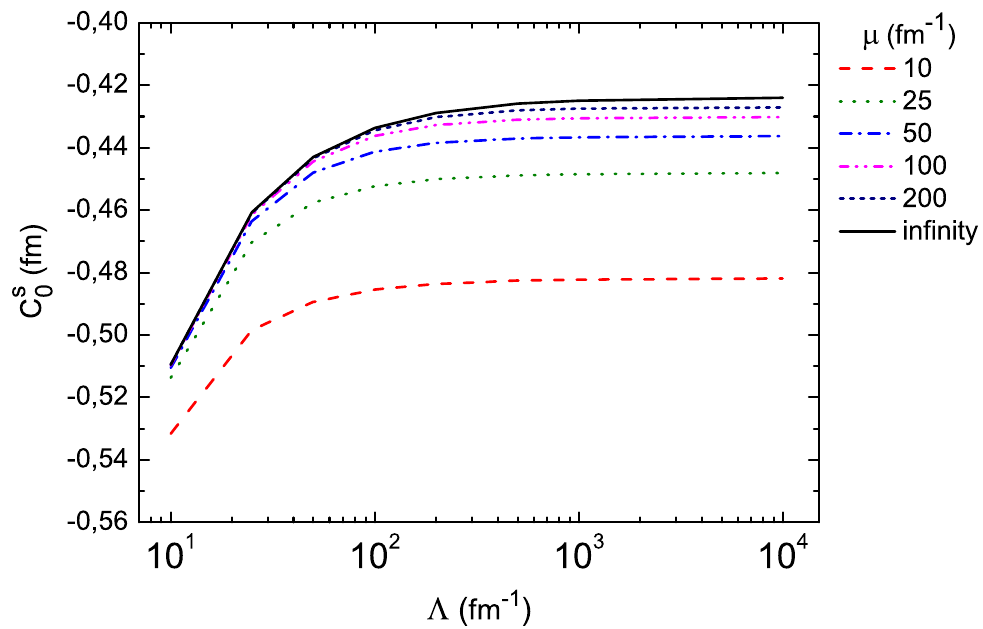}}
\caption{(Color online) Renormalized strength of the contact interaction ${\tilde C}_0^{s}(-\mu^2;\Lambda)$. Left panel: as a function of the scale $\mu$ for several values of the momentum cutoff $\Lambda$; Right panel: as a function of the momentum cutoff $\Lambda$ for several values of the scale $\mu$.}
\label{fig13}
\end{figure}

In Fig. (\ref{fig13}) we show the running of the renormalized strength ${\tilde C}_0^{s}(-\mu^2;\Lambda)$ with the scale $\mu$ for several values of the momentum cutoff $\Lambda$ (left) and with the momentum cutoff $\Lambda$ for several values of the scale $\mu$ (right). For a given value of the momentum cutoff $\Lambda$ the dependence of ${\tilde C}_0^{s}(-\mu^2;\Lambda)$ on the scale $\mu$ vanishes asymptotically in the limit $\mu \rightarrow \infty$ and so the SKM scheme leads to the same results as would be obtained by using the (sharp) cutoff regularization scheme. Similarly, for a given value of the scale $\mu$ the dependence of ${\tilde C}_0^{s}(-\mu^2;\Lambda)$ on the momentum cutoff $\Lambda$ vanishes asymptotically in the limit $\Lambda \rightarrow \infty$, i.e. ${\tilde C}_0^{s}(-\mu^2;\Lambda \rightarrow \infty) \rightarrow C_0^{s}(-\mu^2)$, with $C_0^{s}(-\mu^2)$ absorbing the divergences and leading to the correct value of the scattering length $a_s$. This result clearly shows that the role played by the momentum cutoff $\Lambda$ is purely instrumental, since in the limit $\Lambda \rightarrow \infty$ all ultraviolet divergences in the LS equation are regularized by the subtraction procedure itself, providing a finite $K$-matrix.
\newpage

Although simple, the numerical procedure used in the SKM scheme to fit the scattering length $a_s$ requires a very delicate tuning of the strength ${\tilde C}_0^{s}(-\mu^2;\Lambda)$. As shown in Figs. (\ref{fig11}) and (\ref{fig12}), the resonance width becomes smaller as the scale $\mu$ and the cutoff $\Lambda$ increase (up to the point where the position of the pole stabilizes), such that a higher tuning precision is necessary to achieve a given accuracy.

\begin{figure}[h]
\centerline{\includegraphics[width=11.5 cm]{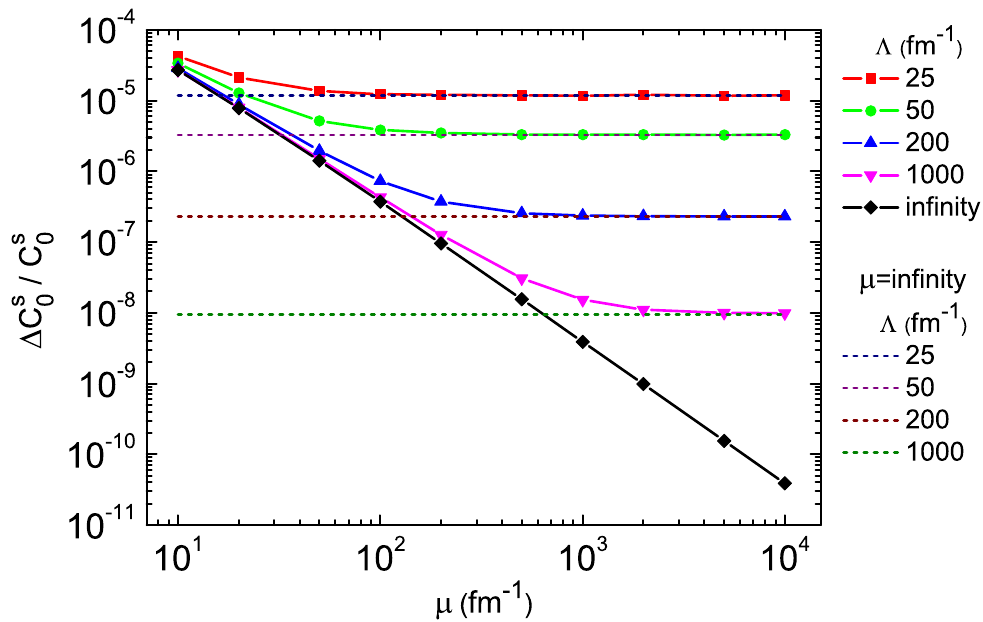}}
\caption{(Color online) Relative precision $\Delta {\tilde C}_0^{s}/{\tilde C}_0^{s}$ to which the strength ${\tilde C}_0^{s}(-\mu^2;\Lambda)$ must be tuned in order to fit the scattering length in the $^1 S_0$ channel with an accuracy of $1 \%$. The solid lines with symbols correspond to the SKM results as a function of the scale $\mu$ for several values of the cutoff $\Lambda$. The horizontal dashed lines correspond to the values for each cutoff $\Lambda$ obtained by using the cutoff regularization scheme ($\mu \rightarrow \infty$).}
\label{fig14}
\end{figure}

In Fig. \ref{fig14} we show the results obtained for the relative precision to which ${\tilde C}_0^{s}(-\mu^2;\Lambda)$ must be tuned, defined as $\Delta {\tilde C}_0^{s}/{\tilde C}_0^{s}$, in order to fit the scattering length in the $^1 S_0$ channel with an accuracy of $1 \%$, as a function of the scale $\mu$ for several values of the cutoff $\Lambda$. We also show the results obtained for each cutoff $\Lambda$ in the limit $\mu \rightarrow \infty$, which correspond to using the (sharp) cutoff regularization scheme.

As expected, for a given cutoff $\Lambda$ the required precision becomes higher as the scale $\mu$ increases (corresponding to a smaller value of $\Delta {\tilde C}_0^{s}/{\tilde C}_0^{s}$), asymptotically approaching the precision required when using the cutoff regularization scheme. One should also note that in the limit $\Lambda \rightarrow \infty$ (no cutoff) the tuning of the strength ${\tilde C}_0^{s}(-\mu^2;\Lambda)$ should become numerically impossible for scales $\mu$ larger then some maximum value.

We now turn to the calculation of the renormalized potential ${\tilde V}_{\cal R}(p,p';\Lambda)$. Once the renormalized strength of the contact interaction ${\tilde C}_0^{s}(-\mu^2;\Lambda)$ is fixed at the scale $\mu$ for a given momentum cutoff $\Lambda$, we can obtain the renormalized potential ${\tilde V}_{\cal R}(p,p';\Lambda)$ from the driving term ${\tilde V}^{(1)}(p,p';-\mu^2;\Lambda)$ by solving an integral equation which is derived from Eq. (\ref{vreneq}) (replacing the $T$-matrix by the $K$-matrix),
\begin{eqnarray}
{\tilde V}_{\cal R}(p,p';\Lambda) = {\tilde V}^{(1)}(p,p';-\mu^2;\Lambda) -
\frac{2}{\pi}{\cal P}~\int_0^\infty dq \; q^2~\theta(\Lambda-q)~\frac{{\tilde V}^{(1)}(p,q;-\mu^2;\Lambda)}{-\mu^2-q^2}
{\tilde V}_{\cal R}(q,p';\Lambda) \; ,
\label{vrint}
\end{eqnarray}
\noindent
Note that we keep the instrumental cutoff $\Lambda$ in place, so as to obtain a SKM renormalized potential which will be numerically more convenient to evolve through the SRG transformation.

In principle, the renormalized potential ${\tilde V}_{\cal R}(p,p';\Lambda)$ obtained from the solution of Eq. (\ref{vrint}) should be independent of the subtraction scale $-\mu^2$, since by construction it corresponds to a renormalization group fixed-point operator. However, a residual dependence on the scale $\mu$ is generated by the fitting procedure used to fix the renormalized strength of the contact interaction ${\tilde C}_0^{s}(-\mu^2;\Lambda)$. As pointed out in subsection (\ref{FP}), in order to obtain a renormalized potential which is rigorously invariant with respect to the change of the subtraction scale $-\mu^2$ the driving term must be evolved via the NRCS equation, with the boundary condition ${\tilde V}^{(1)}(p,p';-\mu^2;\Lambda)|_{\mu \rightarrow {\bar \mu}}= {\tilde V}^{(1)}(p,p';-{\bar \mu}^2;\Lambda)$ specified at some reference scale ${\bar \mu}$ where the renormalized strength ${\tilde C}_0^{s}(-\mu^2;\Lambda)$ is fixed.

\begin{figure}[p]
\centerline{\includegraphics[width=8.0 cm]{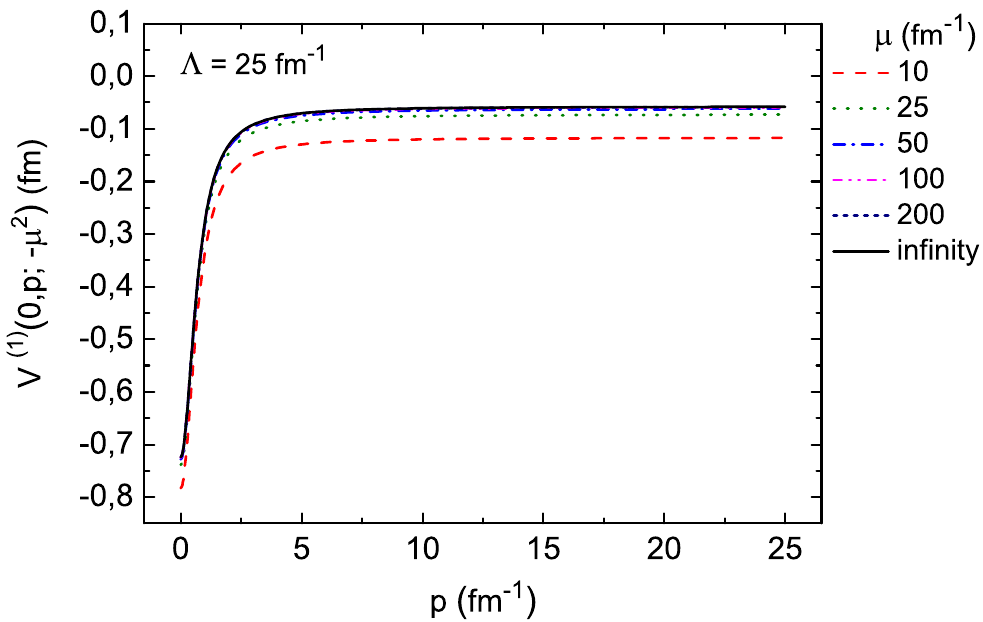}\hspace*{.3cm}\includegraphics[width=8.0 cm]{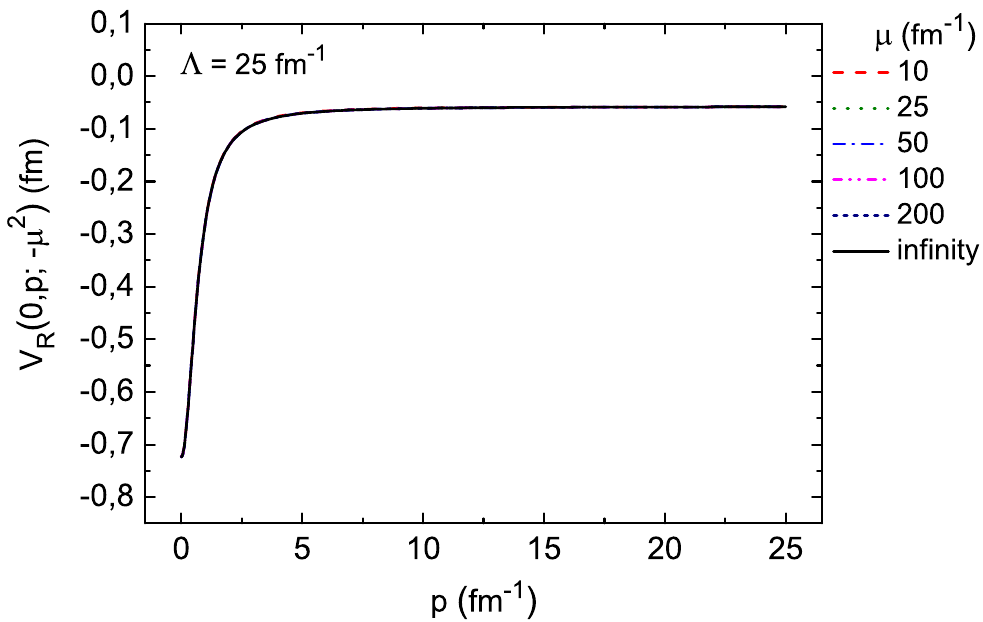}}
\caption{(Color online) Driving term ${\tilde V}^{(1)}(p,0; -\mu^2;\Lambda)$ (left) and renormalized potential ${\tilde V}_{\cal R}(p,0;-\mu^2;\Lambda)$ (right) for several values of the scale $\mu$ with the momentum cutoff fixed at $\Lambda=25~{\rm fm}^{-1}$. }
\label{fig15}
\end{figure}

\begin{figure}[p]
\centerline{\includegraphics[width=8.0 cm]{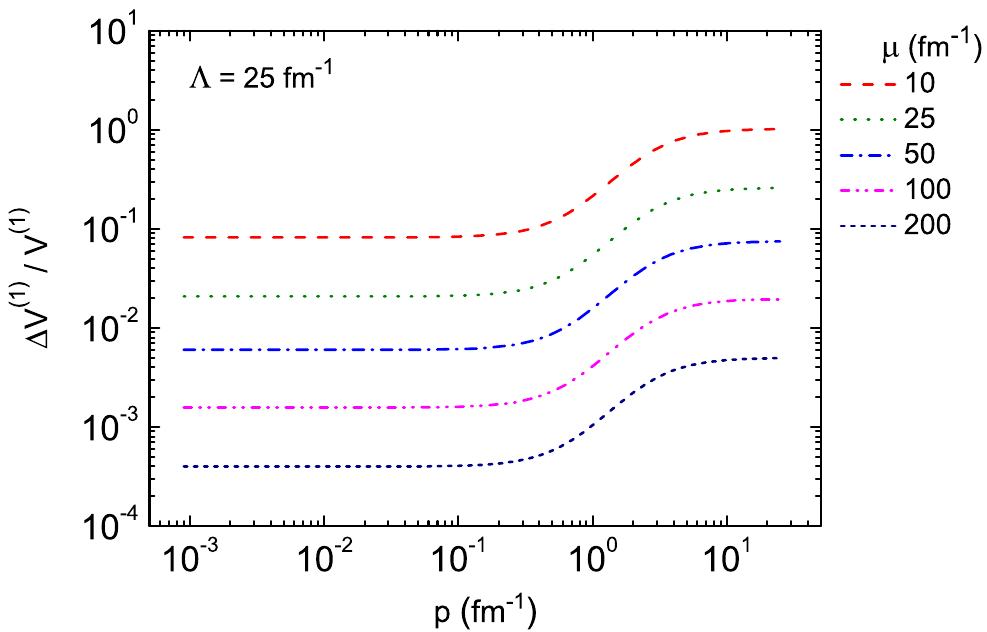}\hspace*{.3cm}\includegraphics[width=8.0 cm]{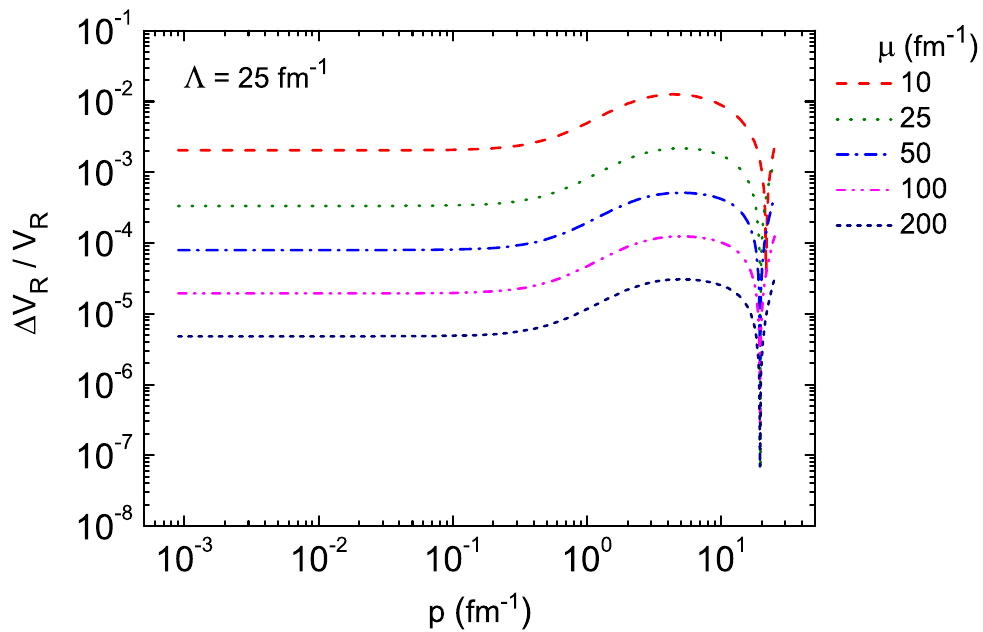}}\vspace*{.8cm}
\centerline{\includegraphics[width=8.0 cm]{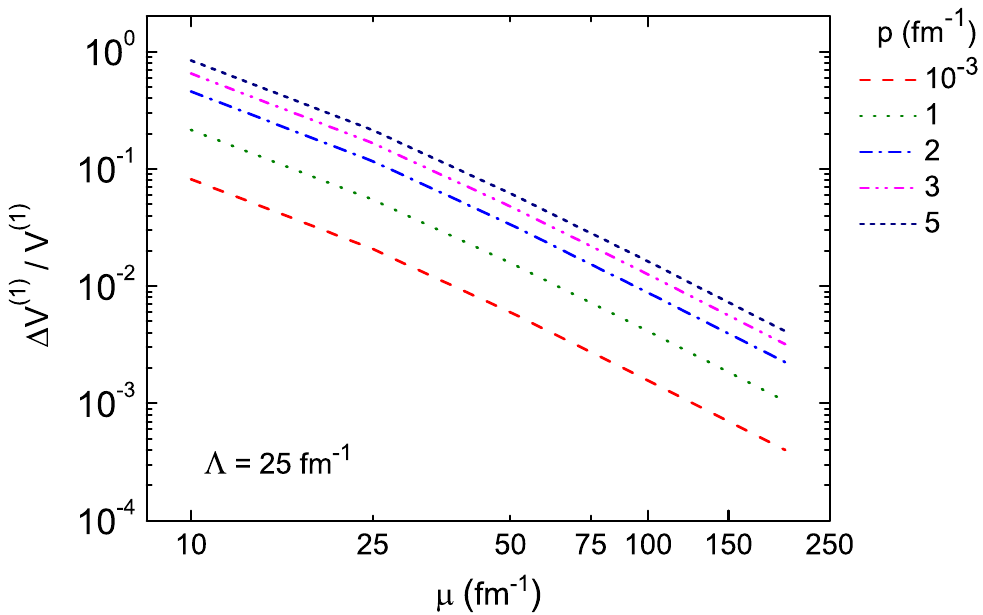}\hspace*{.3cm}\includegraphics[width=8.0 cm]{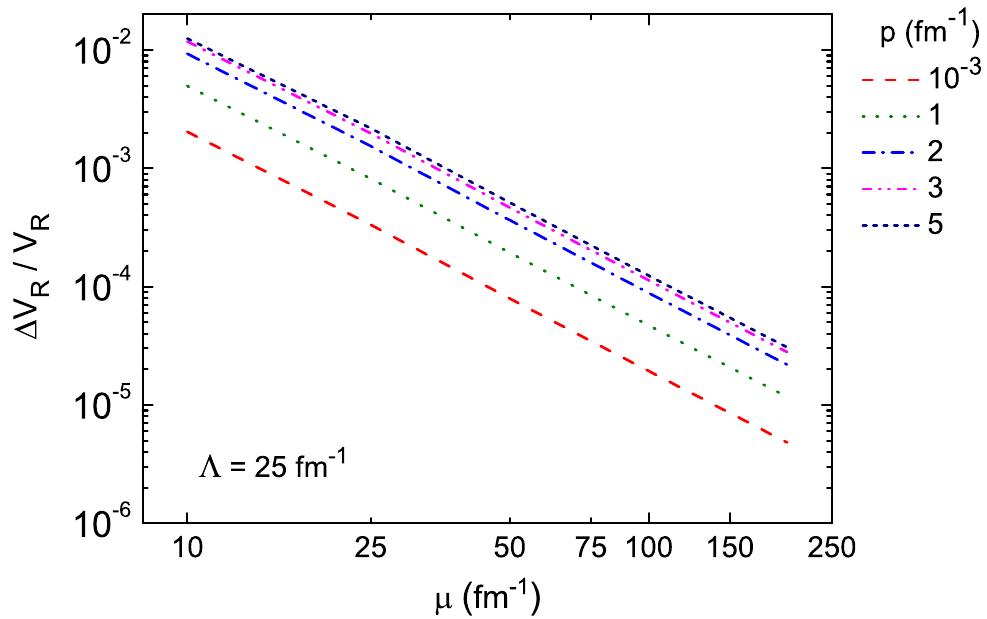}}
\caption{(Color online) Relative differences $\Delta{\tilde V}^{(1)}/{\tilde V}^{(1)}$ for the driving term (left) and $\Delta{\tilde V}_{\cal R}/{\tilde V}_{\cal R}$ for the renormalized potential (right). Top panels: as a function of the momentum $p$ for several values of $\mu$; Bottom panels: as a function of $\mu$ for several values of the momentum $p$.}
\label{fig16}
\end{figure}

In Fig. (\ref{fig15}) we show the driving terms ${\tilde V}^{(1)}(p,0;-\mu^2;\Lambda)$ (left) and the corresponding renormalized potentials ${\tilde V}_{\cal R}(p,0;-\mu^2;\Lambda)$ (right) at $p'=0$, for several values of the scale $\mu$ with the momentum cutoff fixed at $\Lambda=25~{\rm fm}^{-1}$. At each value of $\mu$, the driving term was obtained by tuning the renormalized strength ${\tilde C}_0^{s}(-\mu^2;\Lambda)$ to fit the scattering length $a_s$ in the $^1S_0$ channel with the same accuracy, corresponding to an absolute error of $10^{-5}$. In Fig. (\ref{fig16}) we show the relative differences between the driving term at a scale $\mu$ and at $\mu \rightarrow \infty$ (left),
\begin{equation}
\frac{\Delta{\tilde V}^{(1)}}{{\tilde V}^{(1)}} \equiv \frac{[{\tilde V}^{(1)}(p,0;-\mu^2;\Lambda)-{\tilde V}^{(1)}(p,0;-\infty;\Lambda)]}{{\tilde V}^{(1)}(p,0;-\infty;\Lambda)} \; ,
\end{equation}
\noindent
and between the renormalized potential at a scale $\mu$ and at $\mu \rightarrow \infty$ (right),
\begin{equation}
\frac{\Delta{\tilde V}_{\cal R}}{{\tilde V}_{\cal R}} \equiv \frac{[{\tilde V}_{\cal R}(p,0;-\mu^2;\Lambda)-{\tilde V}_{\cal R}(p,0;-\infty;\Lambda)]}{{\tilde V}_{\cal R}(p,0;-\infty;\Lambda)} \; ,
\end{equation}
\noindent
as a function of the momentum $p$ for several values of the scale $\mu$ (top panels) and as a function of the scale $\mu$ for several values of the momentum $p$ (bottom panels). As expected, the driving term ${\tilde V}^{(1)}(p,0;-\mu^2;\Lambda)$ changes with the scale $\mu$ in such a way that the renormalized potential ${\tilde V}_{\cal R}(p,0;-\mu^2;\Lambda)$  remains approximately invariant. In the limit $\mu \rightarrow \infty$, the driving term matches the renormalized potential, both becoming independent of the scale $\mu$. One can also observe that for a fixed momentum $p$ the relative differences scale approximately as $1/\mu^2$ in the range of scales $\mu$ considered in the calculations, both for the driving term and the renormalized potential.

\begin{figure}[p]
\centerline{\includegraphics[width=8.0 cm]{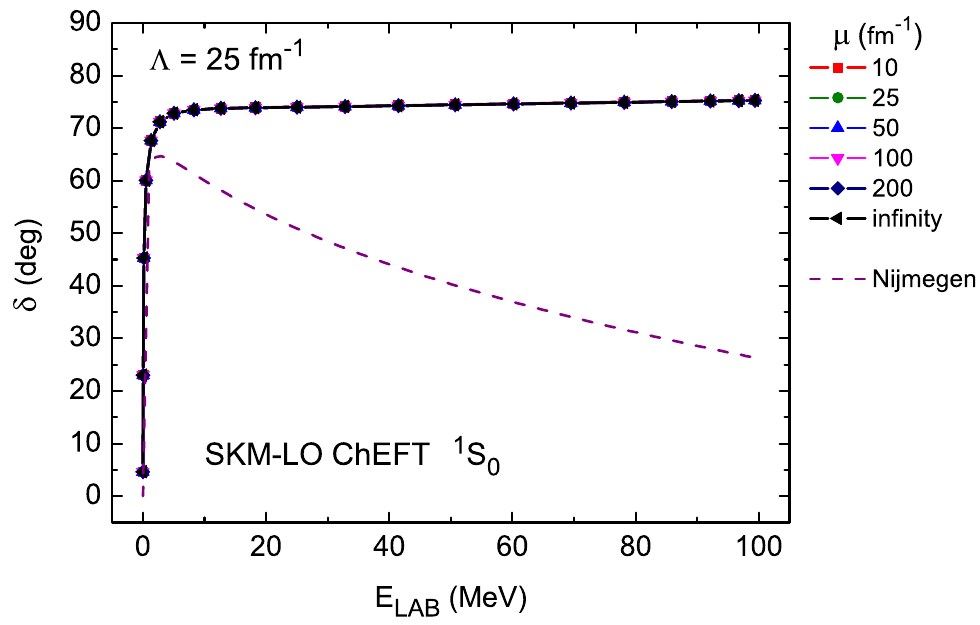}\hspace*{.3cm}\includegraphics[width=8.0cm]{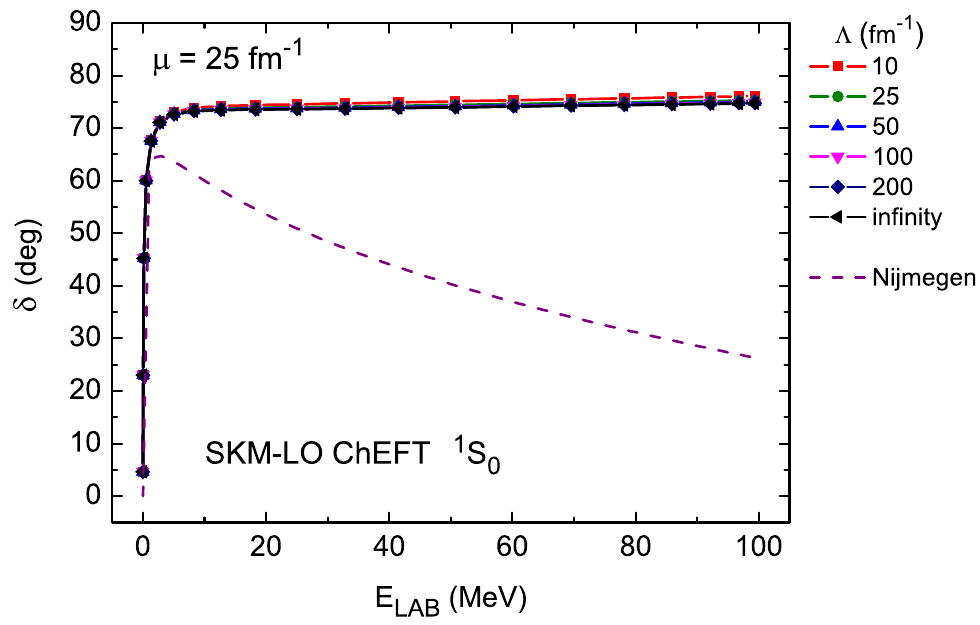}}
\caption{(Color online) Phase-shifts in the $^1 S_0$ channel as a function of the laboratory energy $E_{\rm LAB}$ for the LO ChEFT potential evaluated from the subtracted kernel LS equation for the $K$-matrix. Left panel: for several values of the scale $\mu$ with the momentum cutoff $\Lambda$ fixed at $25~{\rm fm}^{-1}$; Right panel: for several values of the momentum cutoff $\Lambda$ with the scale $\mu$ fixed at $25~{\rm fm}^{-1}$.}
\label{fig17}
\end{figure}

Once the driving term ${\tilde V}^{(1)}(p,p';-\mu^2;\Lambda)$ is fixed at the subtraction scale $-\mu^2$ for a given momentum cutoff $\Lambda$, we can obtain the $NN$ phase-shifts for the LO ChEFT potential from the numerical solution of the subtracted kernel LS equation for the partial-wave $K$-matrix, Eq. (\ref{KLS1SO}). In Fig. (\ref{fig17}) we show the results for the $NN$ phase-shifts in the $^1S_0$ channel as a function of the laboratory energy  $E_{\rm LAB}$ calculated for several values of the scale $\mu$ with the momentum cutoff $\Lambda$ fixed at $25~{\rm fm}^{-1}$ (left) and for several values of the momentum cutoff $\Lambda$ with the scale $\mu$ fixed at $25~{\rm fm}^{-1}$ (right). The results obtained for the phase-shifts from the subtracted kernel LS equation strongly deviates from the Nijmegen PWA, similar to the well known results obtained for the LO ChEFT interaction in the $^1S_0$ channel renormalized through other schemes.

\begin{figure}[p]
\centerline{\includegraphics[width=8.0 cm]{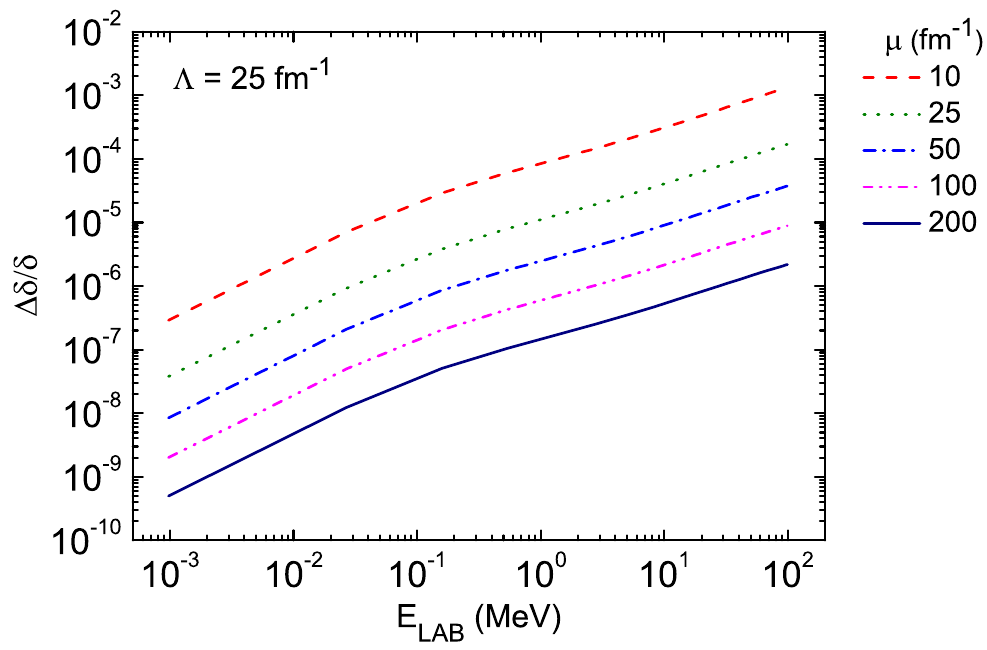}\hspace*{.5cm}\includegraphics[width=8.0 cm]{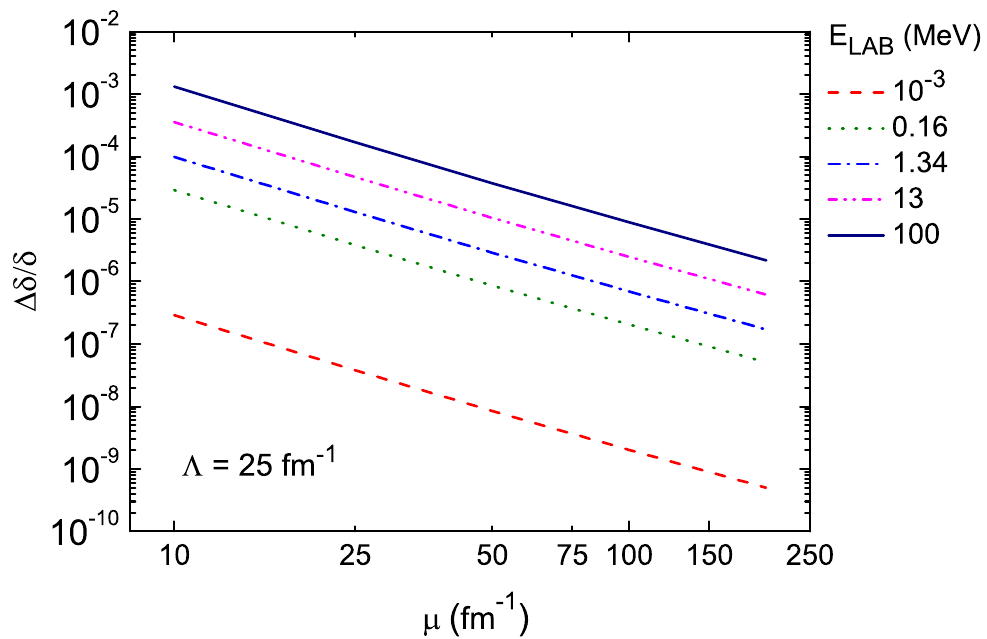}}
\caption{(Color online) Relative differences $\Delta \delta[\mu]/\delta[\mu]$ (with the momentum cutoff $\Lambda$ fixed at $25~{\rm fm}^{-1}$). Left panel: as a function of $E_{\rm LAB}$ for several values of the scale $\mu$; Right panel: as a function of the scale $\mu$ for several values of $E_{\rm LAB}$.}
\label{fig18}
\end{figure}

In Fig. (\ref{fig18}) we show the relative differences between the phase-shifts in the $^1S_0$ channel calculated at a scale $\mu$ and at $\mu \rightarrow \infty$ (with the momentum cutoff $\Lambda$ fixed at $25~{\rm fm}^{-1}$),
\begin{equation}
\frac{\Delta \delta[\mu]}{\delta[\mu]} \equiv \frac{[\delta(E_{\rm LAB};-\mu^2;\Lambda)-\delta(E_{\rm LAB};-\infty;\Lambda)]}{\delta(E_{\rm LAB};-\infty;\Lambda)} \; ,
\end{equation}
\noindent
as a function of $E_{\rm LAB}$ for several values of $\mu$ (left) and as a function of $\mu$ for several values of $E_{\rm LAB}$ (right). In the range of scales $\mu$ considered in the calculations, the relative differences $\Delta \delta[\mu]/\delta[\mu]$ scale approximately as $E_{\rm LAB} (\propto k^2)$ for fixed $\mu$ and as $1/\mu^2$ for fixed $E_{\rm LAB}$. In the limit $\mu \rightarrow \infty$, the phase-shifts should become independent of $\mu$, matching the results obtained by using the cutoff regularization scheme.

\begin{figure}[p]
\centerline{\includegraphics[width=8.0 cm]{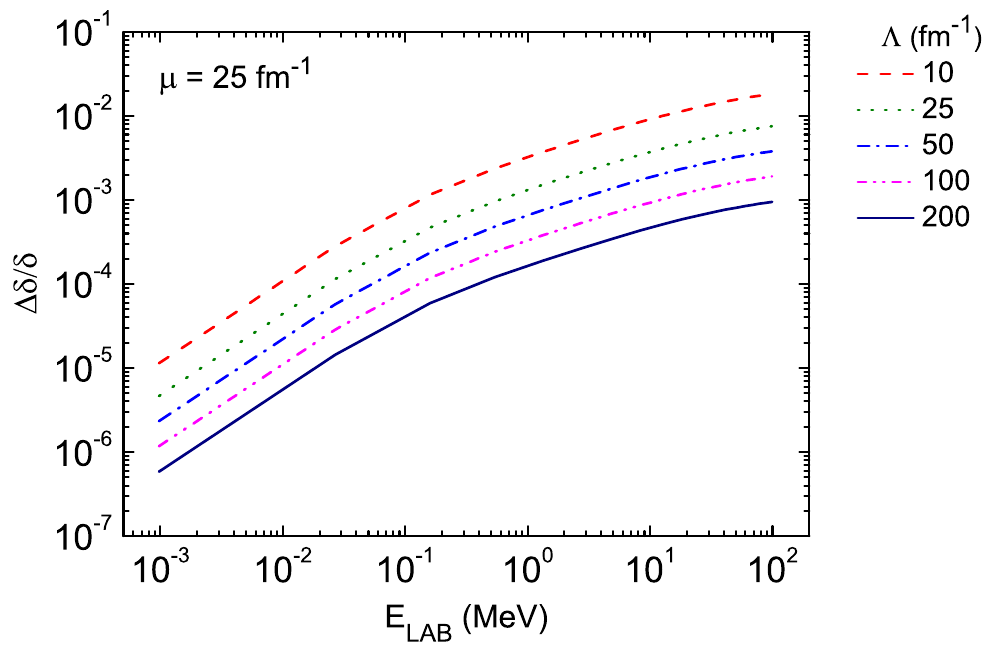}\hspace*{.5cm}\includegraphics[width=8.0 cm]{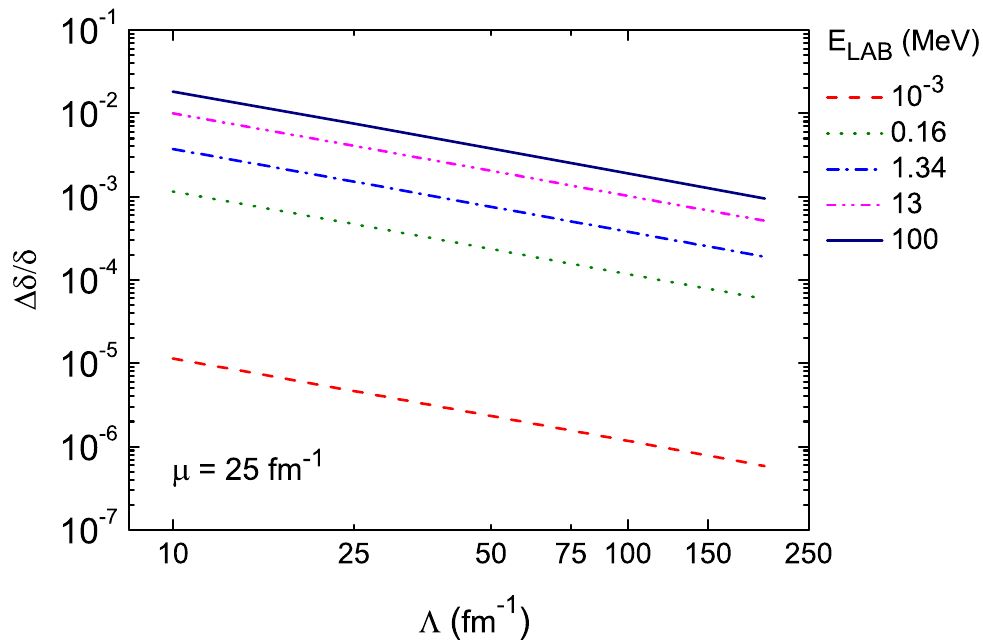}}
\caption{(Color online) Relative differences $\Delta \delta[\Lambda]/\delta[\Lambda]$ (with the scale $\mu$ fixed at $25~{\rm fm}^{-1})$. Left panel: as a function of $E_{\rm LAB}$ for several values of the momentum cutoff $\Lambda$; Right panel: as a function of the momentum cutoff $\Lambda$ for several values of $E_{\rm LAB}$.}
\label{fig19}
\end{figure}

In Fig. (\ref{fig19}) we show the relative differences between the phase-shifts in the $^1S_0$ channel calculated at a momentum cutoff $\Lambda$ and at $\Lambda \rightarrow \infty$ (with the scale $\mu$ fixed at $25~{\rm fm}^{-1}$),
\begin{equation}
\frac{\Delta \delta[\Lambda]}{\delta[\Lambda]} \equiv \frac{[\delta(E_{\rm LAB};-\mu^2;\Lambda)-\delta(E_{\rm LAB};-\mu^2;\infty)]}{\delta(E_{\rm LAB};-\mu^2;\infty)} \; ,
\end{equation}
\noindent
as a function of $E_{\rm LAB}$ for several values of $\Lambda$ (left) and as a function of $\Lambda$ for several values of $E_{\rm LAB}$ (right). In the range of momentum cutoffs $\Lambda$ considered in the calculations, the relative differences $\Delta \delta[\Lambda]/\delta[\Lambda]$ scale approximately as $E_{\rm LAB} (\propto k^2)$ for fixed $\Lambda$, similar to the scaling of $\Delta \delta[\mu]/\delta[\mu]$. On the other hand, for fixed $E_{\rm LAB}$  they scale as $1/\Lambda$. As expected, in the limit $\Lambda \rightarrow \infty$ the phase-shifts should become independent of $\Lambda$. It is important to remind that the cutoff $\Lambda$ is just an instrumental regulator, whose effects on the phase-shifts could be eliminated by using standard methods to take the limit $\Lambda \rightarrow \infty$, such as the Richardson extrapolation \cite{richardson}.

We emphasize again that by evolving the driving term ${\tilde V}^{(1)}(p,p';-\mu^2;\Lambda)$ from a reference scale ${\bar \mu}$ to the scale $\mu$ via the NRCS equation, instead of using the fitting procedure to fix the renormalized strength ${\tilde C}_0^{s}(-\mu^2;\Lambda)$, the residual dependence of the renormalized potential ${\tilde V}_{\cal R}(p,p';;-\mu^2;\Lambda)$ and the phase-shifts on the scale $\mu$ should vanish. In this case, a non-trivial dependence of the driving term on the scale $\mu$ is expected in order to ensure the invariance of the subtracted $K$-matrix, which would lead to a more complicated scaling behavior.

\subsection{SRG Evolution of the SKM-LO ChEFT potential}
\label{SRGSKMNN}

When considering the application of the SRG approach to evolve the effective $NN$ potential in LO ChEFT, we use as an initial potential the fixed-point renormalized potential $V_{\cal R}$ derived by implementing the SKM scheme for the LO ChEFT interaction. We solve Eq. (\ref{flowNN}) numerically, with the boundary condition set at $s=0$ ($\lambda \rightarrow \infty$) such that the initial potential $V_{s = 0}(p,p')$ is given by $V_{\cal R}(p,p')$ (as pointed before, for numerical convenience the calculations are performed by using an ultraviolet momentum cutoff $\Lambda$ that acts just as an  instrumental regulator).

\begin{figure}[h]
\centerline{\includegraphics[width=11.5 cm]{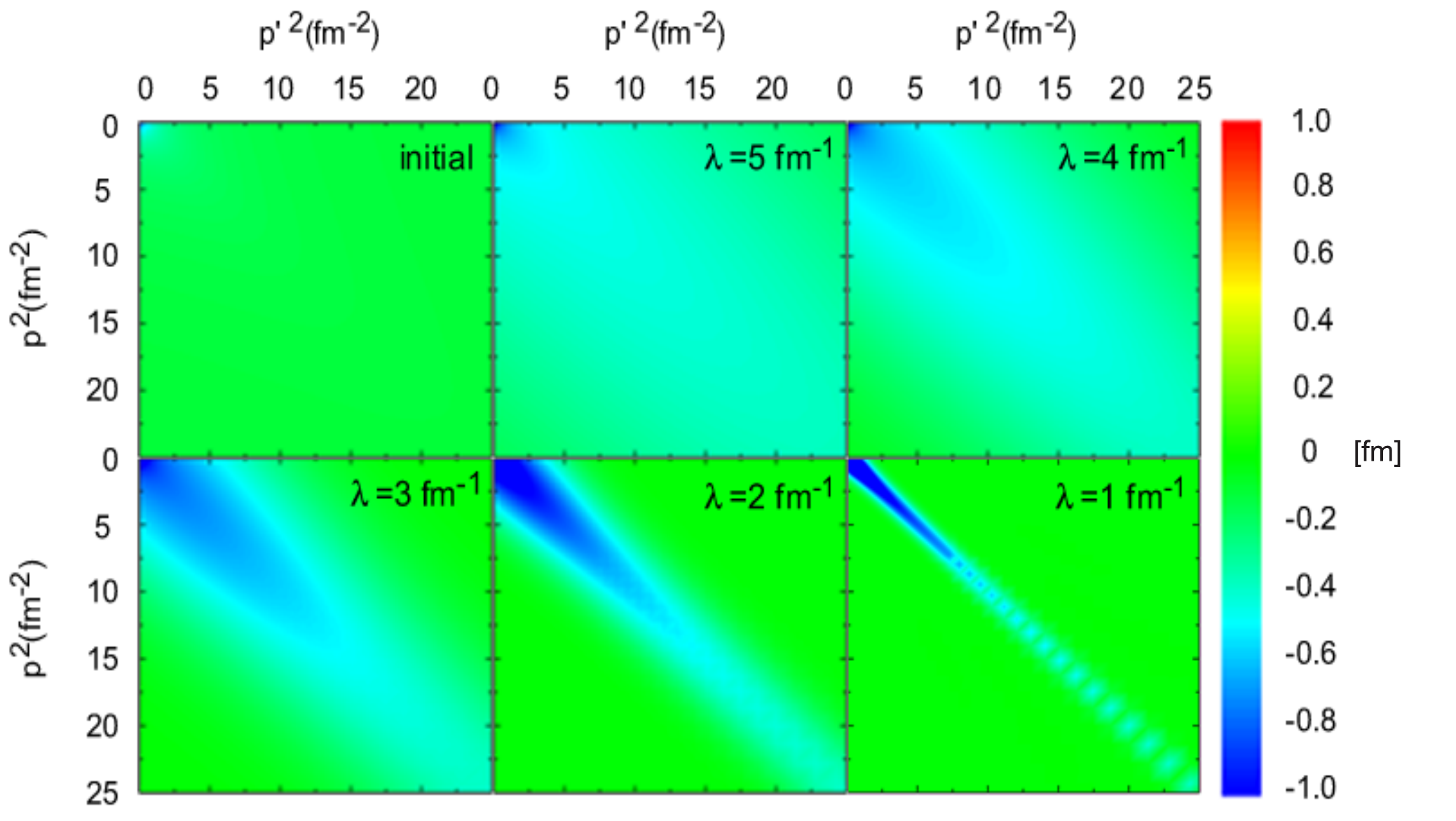}}
\caption{(Color online) SRG evolution of the SKM-LO ChEFT potential in the $^1 S_0$ channel (contour plots).}
\label{fig20}
\end{figure}

\begin{figure}[h]
\centerline{\includegraphics[width=4.6 cm]{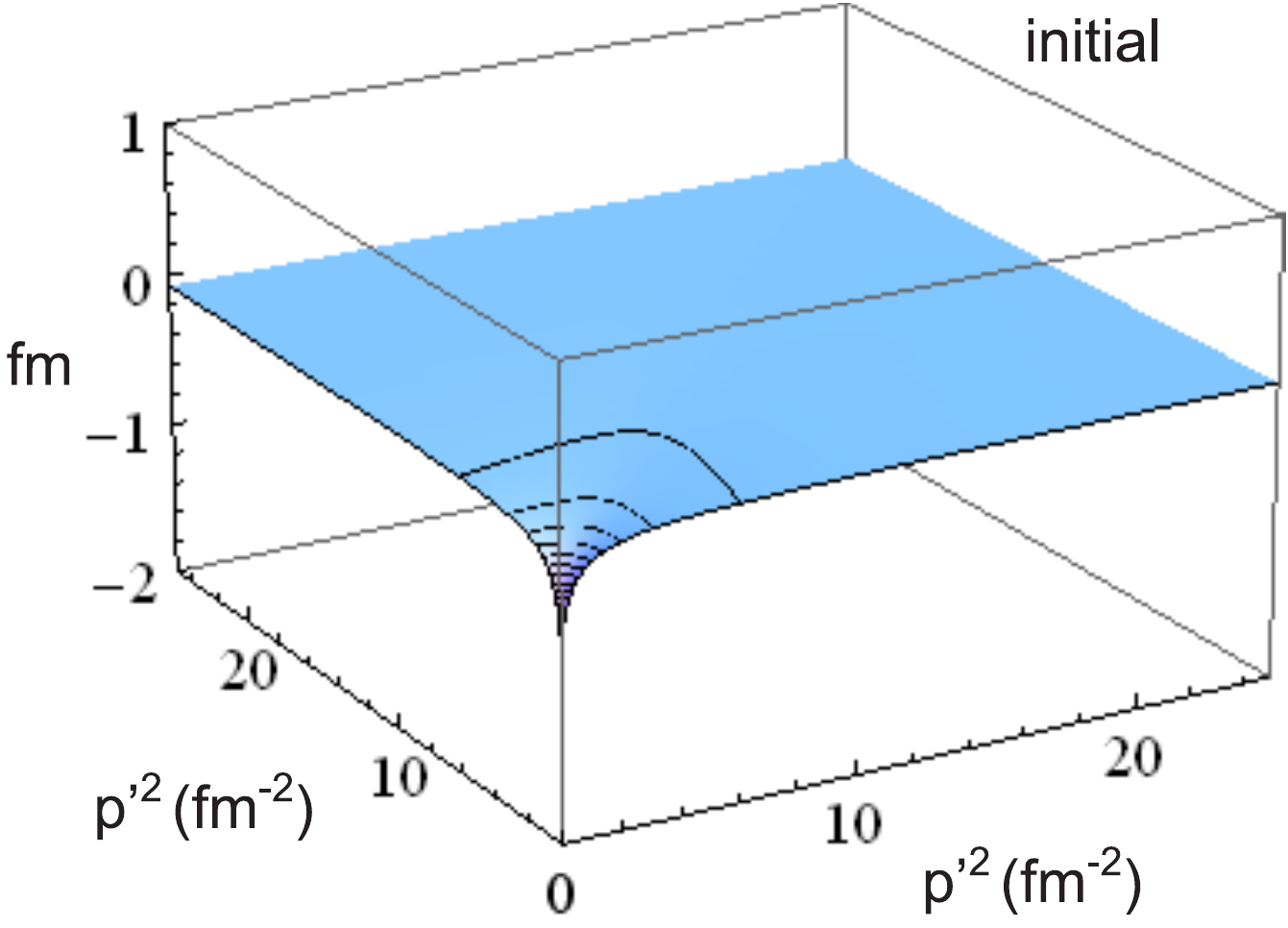}\hspace*{.5cm}\includegraphics[width=4.6 cm]{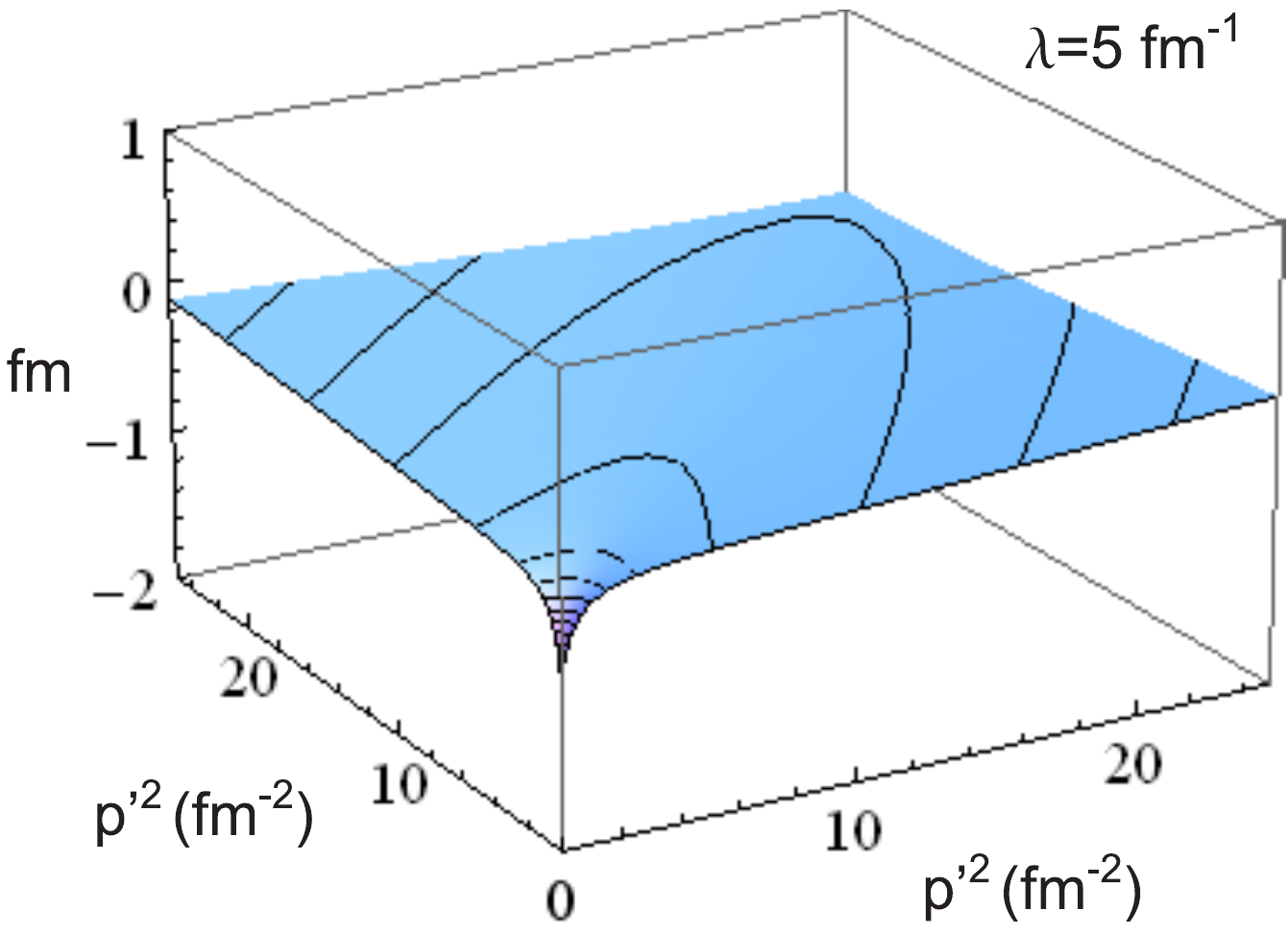}\hspace*{.5cm}\includegraphics[width=4.6 cm]{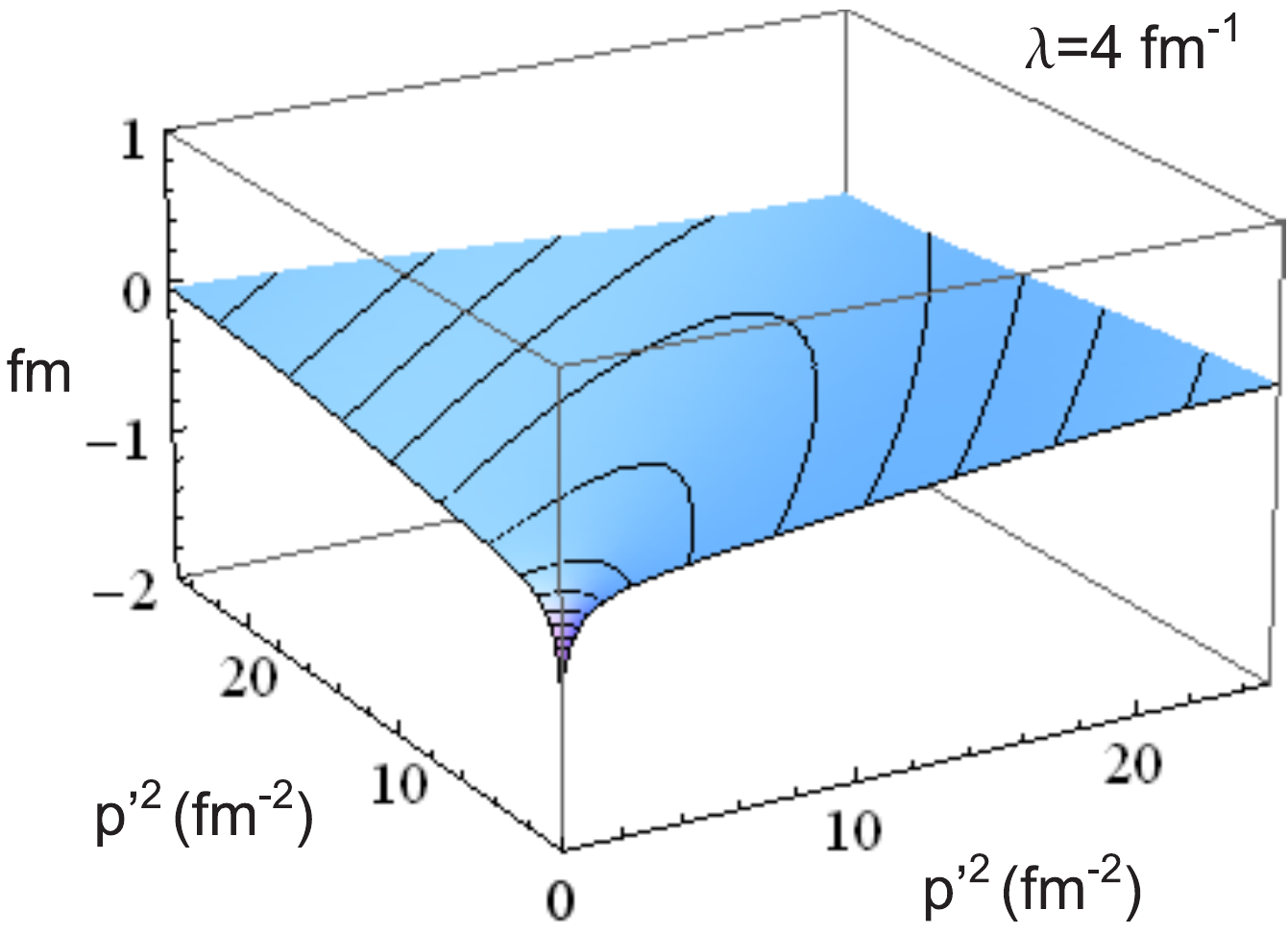}}\vspace*{.3cm}
\centerline{\includegraphics[width=4.6 cm]{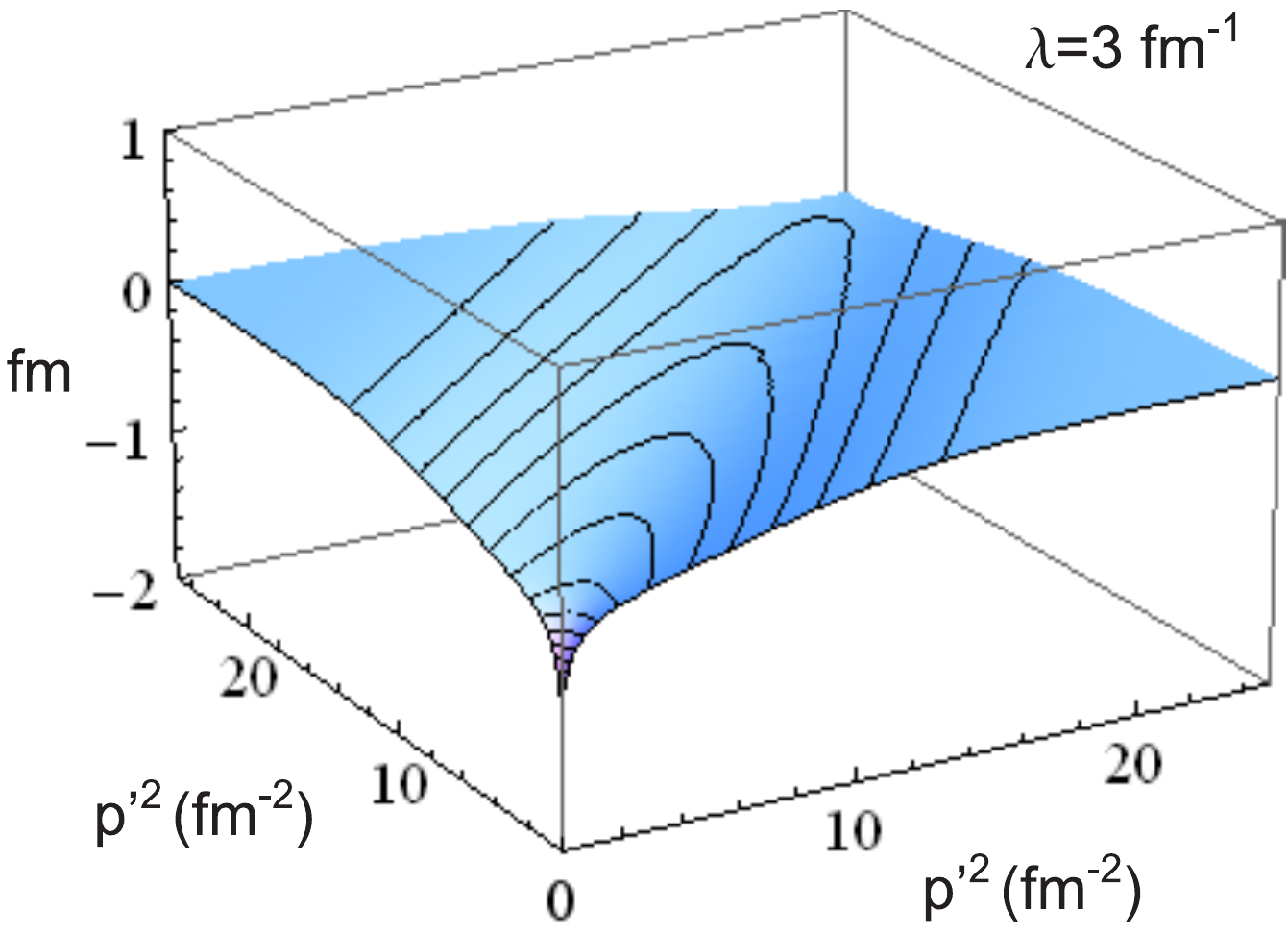}\hspace*{.5cm}\includegraphics[width=4.6 cm]{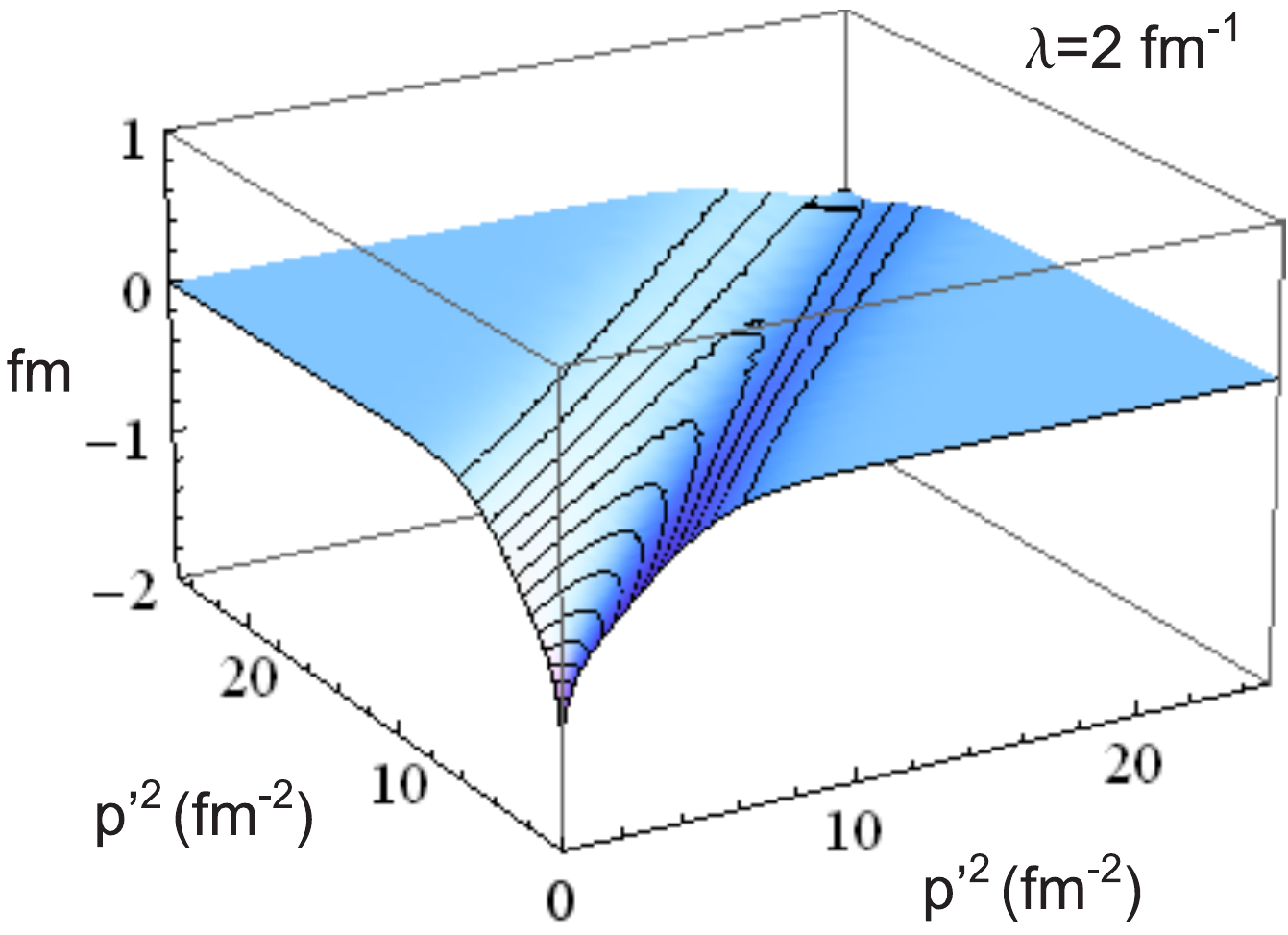}\hspace*{.5cm}\includegraphics[width=4.6 cm]{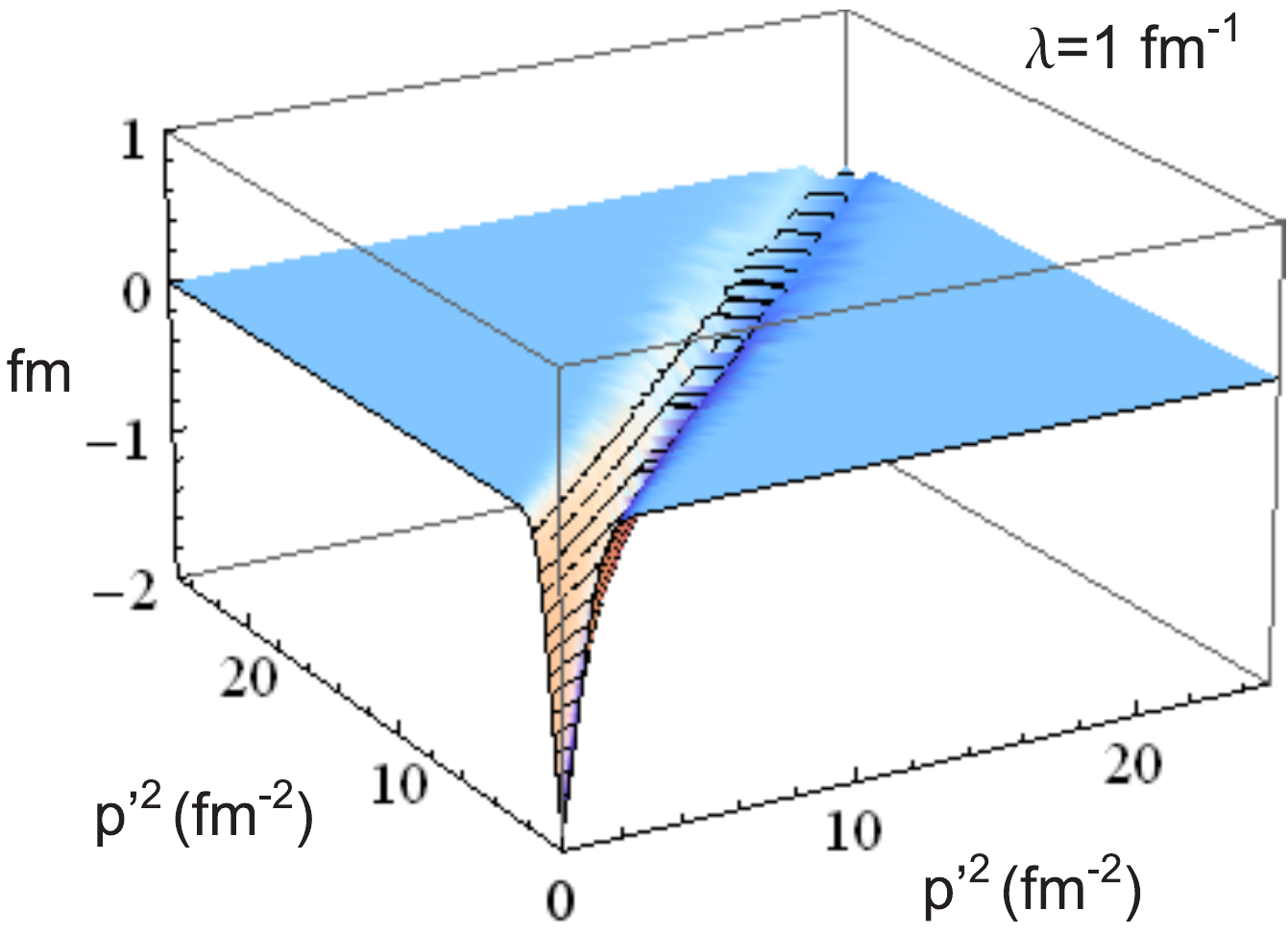}}
\caption{(Color online) SRG evolution of the SKM-LO ChEFT potential in the $^1 S_0$ channel (surface plots).}
\label{fig21}
\end{figure}

In Figs. (\ref{fig20}) and (\ref{fig21}) we show respectively the contour and the surface plots for the SRG evolution of the SKM-LO ChEFT potential in the $^1 S_0$ channel. The initial potential $V_{\cal R}(p,p')$ was calculated by fixing $C_0(-\mu^2)$ at $\mu=25~{\rm fm^{-1}}$ and using a momentum cutoff $\Lambda= 25~{\rm fm^{-1}}$. As one can observe, the initial SKM-LO ChEFT potential has non-zero off-diagonal matrix elements which extend up to high momenta. As the similarity cutoff $\lambda$ is lowered, the off-diagonal matrix elements are systematically suppressed while the low-momentum components are enhanced, similar to what happens for the Nijmegen potential. Thus, the result clearly shows that the SKM-LO ChEFT potential evolved through the SRG transformation is driven towards a band-diagonal form.

\begin{figure}[h]
\centerline{\includegraphics[width=8.2cm]{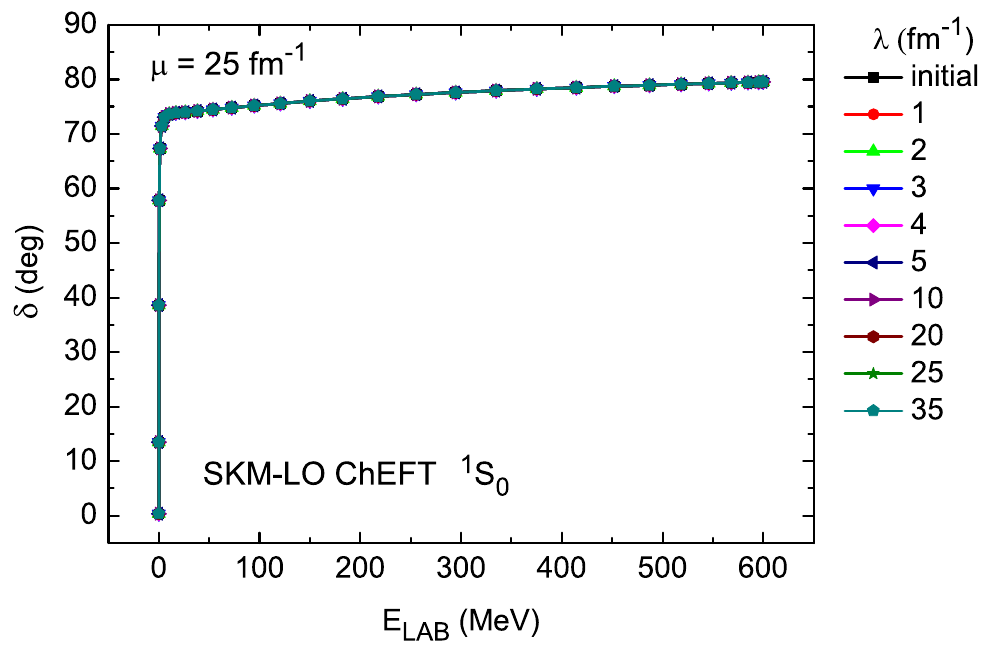}}
\caption{(Color online) Phase-shifts in the $^1 S_0$ channel as a function of the laboratory energy $E_{\rm LAB}$ for the initial SKM-LO ChEFT potential and the corresponding SRG potentials evolved to several values of $\lambda$.}
\label{fig22}
\end{figure}

In Fig. (\ref{fig22}) we show the $NN$ phase-shifts in the $^1 S_0$ channel as a function of $E_{\rm LAB}$ calculated from the numerical solution of the (formal) LS equation for the partial-wave $K$-matrix with the initial SKM-LO ChEFT potential and the corresponding potentials evolved through the SRG transformation to several values of the similarity cutoff $\lambda$. As expected, the SRG evolved potentials yield the same results as the initial potential for all energies, apart from relative differences smaller than $10^{-9}$ due to numerical errors.

\begin{figure}[h]
\centerline{\includegraphics[width=8.0cm]{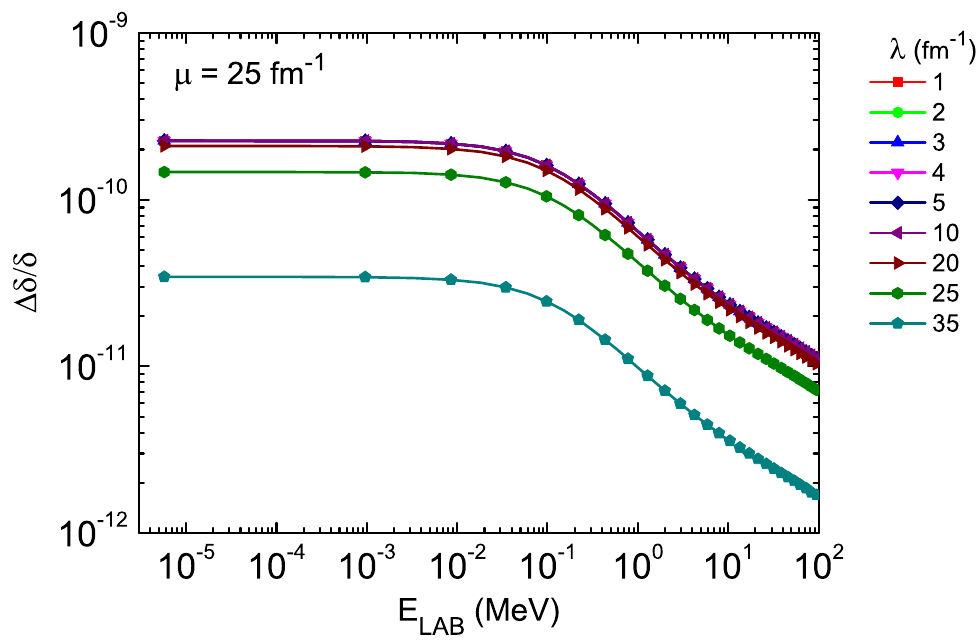}\hspace*{0.3cm}\includegraphics[width=8.2 cm]{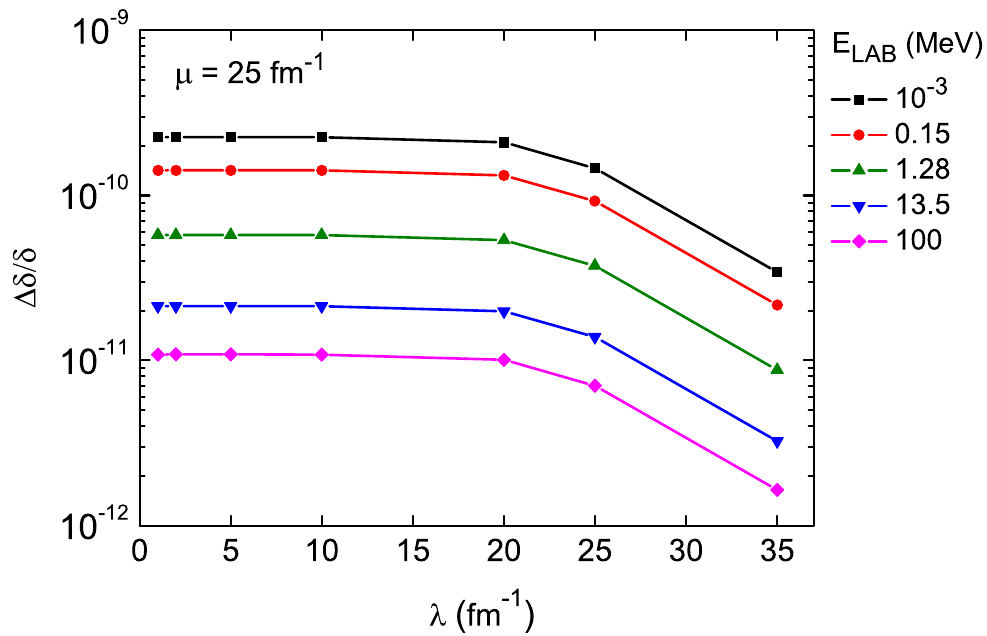}}
\caption{(Color online) Relative errors in the phase-shifts for the SRG potentials evolved from the initial SKM-LO ChEFT potential. Left panel: as a function of $E_{\rm LAB}$ for several values of $\lambda$; Right panel: as a function of $\lambda$ for several values of $E_{\rm LAB}$. }
\label{fig23}
\end{figure}

In Fig. (\ref{fig23}) we show the relative errors in the phase-shifts for the SRG evolved potentials (with respect to the results for the initial potential) as a function of $E_{\rm LAB}$ for several values of the similarity cutoff $\lambda$ (left) and as a function of the similarity cutoff $\lambda$ for several values of $E_{\rm LAB}$ (right). Similar to what is obtained for the SRG evolution of the two-dimensional Dirac-delta potential, two different scaling regions can be observed in the error plots for fixed $E_{\rm LAB}$ (right), with a crossover at $\lambda \sim \Lambda= 25~{\rm fm^{-1}}$.

\begin{figure}[t]
\centerline{\includegraphics[width=8.0cm]{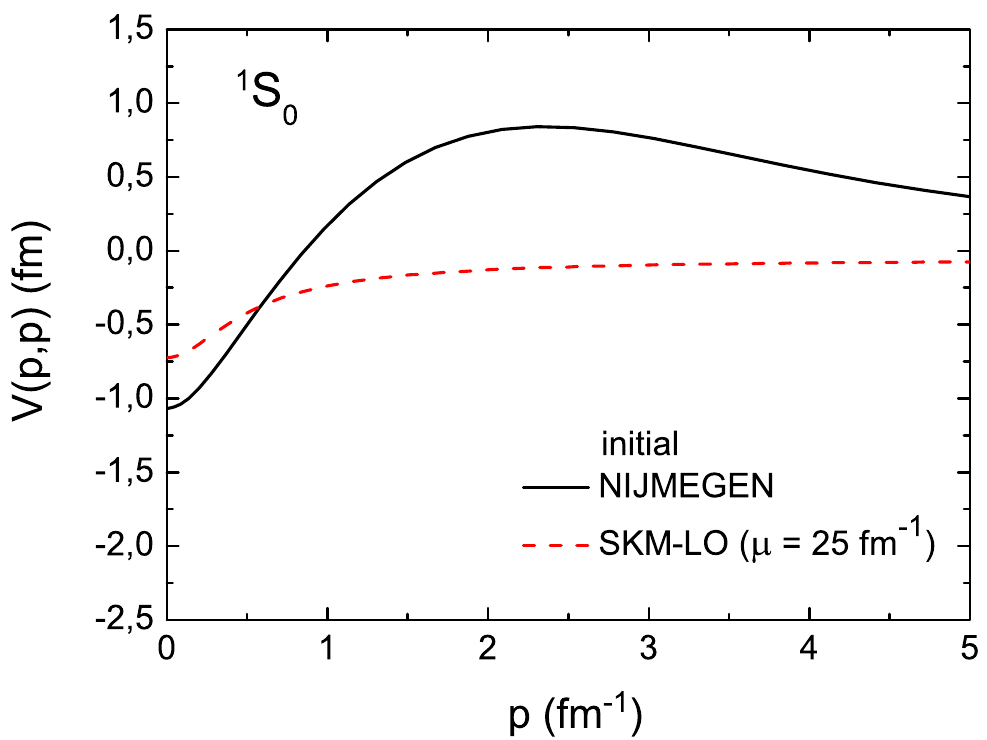}\hspace*{0.5cm}\includegraphics[width=8.0cm]{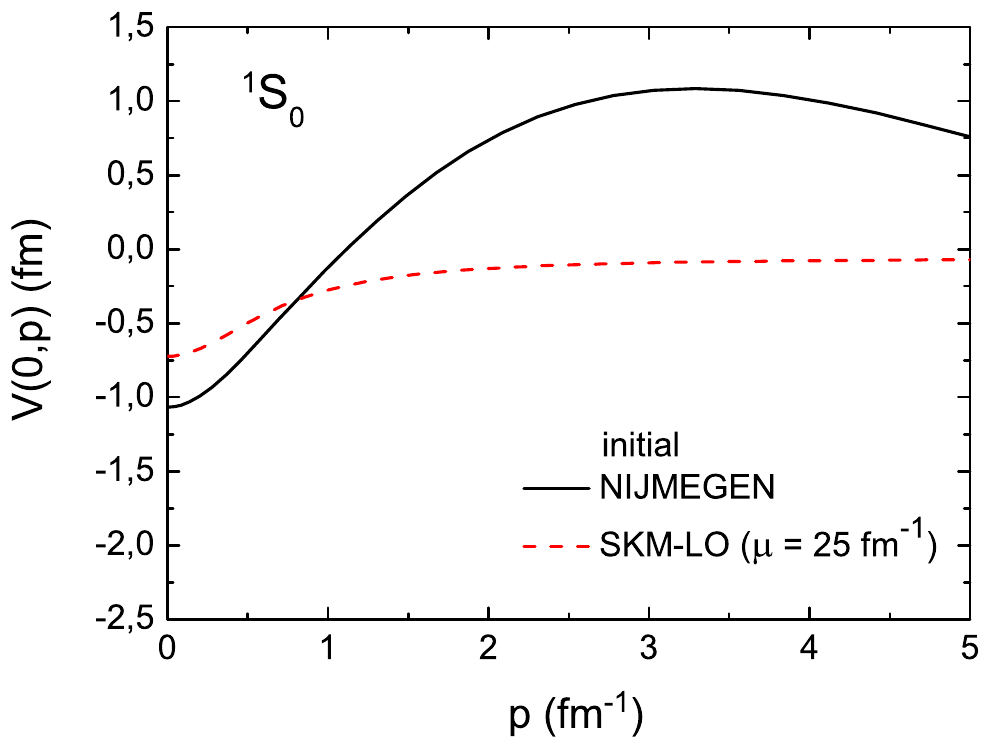}}\vspace*{.8cm}
\centerline{\includegraphics[width=8.0cm]{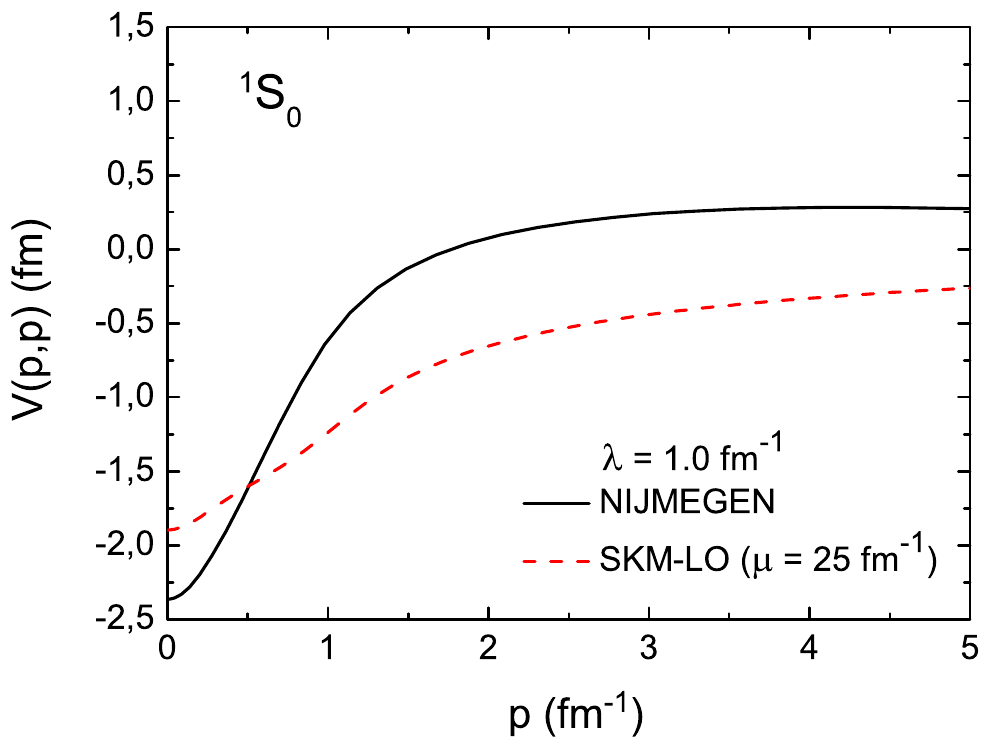}\hspace*{0.5cm}\includegraphics[width=8.0cm]{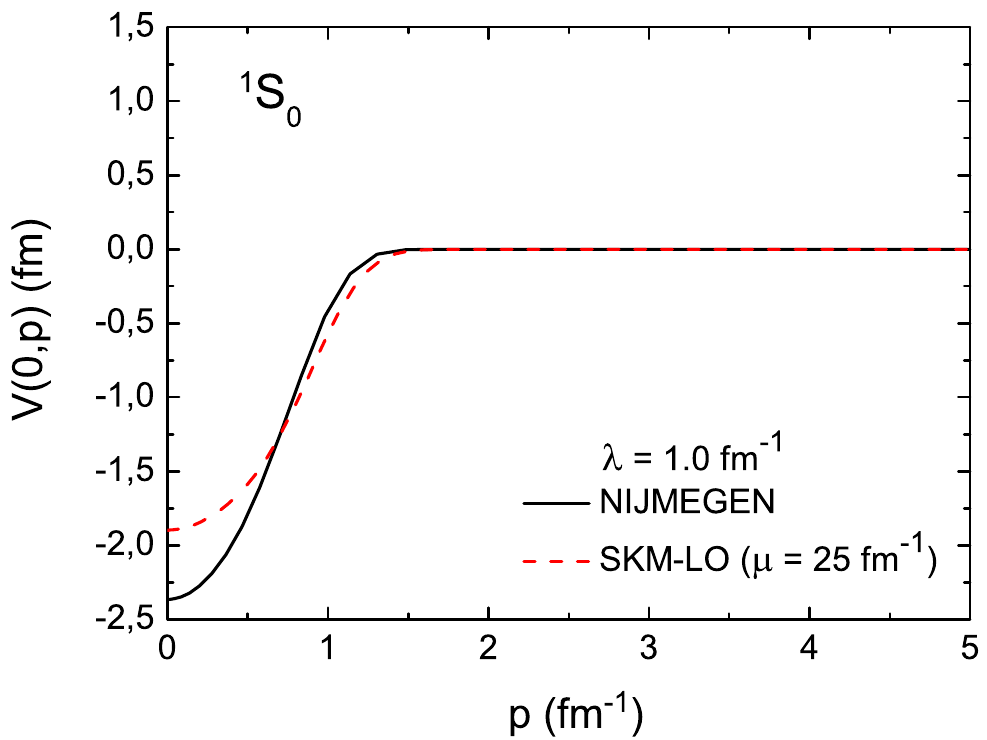}}
\caption{(Color online) SRG evolution of the Nijmegen and the SKM-LO ChEFT potentials in the $^1 S_0$ channel. Left: Diagonal matrix-elements; Right: Off-diagonal matrix-elements.}
\label{fig24}
\end{figure}

In Fig. (\ref{fig24}) we show a comparison between the momentum-space matrix-elements of the Nijmegen and the SKM-${\rm LO}$ ChEFT potentials in the $^1 S_0$ channel evolved through the SRG transformation. As one can observe, the potentials remain clearly distinct when the similarity cutoff is lowered to $\lambda=1~{\rm fm^{-1}}$. Such a behavior is expected, since the two initial potentials are not phase-shift equivalent (as shown before, the $NN$ phase-shifts obtained for the SKM-${\rm LO}$ ChEFT potential strongly deviates from those obtained for the Nijmegen potential). Nevertheless, one can see a tendency of the matrix-elements of the two SRG evolved potentials to become similar at low-momentum, in particular the off-diagonal matrix-elements.

Since the description of low-energy $NN$ observables provided by ChEFT potentials becomes systematically more accurate (i.e., approach the Nijmegen PWA results) at higher-orders in the chiral expansion (e.g., ${\rm NLO}$, ${\rm NNLO}$ and ${\rm N^3LO}$), low-momentum universality as described by Bogner et. al. \cite{srg1,vlowsrg} is expected to emerge as higher-order contributions are included in the initial SKM ChEFT potential, i.e. the low-momentum parts of the Nijmegen and the SKM ChEFT potentials evolved through the SRG transformation should flow to quantitatively similar forms as the similarity cutoff is lowered. This must be verified by explicit calculations of SRG evolved higher-order SKM ChEFT potentials, which we will perform in a future work.

\begin{figure}[ht]
\centerline{\includegraphics[width=7.7cm]{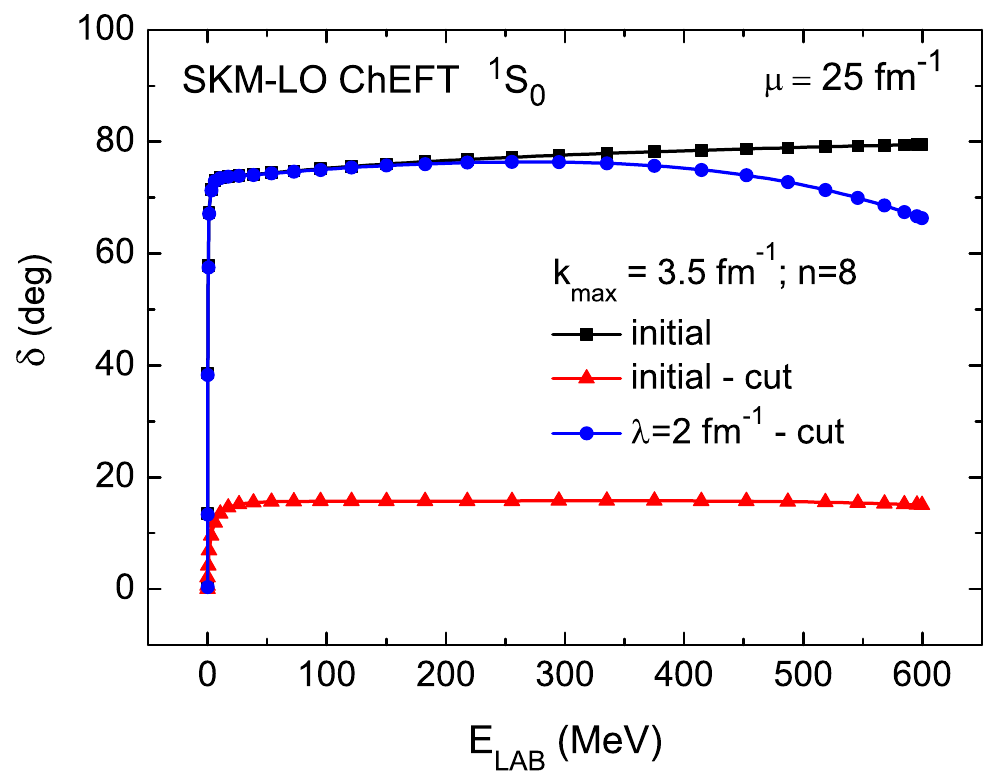}\hspace*{.8cm}\includegraphics[width=7.7cm]{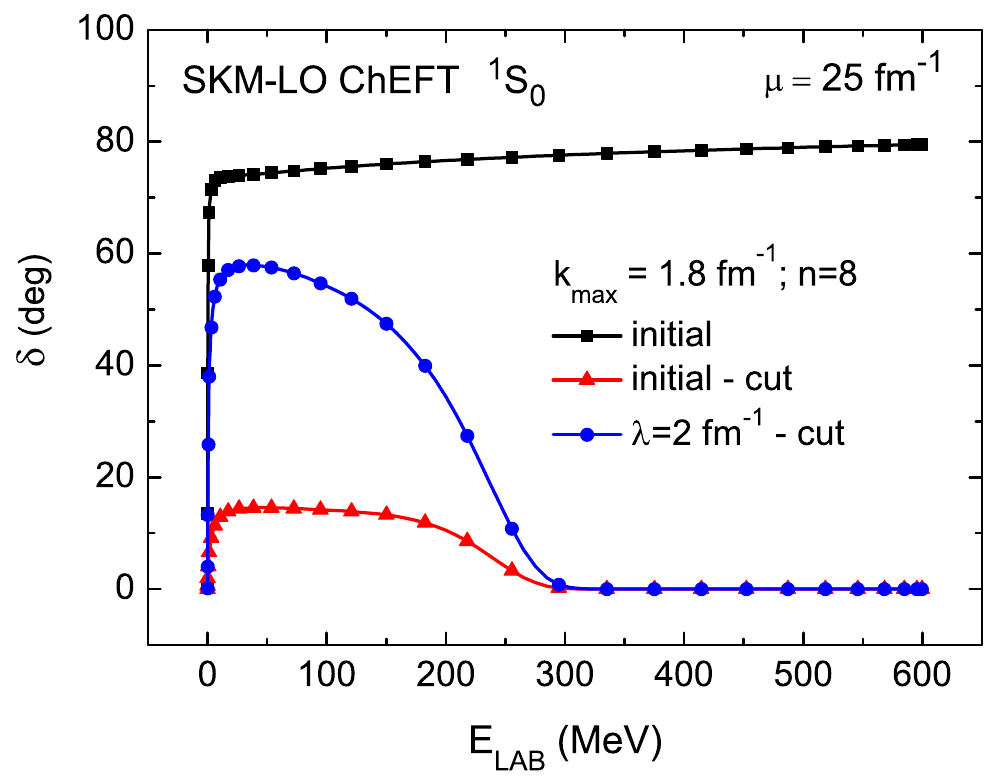}}
\caption{(Color online) Phase-shifts in the $^1 S_0$ channel for the initial SKM-LO ChEFT potential and the corresponding SRG potential evolved to $\lambda=2 \; {\rm fm}^{-1}$ with the application of an exponential regularizing function. Left: n=8 and $k_{max}=3.5 ~{\rm fm^{-1}}$; Right: n=8 and $k_{max}=1.8 ~{\rm fm^{-1}}$.}
\label{fig25}
\end{figure}

In order to analyze the decoupling between low- and high-momentum components in the SRG evolved SKM-LO ChEFT potential, we use the exponential regularizing function defined by Eq. (\ref{smooth}). In Fig. (\ref{fig25}) we show the results for the $NN$ phase-shifts in the $^1 S_0$ channel as a function of $E_{\rm LAB}$ obtained by cutting the initial SKM-LO ChEFT potential and the corresponding SRG potential evolved to a similarity cutoff $\lambda=2 \; {\rm fm}^{-1}$ at $k_{\rm max}=3.5 \; {\rm fm}^{-1}$ (left) and $k_{\rm max}=1.8 \; {\rm fm}^{-1}$ (right), with $n=8$. For the initial SKM-LO ChEFT potential, the phase-shifts obtained for the cut potential are completely modified in comparison to those obtained for the uncut potential.

The qualitative decoupling pattern for the SRG potential evolved from the SKM-LO ChEFT potential is similar to that obtained for the SRG potential evolved from the Nijmegen potential. The phase-shifts remain practically unchanged at low energies when the SRG evolved potential is cut above $\lambda$ and significantly change for all energies when it is cut below $\lambda$.

\begin{figure}[p]
\centerline{\includegraphics[width=8.0cm]{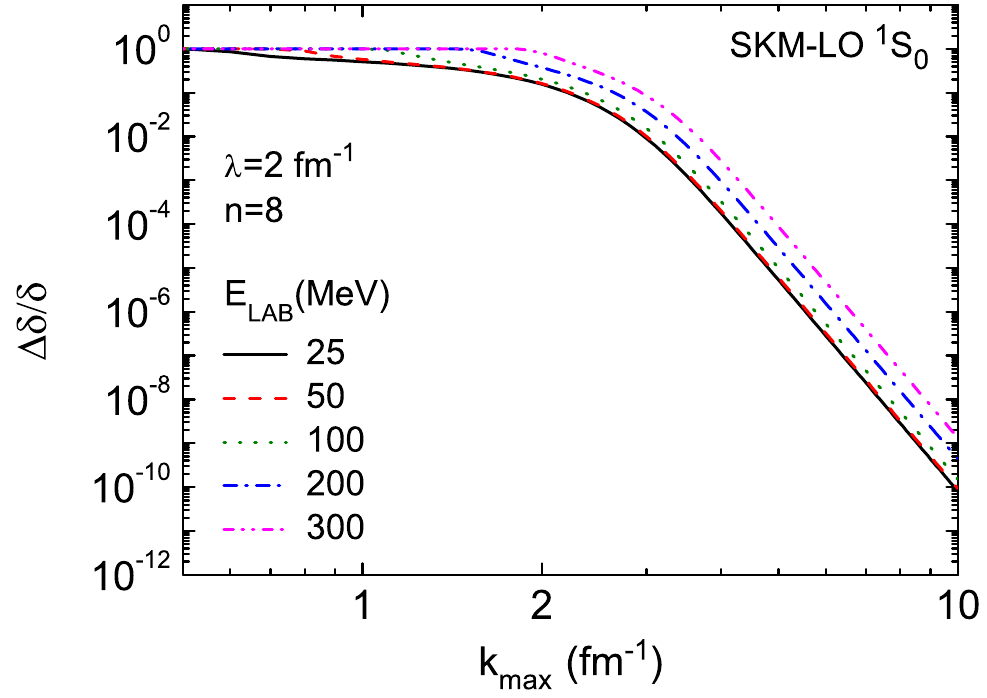}\hspace*{.5cm}\includegraphics[width=8.0cm]{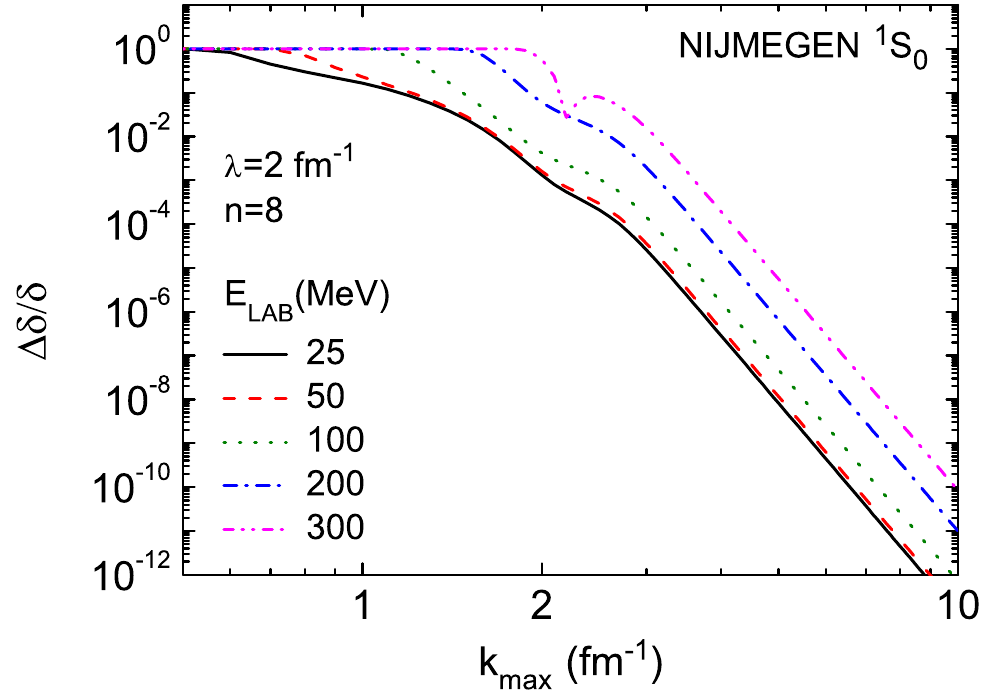}}\vspace*{.8cm}
\centerline{\includegraphics[width=8.0cm]{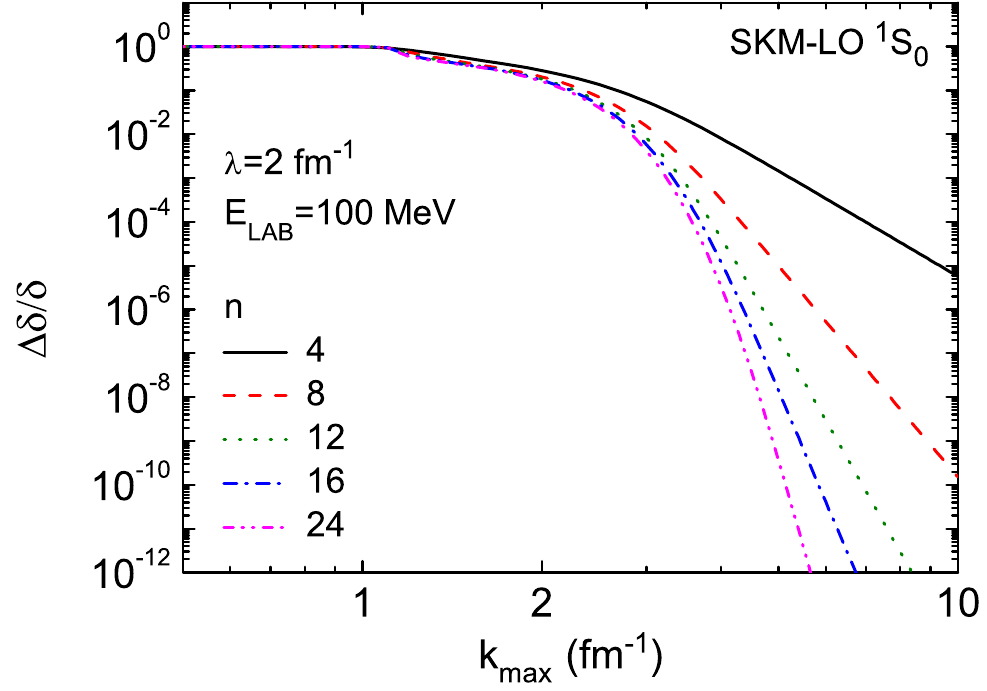}\hspace*{.5cm}\includegraphics[width=8.0cm]{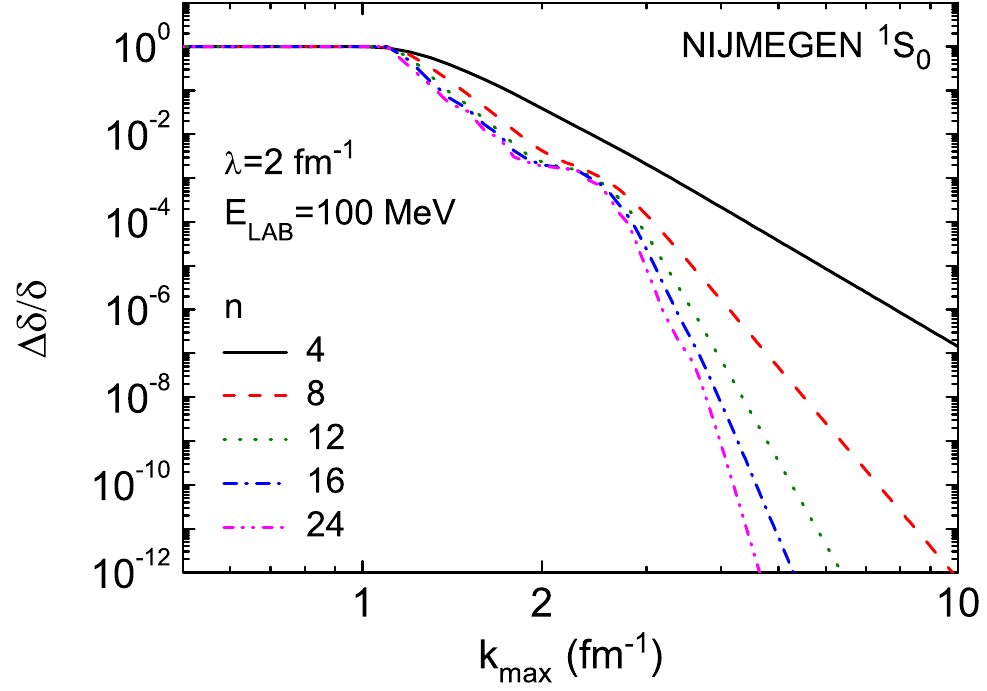}}\vspace*{.8cm}
\centerline{\includegraphics[width=8.0cm]{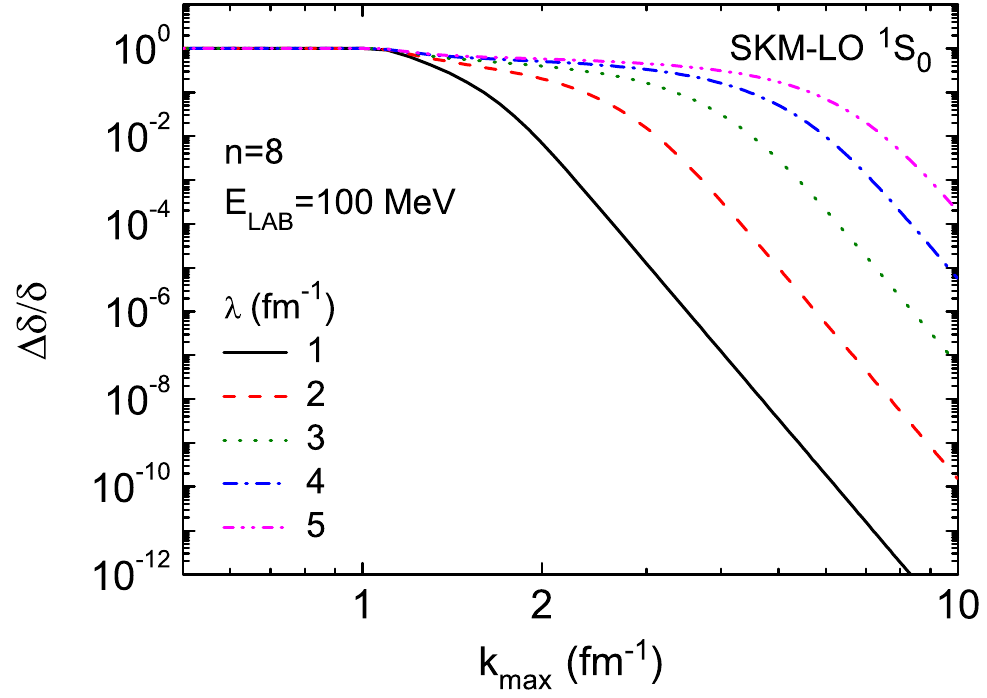}\hspace*{.5cm}\includegraphics[width=8.0cm]{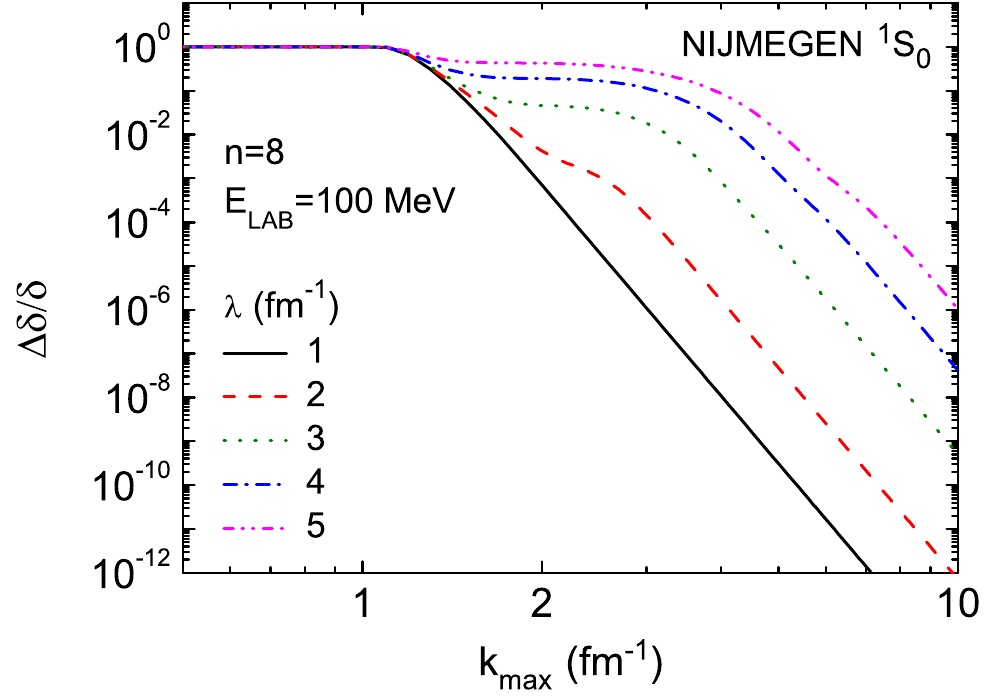}}
\caption{(Color online) Relative errors in the phase-shifts for the cut SRG evolved potentials as a function of the momentum cut, $k_{\rm max}$, for several values of $E_{\rm LAB}$, $n$ and $\lambda$. Left: SKM-LO ChEFT potential; Right: Nijmegen potential.}
\label{fig26}
\end{figure}

A more quantitative analysis of the decoupling pattern can be performed by evaluating the relative errors in the phase-shifts for the cut SRG evolved potential (with respect to the results for the corresponding uncut potential) as a function of the momentum cut $k_{\rm max}$, which provide a measurement of the degree of decoupling \cite{srg3}. In Fig. (\ref{fig26}) we show the log-log plots for the relative errors obtained for the cut SRG potentials evolved from the SKM-LO ChEFT potential (left panels) and from the Nijmegen potential (right panels), respectively for several values of $E_{\rm LAB}$, the regularizing function parameter $n$ and the similarity cutoff $\lambda$. As one can observe, there are three different regions in all error plots, both for the SRG potentials evolved from the SKM-LO ChEFT and from the Nijmegen potential. As expected, when $k_{\rm max} < \sqrt{M E_{\rm LAB}/2}$ the relative errors go to one, since in this case the matrix-elements of the cut potentials vanish such that the corresponding phase-shifts go to zero. When $k_{\rm max} > {\lambda}$ the relative errors scale as a power-law $(1/k_{\rm max})^{2n}$, indicating a perturbative decoupling regime. This power-law scaling behavior is clearly evidenced by the straight lines with slope $- 2n$ in the error plots. It is important to note that the crossover to the power-law scaling region typically occurs at $k_{\rm max}$ slightly above $\lambda$, independently of the values of $E_{\rm LAB}$ and $n$. In the intermediate region the scaling behavior is not definite, becoming different for the SRG potentials evolved from the SKM-LO ChEFT and from the Nijmegen potential.

\begin{figure}[t]
\centerline{\includegraphics[width=8.0cm]{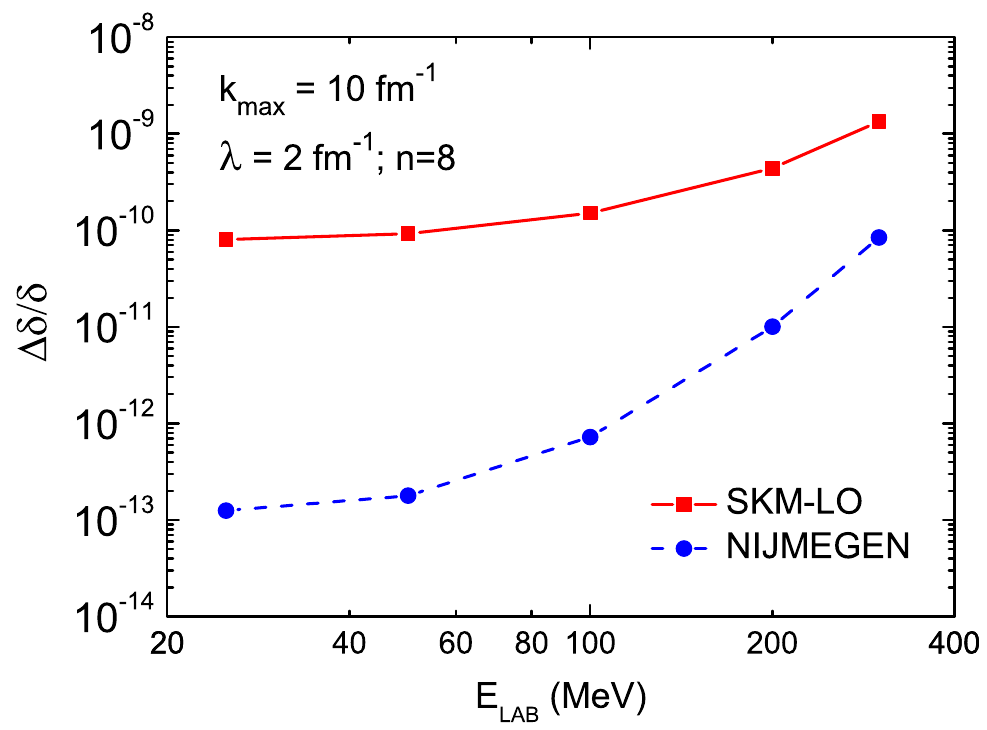}\hspace*{.6cm}\includegraphics[width=7.8cm]{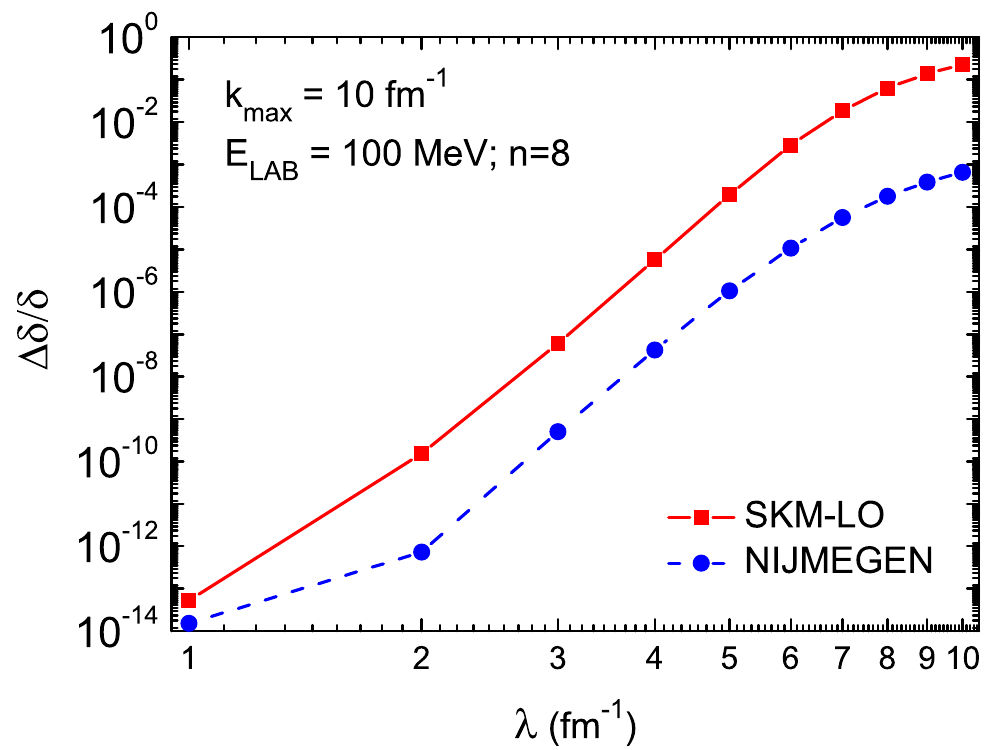}}\vspace*{.4cm}
\caption{(Color online) Relative errors in the phase-shifts for the SRG evolved potentials cut at $k_{\rm max}=10~{\rm fm^{-1}}$ with $n=8$. Left: as a function of $E_{\rm LAB}$ for fixed $\lambda=2 ~{\rm fm^{-1}}$; Right: as a function of $\lambda$ for fixed $E_{\rm LAB}={\rm 100 ~MeV}$.}
\label{fig27}
\end{figure}

In Fig. (\ref{fig27}) we show the relative errors in the phase-shifts for the SRG evolved potentials cut at $k_{\rm max}=10~{\rm fm^{-1}}$ with $n=8$ as a function of $E_{\rm LAB}$ for fixed $\lambda=2 ~{\rm fm^{-1}}$ (left) and as a function of $\lambda$ for fixed $E_{\rm LAB}={\rm 100~MeV}$ (right). As one can observe in the left panel, the dependence of the relative errors on $E_{\rm LAB}$ is weak for $E_{\rm LAB} \leq {\rm 100~MeV}$. In the right panel one can observe that the dominant relative errors scale as a power-law $\lambda^{2n}$ in the range $2~{\rm fm^{-1}}< \lambda < 6 ~{\rm fm^{-1}}$ for the SRG potentials evolved from the SKM-LO ChEFT and in the range $2~{\rm fm^{-1}}< \lambda < 5~{\rm fm^{-1}}$ for the SRG potentials evolved from the Nijmegen potential.

The general results we have obtained in the quantitative analysis of the decoupling pattern for the SRG potentials evolved from the SKM-LO ChEFT potential and from the Nijmegen potential in the $^1 S_0$ channel are similar to those described in Ref. \cite{srg3} for the SRG potentials evolved from the chiral ${\rm N^3LO}$ potential with a $500~{\rm MeV}$ cutoff \cite{cheft23}. However, there is a difference which is worth to mention. As shown in Ref. \cite{srg3}, the change in the phase-shift relative errors for the cut SRG evolved potentials saturates when the similarity cutoff $\lambda$ approaches the underlying ${\rm 500~MeV}~(\sim 2.54~{\rm fm^{-1}})$ momentum cutoff of the initial chiral ${\rm N^3LO}$ potential, in particular the change in the position of the shoulder in the error plots indicating the crossover to the power-law scaling region. Here, as one can observe in the bottom panels of Fig. (\ref{fig26}), there is much less saturation for similarity cutoffs $\lambda$ up to $\sim 5~{\rm fm^{-1}}$, because the initial SKM-LO ChEFT potential has no underlying cutoff (except for the large instrumental cutoff $\Lambda= 25~{\rm fm^{-1}}$), going to a small constant value for momenta larger than $\sim 6~{\rm fm^{-1}}$, and the initial Nijmegen potential has an effective momentum cutoff $\sim 5~{\rm fm^{-1}}$ (which results from the combination of the several gaussian cutoff parameters used to regularize the terms associated to the different meson exchange processes included in the potential \cite{nijmegen}). Indeed, as one can see in the right panel of Fig. (\ref{fig27}), the saturation effect is observed only for values of $\lambda$ larger than $\sim 5-6~{\rm fm^{-1}}$.

\section{Summary and Conclusions}
\label{CONCL}

So far, we have performed a detailed investigation of the similarity renormalization group (SRG) evolution of the $NN$ interaction in leading order chiral effective field theory (ChEFT) within the subtractive kernel method (SKM) renormalization approach. In the following, we summarize the results of our study and present our main conclusions.

First, we considered the two-dimensional Dirac-delta contact interaction, aiming to illustrate the application of the SRG through a simple example and to fine-tune the parameters of the Runge-Kutta solver used in our code to obtain an accurate numerical solution of Wegner's SRG flow equation. The matrix elements of the two-dimensional Dirac-delta potential in momentum space are constants, thus generating ultraviolet logarithmic divergences when iterated in the LS or the Schr{\"o}dinger equation. Nevertheless, exact analytic solutions can be obtained by introducing a momentum cutoff and renormalizing the coupling constant that characterizes the strength of the contact interaction in order to remove cutoff dependence in the observables, leading to dimensional transmutation and asymptotic freedom. We solved Wegner's flow equation for the two-dimensional Dirac-delta potential renormalized by demanding that the binding-energy remains fixed as the momentum cutoff is removed and verified that the SRG evolved potential is driven to a band-diagonal form as the similarity cutoff $\lambda$ is reduced. From the numerical solution of the LS equation for the $K$-matrix we calculated the phase-shifts as a function of $E_{\rm LAB}$ for the initial two-dimensional Dirac-delta potential and for the corresponding SRG potentials evolved to several values of the similarity cutoff $\lambda$ and verified the unitarity of the SRG transformation up to relative numerical errors smaller than $10^{-9}$. Two scaling regions were observed in the error plots for fixed $E_{\rm LAB}$: the relative errors are nearly constant for $\lambda < \Lambda$ and significantly decrease for $\lambda > \Lambda$. Such a change in the scaling behavior of the errors is a numerical artifact related to the number of steps required by the Runge-Kutta solver to evolve the potential.

Next, we considered the SRG evolution of the high-precision Nijmegen $NN$ potential in the $^1 S_0$ channel. We also verified the band-diagonalization of the SRG evolved potential and, by evaluating the phase-shifts, the unitarity of the SRG transformation (up to relative numerical errors smaller than $10^{-9}$). Then, following the method described by Bogner et al. \cite{srg2,srg3}, we used an exponential regularizing function to smoothly cut the SRG evolved potentials above a given momentum $k_{\rm max}$. We verified that, as a consequence of the decoupling of low-energy observables from high-energy degrees of freedom, the phase-shifts calculated with the SRG evolved potentials cut at a momentum $k_{\rm max}$ above the similarity cutoff $\lambda$  agree with those obtained for the corresponding uncut potentials up to $E_{\rm LAB}\sim 2 \; k_{\rm max}^2/M$.

Then, we presented a discussion of the SKM renormalization approach. As a first example, we considered the $NN$ interaction in ${\rm LO}$ pionless EFT, which simply consists of a pure Dirac-delta contact term. In this case, only one subtraction in enough to renormalize the interaction and the subtracted kernel LS equation can be solved analytically to obtain the $T$-matrix. The renormalized strength of the  contact interaction is fixed by fitting the experimental value of the scattering length. Our result for the scattering amplitude in the $^1 S_0$ channel, obtained from the subtracted kernel LS equation, is independent of the subtraction scale $\mu$ and exactly matches the one obtained by solving the formal LS equation regularized using a momentum cutoff $\Lambda$ and then taking the limit when such a cutoff goes to infinity. Next, as an example to illustrate the application of the SKM renormalization approach when multiple subtractions are performed in the kernel of the LS equation, we considered the $NN$ interaction in ${\rm NLO}$ pionless EFT. In this case three subtractions are required, since the ${\rm NLO}$ interaction also includes a first-derivative contact term, and so a 3-fold subtracted kernel LS equation must be recursively constructed. Using the method described in Refs. \cite{birse1,cohen6,gegelia2}, the 3-fold subtracted kernel LS equation can still be solved analytically to obtain the scattering amplitude. The renormalized strengths of the two contact interactions are determined by matching the calculated scattering amplitude and the ERE to order $k^2$ (where $k$ is the on-shell momentum), such that their values at a given subtraction scale $\mu$ are fixed by fitting the scattering length and the effective range in $^1S_0$ channel. We found that in order to get real values for the renormalized strengths, thus preserving the unitarity of the $T$-matrix, the value of subtraction scale $\mu$ must be less than $\sim 0.8~{\rm fm}^{-1}$, a result which is consistent with the Wigner bound on the effective range. Moreover, the dominant relative errors in the phase-shifts for the ${\rm NLO}$ potential with respect to the results obtained from the ERE to order $k^2$ scale, as expected, like $(k/\mu)^4$.

We then addressed the SRG evolution of chiral $NN$ potentials renormalized within the framework of the SKM. In our calculations, we implemented the SKM procedure by using the $K$-matrix instead of the $T$-matrix. For numerical convenience when evolving the potential through the SRG transformation, we introduced an instrumental momentum cutoff $\Lambda$ in order to work with a finite-range gaussian grid of integration points.

First, we performed a detailed analysis of the renormalization of the $NN$ interaction in ${\rm LO}$ ChEFT through the SKM approach. The ${\rm LO}$ ChEFT potential consists of the OPEP plus a Dirac-delta contact interaction and so only one subtraction is required. From the solution of the subtracted kernel LS equation in the $^1S_0$ channel, we evaluated the on-shell $K$-matrix at $E=k^2=0$ and by fitting the scattering length in the $^1S_0$ channel we investigated the running of the renormalized strength of the contact interaction with the subtraction scale $\mu$ and with the instrumental cutoff $\Lambda$. For a fixed $\Lambda$ the renormalized strength increases with $\mu$ and in the limit $\mu\rightarrow \infty$ it converges asymptotically to the same value that is obtained by using the sharp cutoff regularization scheme. Similarly, for a fixed $\mu$ the renormalized strength increases with $\Lambda$ and in the limit $\Lambda\rightarrow \infty$ its dependence on $\Lambda$ vanishes asymptotically, evidencing the instrumental nature of such a regulator. We also found that the precision to which the renormalized strength must be tuned in order to fit the scattering length with a given accuracy becomes higher as both $\mu$ and $\Lambda$ increase.

Next, we evaluated the driving term and the corresponding fixed-point renormalized potential for several values of the subtraction scale $\mu$ with the instrumental cutoff $\Lambda$ fixed. As expected, the driving term runs with $\mu$ such that the fixed-point renormalized potential remains approximately invariant. In the limit $\mu\rightarrow \infty$ the driving term becomes independent of $\mu$, matching the fixed-point renormalized potential. We found that the relative differences between the driving term at a given $\mu$ and at $\mu\rightarrow \infty$ scale like $1/\mu^2$ for fixed momenta. The relative differences for the corresponding fixed-point renormalized potentials are about two orders of magnitude smaller, also scaling like $1/\mu^2$. The residual dependence of the fixed-point renormalized potential on $\mu$ is a consequence of the fitting procedure used to fix the renormalized strength of the contact interaction and should be eliminated by evolving the driving term from a reference subtraction scale to the subtraction scale $\mu$ via the NRCS equation. Then, we calculated the phase-shifts in the $^1S_0$ channel as a function of $E_{\rm LAB}$ for several values of the subtraction scale $\mu$ and the instrumental cutoff $\Lambda$. The relative differences between the phase-shifts calculated at a given $\mu$ and at $\mu\rightarrow \infty$ (with $\Lambda$ fixed) scale like $E_{\rm LAB} (\propto k^2)$ for fixed $\mu$ and like $1/\mu^2$ for fixed $E_{\rm LAB}$. The relative differences between the phase-shifts calculated at a given $\Lambda$ and at $\Lambda\rightarrow \infty$ (with $\mu$ fixed) scale like $E_{\rm LAB} (\propto k^2)$ for fixed $\Lambda$ and like $1/\Lambda$ for fixed $E_{\rm LAB}$.

Finally, we moved to the calculations of the SRG evolution of the SKM-${\rm LO}$ ChEFT potential in the $^1S_0$ channel. We solved Wegner's flow equation using as an input the fixed-point renormalized potential derived through the SKM approach. As expected, the SRG evolution systematically suppresses the off-diagonal matrix elements as the similarity cutoff $\lambda$ is lowered, driving the potential towards a band diagonal form. By evaluating the phase-shifts, we also verified the unitarity of the SRG transformation (apart from relative numerical errors smaller than $10^{-9}$). Like in the case of the SRG evolution of the two-dimensional Dirac-delta potential, two scaling regions were observed in the relative error plots for fixed $E_{\rm LAB}$, with a crossover at $\lambda \sim \Lambda$. By comparing the matrix-elements of the Nijmegen and the SKM-${\rm LO}$ ChEFT potentials evolved through the SRG transformation, we verified that the potentials remain distinct as the similarity cutoff $\lambda$ is lowered, showing only a tendency to become similar at low-momentum. This result is expected, since the two initial potentials are not phase-shift equivalent. Nevertheless, universality is expected to emerge as higher-order contributions in the chiral expansion are included in the initial SKM ChEFT potential. As previously described for the Nijmegen potential, we analyzed the decoupling of low-energy observables from high-energy degrees of freedom by using an exponential regularizing function to cut the potential above a given momentum $k_{\rm max}$. We found a qualitative decoupling pattern for the SRG evolved SKM-${\rm LO}$ ChEFT potential similar to that obtained for the SRG evolved Nijmegen potential. In order to perform a quantitative analysis, we evaluated the relative errors in the phase-shifts for the cut SRG evolved potentials as a function of $k_{\rm max}$. From the log-log error plots we verified that for $k_{\rm max} > \lambda$ the relative errors scale like a power-law $(1/k_{\rm max})^{2n}$ both for the SKM-${\rm LO}$ ChEFT and the Nijmegen potentials, indicating a perturbative decoupling regime.

The main purpose of this work was to set up a basis for a comparative study of ChEFT $NN$ potentials using the SRG transformation, particularly through the SRG decoupling pattern analysis introduced by Bogner et al. \cite{srg2,srg3}. The simple examples considered here indicate that the SRG provides a powerful tool for analyzing the scale dependence of effective $NN$ interactions and for estimating the uncertainties in calculations of $NN$ systems.

Our next steps, to be pursued in forthcoming works, will be to study the SRG evolution of ChEFT $NN$ potentials renormalized via the SKM approach up to higher-orders in the chiral expansion (${\rm NLO}$, ${\rm NNLO}$ and ${\rm N^3LO}$) and in other partial-wave channels (including calculations of phase-shifts and deuteron observables) and to consider the effects of using different generators for the SRG transformation. We will compare the results for the SKM ChEFT $NN$ potentials with those for ChEFT $NN$ potentials renormalized using cutoff regularization and investigate the interplay between the power counting (and hence the systematic improvement) and the non-perturbative renormalization using these two schemes. Another important issue to be addressed in such a study will be to verify how low-momentum universality of the SRG evolved potentials emerges as the higher-order contributions in the chiral expansion are included in the initial potentials and what are the effects of using different regularization schemes. We will also investigate the renormalization group invariance in the SKM formalism, by solving the non-relativistic Callan-Symanzik equation (NRCS) for the driving term of the subtracted LS equations (both in pionless EFT and ChEFT), and its possible extension to renormalize $3N$ interactions. Then, we intend to perform a detailed comparative analysis of several accurate ChEFT $NN$ potentials, including the chiral ${\rm N^3LO}$ potentials of Epelbaum, Gl\"ockle and Mei{\ss}ner \cite{cheft20} and of Entem and Machleidt \cite{cheft23}, the chiral ${\rm N^3LO}$ potential renormalized using the SKM approach and the brazilian chiral potential \cite{cheft34}, as well as the Argonne V18 potential \cite{argonne} and the Nijmegen potential \cite{nijmegen} which are to be used as a baseline for the analysis. One of our motivations in this study is to investigate the effects on $NN$ observables due to the mid-range correlated two-pion exchange terms present in the brazilian chiral potential, which we expect may be better understood through the scale-dependence analysis provided by the SRG.

\section*{Acknowledgments}
This work was supported by CNPq, FAPESP and Instituto Presbiteriano Mackenzie through Fundo Mackenzie de Pesquisa. We would like to thank M. R. Robillota for many useful discussions and E. Epelbaum for a careful reading of the manuscript.

\end{document}